 \documentclass[sigconf]{acmart}

\AtBeginDocument{%
  \providecommand\BibTeX{{%
    Bib\TeX}}}

\usepackage{amsmath,amsfonts}
\usepackage{algorithmicx}
\usepackage{algorithm}
\usepackage{algpseudocode}
\usepackage{graphicx}
\usepackage{textcomp}
\usepackage{xcolor}
\usepackage{adjustbox}
\usepackage{enumitem}
\usepackage{MnSymbol}
\usepackage{adjustbox}
\usepackage{varwidth}
\usepackage[normalem]{ulem}
\usepackage{bm}
\usepackage{makecell}
\usepackage{marginnote}
\usepackage{lineno}

\newcommand{\from}[2]{
	  \textcolor{red}
	{\bf [{\sc from #1:} #2]}}

\newcommand{\note}[1]{%
}

\newcommand{\noteA}[1]{%
}

\newcommand{\noteB}[1]{%
}

\newcommand{\noteC}[1]{%
}

\newcommand{\replace}[2]{{}{\color{black}#2}}

\newcommand{\replaceA}[2]{{}{\color{black}#2}}

\newcommand{\replaceB}[2]{{}{\color{black}#2}}

\newcommand{\replaceC}[2]{{}{\color{black}#2}}

\renewcommand{\times}{\cdot}

\newcommand{\dpath}{\rightarrow \ldots \rightarrow}

\newtheorem{definition}{Definition}
\newtheorem{observation}{Key Observation}
\newtheorem{example}{Example}

\newcommand{\maybe}[1]{}

\newcommand{\nsbp}{\nobreak\kern\fontdimen2\font\relax}

\newcommand{\default}[1]{#1}

\def\diaend{\hspace*{\fill} $\diamond$}
\def\circend{\hspace*{\fill} $\circ$}

\def\BibTeX{{\rm B\kern-.05em{\sc i\kern-.025em b}\kern-.08em
    T\kern-.1667em\lower.7ex\hbox{E}\kern-.125emX}}

\setcopyright{none}
\renewcommand\footnotetextcopyrightpermission[1]{}

\begin{document}
\pagestyle{plain}

\title{Causal Search for Skylines (CSS): Causally-Informed Selective Data De-Correlation}


\author{Pratanu Mandal}
\orcid{0009-0007-0529-2179}
\email{pmandal5@asu.edu}
\affiliation{
    \institution{Arizona State University}
    \city{Tempe}
    \state{Arizona}
    \country{USA}
}

\author{Abhinav Gorantla}
\orcid{0009-0003-1242-5671}
\email{agorant2@asu.edu}
\affiliation{
    \institution{Arizona State University}
    \city{Tempe}
    \state{Arizona}
    \country{USA}
}

\author{K. Sel\c{c}uk Candan}
\orcid{0000-0003-4977-6646}
\email{candan@asu.edu}
\affiliation{
    \institution{Arizona State University}
    \city{Tempe}
    \state{Arizona}
    \country{USA}
}

\author{Maria Luisa Sapino}
\orcid{0000-0002-7621-3753}
\email{mlsapino@di.unito.it}
\affiliation{
    \institution{University of Turin}
    \city{Turin}
    \country{Italy}
}

\renewcommand{\shortauthors}{Mandal et al.}

 




\begin{abstract}
Skyline queries are popular and effective tools in multi-criteria decision support as they extract \textit{interesting} (pareto-optimal) points that help summarize the available data  with respect to a given set of preference attributes.
Unfortunately, the efficiency of the skyline algorithms depends heavily on the underlying data statistics.
%
In this paper, we argue that the efficiency of the skyline algorithms could be significantly boosted if one could erase any attribute correlations that do not agree with the preference criteria, while preserving (or even boosting) correlations that agree with the user provided criteria.
Therefore, we propose a {\em causally-informed selective de-correlation} mechanism to enable
skyline algorithms to better leverage the pruning opportunities provided by the positively-aligned data distributions, without having to suffer from the mis-alignments.
In particular, we show that, given a causal graph that describes the underlying causal structure of the data, one can identify a subset of the attributes that can be used to selectively de-correlate the preference attributes.
Importantly, the proposed {\em causal search for skylines (CSS)} approach is agnostic to the underlying candidate enumeration and pruning strategies and, therefore, can be leveraged to improve any popular skyline discovery algorithm.
%
Experiments on multiple real and synthetic data sets and for different skyline discovery algorithms  show that the proposed causally-informed selective de-correlation technique significantly reduces 
both the number of dominance checks as well as the overall time needed to 
locate skyline points.
\end{abstract}

\maketitle


\section{Introduction}\label{sec:intro}
Skyline queries are popular and effective tools in decision support. Intuitively, a skyline query extracts \textit{interesting} (pareto-optimal) points that help summarize the available data question with respect to a given set of preference attributes, providing insights into the diversity of the data across these attributes. Given a set, $D$, of data points in an attribute/feature\footnote{We use the terms "attribute", "feature" , and "variable" interchangeably.
} space, $A$, and a set, $P \subseteq A$ of preference attributes, the {\em skyline} of $D$ consists of the subset of points that are not dominated\footnote{A point dominates another point if it is as good or better in all preference attributes, and better in at least one attribute.} by any other data point in
$D$~\cite{Borzsonyi01theskyline}.  Searching for such non-dominated data is valuable in many applications that involve multi-criteria decision making~\cite{steuer86a}.

\begin{example}[House hunting]\label{ex:house}
Let us consider a user who aims to identify a subset of homes 
in the market with low ${\tt Price}$  and low $\tt Commute$. This task can be formulated as searching a non-dominated subset of houses in the market with respect to the preference attributes, $P= \{{\tt Price}, {\tt Commute}\}$ and the corresponding preference criteria, $\Theta= \{\theta_{\tt Price}, \theta_{\tt Commute}\}$, where $\theta_{\tt Price} = min$ and $\theta_{\tt Commute} = min$.
\circend
\end{example}

\begin{figure}[t]
\centering{%
   \begin{tabular}{cc}
    \includegraphics[width=0.3\columnwidth]{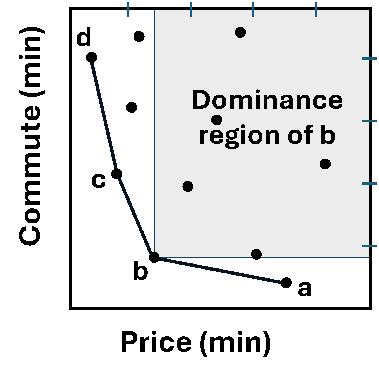}  &
    \includegraphics[width=2in]{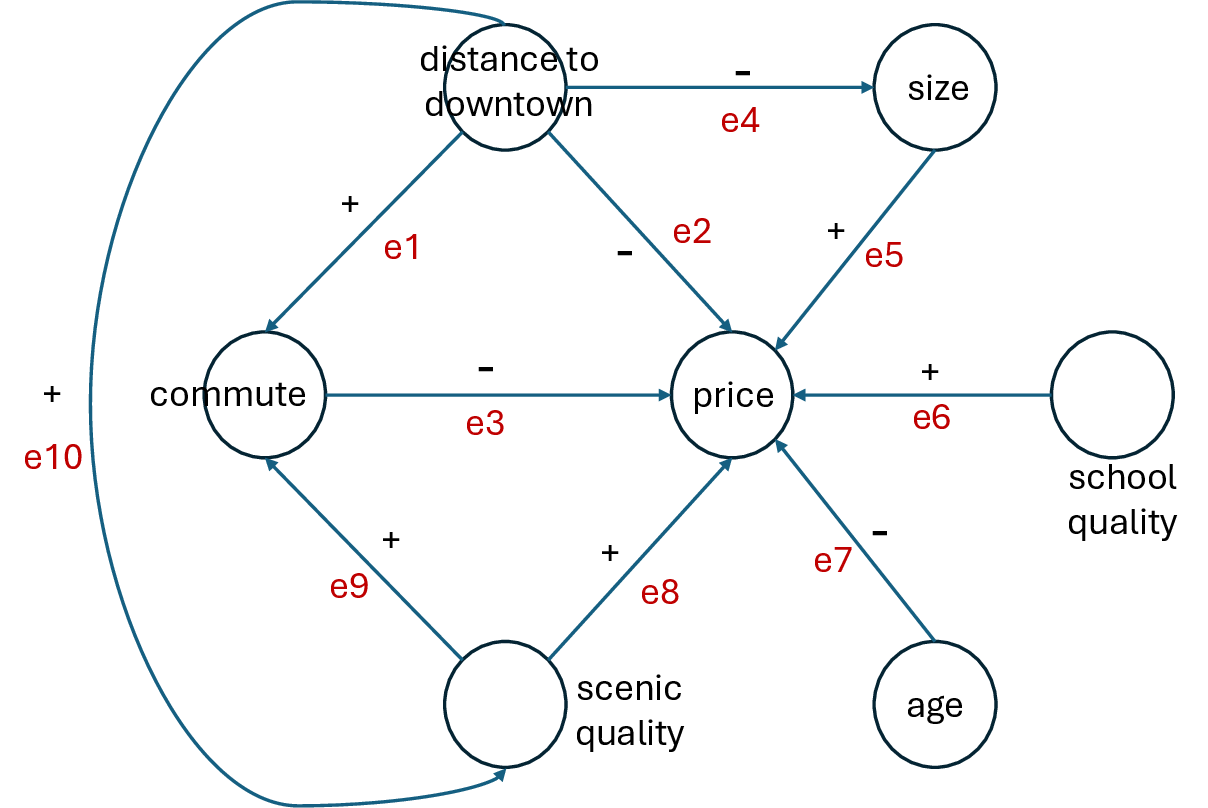}\\
    (a) skyline& 
    (b) causal graph\\
   \end{tabular}
}
\caption{Running example: skylines for house hunting }\label{fig:cg_houses}
\end{figure}

Based on these specifications, any house that is in the skyline subset will either have a lower ${\tt Price}$ or lower ${\tt Commute}$ than all the other houses in the data set; in other words, the user cannot find any houses that are not in the skyline but have both lower ${\tt Price}$ and lower ${\tt Commute}$ than the houses in the skyline.
In Figure~\ref{fig:cg_houses}(a), the shaded area indicates the dominance region of the house $b$: for any house in this preference range, the house $b$ is either cheaper or has lesser commute. Therefore, $b$ can be said to be more {\em useful} (with respect to the preference criteria) than all houses that it dominates.

\subsection{Challenge: Impact of the Data Distribution on Skyline Discovery Performance}

\begin{figure}[t]
\centering{
   \begin{tabular}{cc}
    \includegraphics[width=0.35\columnwidth]{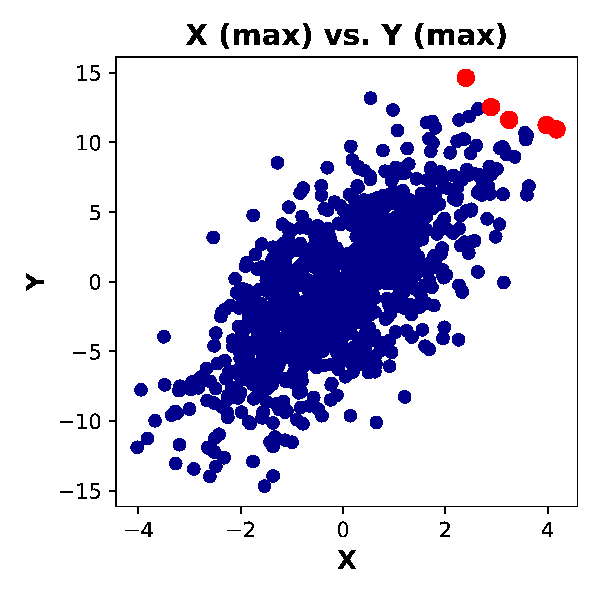}  &
    \includegraphics[width=0.35\columnwidth]{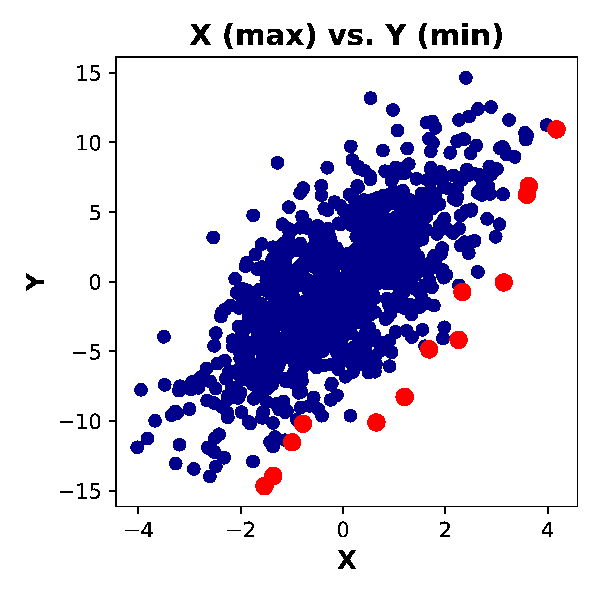}  \\
    (a) max-max with '+' corr& 
    (b) max-min with '+' corr \\
   \end{tabular}
   }
\caption{Impact of the alignment between data distribution and preference criteria:
in 
(a) the dominance region is determined by a few extreme points, whereas in 
(b) there are a large number of skyline points on the \replace{p}{P}areto \replace{border}{front}
}\label{fig:alignment}
\end{figure}

As we will further discuss in Section~\ref{sec:rel}, in the literature there are  several algorithms developed for identifying skyline data sets. These include nested-loop skylines~\cite{Borzsonyi01theskyline}, sort-first skylines~\cite{Chomicki03skylinewith}, and index-supported skylines~\cite{TanEO01}. Intuitively, all of these algorithms rely on an enumerate-and-prune strategy: they enumerate candidate skyline objects in the data set and they use dominance-checks to prune-away those objects that should not be included in the skyline set. 
The primary goal of these algorithms is to enumerate the skyline candidates in such a way that, the  data set can be pruned away in the most efficient manner (i.e., using the fewest dominance checks and with the least amount of memory to store the candidate objects).
Unfortunately, the efficiency of the skyline algorithms depends heavily on the underlying data statistics:  as illustrated in Figure~\ref{fig:alignment}, for data sets where the attribute correlations mirror the corresponding preference criteria (positively-correlated attributes for {\tt max-max} and {\tt min-min} preference criteria and negatively-correlated for {\tt max-min} and {\tt min-max} criteria), the number of skyline objects are fewer; whereas for data where the attribute correlations are mis-aligned with the preference criteria, the skyline set may contain large numbers of non-dominated data objects. 


\begin{figure}[t]
\noteB{[R2.O1]}
\centering{
   \begin{tabular}{cc}
   \multicolumn{2}{c}{
   \setlength{\fboxsep}{0.85pt}
	\fbox{\includegraphics[width=0.35\columnwidth]{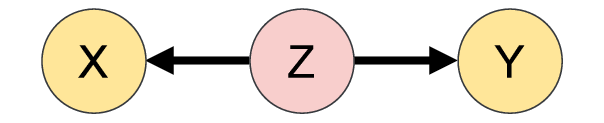}}
    } \\
\multicolumn{2}{c}{\replaceB{}{(a) $Z$ is a common cause of $X$ and $Y$}} \\
    \includegraphics[width=0.35\columnwidth]{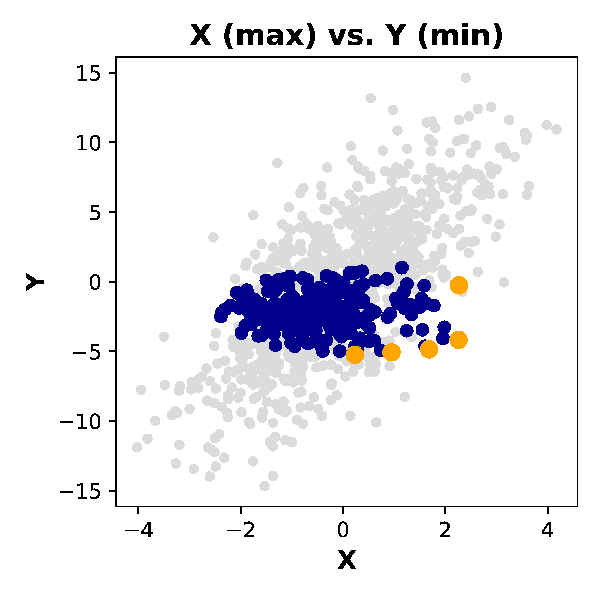}  &
    \includegraphics[width=0.35\columnwidth]{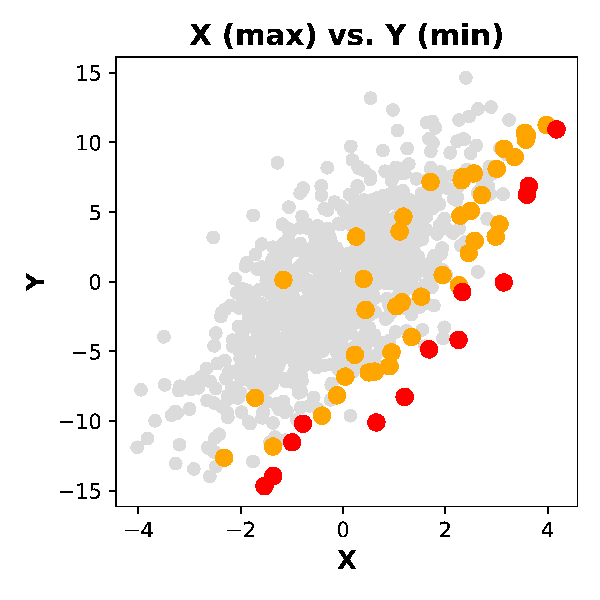}  \\
    \replaceB{}{(b) skyline of one cluster of Z} & 
    \replaceB{}{(c) skyline of  skylines}\\
   \end{tabular}
   }
\caption{\replaceB{}{Impact of clustering on the (a) confounder Z:
(b) within each cluster, the correlation between $X$ and $Y$ is close to zero, and
(c) seeking the skyline among the merged Pareto fronts (orange dots) across clusters is much less expensive, as many tuples have already been pruned}
}\label{fig:clustering}
\end{figure}


More critically \noteB{[R2.O1]}though, when data correlation is mis-aligned with the preference criteria, it becomes especially difficult to enumerate the data objects in a way that helps prune non-skyline objects without having to rely on large numbers of explicit dominance checks. 
%
%
As a result, the cost of enumerating skylines explodes in situations where there is a mis-alignment between the preference criteria and the underlying data distribution.
\replaceB{}{
In this paper,  we point out that we can potentially exploit underlying causal structures in the data to optimize skyline computation. In Figure~\ref{fig:clustering}(a), attribute $Z$  
causally influences both $X$ and $Y$, creating an apparent correlation between them 
and, since the preference criterion is max-min, the resulting misalignment would render the skyline search expensive as we see in Figure~\ref{fig:alignment}(b). 
As we discuss in Section~\ref{sec:cond}, clustering on $Z$ would reduce the correlation between $X$ and $Y$ within each cluster to approximately zero, making the skyline computation for each cluster much less expensive (Figure~\ref{fig:clustering}(b)). If we then merge the resulting per-cluster pareto fronts and seek the skyline of the merged set, since the number of considered tuples in the merged set is far smaller than in the original data set, the negative impact of misalignment on the cost of the skyline search can be significantly reduced
(Figure~\ref{fig:clustering}(c)).
}

\subsection{Our Contributions: Causally-Informed Selective Data De-Correlation}

\replace{I}{As outlined above, i}n this paper, we argue that the efficiency of the skyline algorithms could be significantly boosted if one could erase any attribute correlations that do not agree with the preference criteria (i.e., {\em negatively-aligned} data distributions), while preserving (or even boosting) correlations that agree with the user provided criteria (i.e., {\em positively-aligned} data distributions).
If such {\bf selective de-correlation} were possible,  the skyline algorithms would benefit from the  effective pruning opportunities provided by the positively-aligned data distributions, without having to suffer from the mis-alignments.
In the rest of the paper, we show that this challenge can be tackled effectively in situations where we have {\em a priori} causal-knowledge\footnote{When such causal knowledge is not  available, we can rely on causal discovery algorithms~\cite{chickering2002optimal, spirtes2000causation, shimizu2006linear} to learn causal graphs, \replaceB{but these are out of the scope of this paper}{as shown in Section~\ref{sec:inferred}}. \noteB{[R2.O2]}} that explains the underlying data distributions. In particular, 
\begin{itemize}[leftmargin=*]
\item we show that, given a causal graph that describes the underlying causal structure of the data, one can identify a subset, $\mathcal{Z}$, of the attributes that can be used to selectively de-correlate the preference attributes and
\item since positively-aligned attribute pairs are desirable for efficient skyline discovery, whereas negatively-aligned attribute pairs need to be avoided, we present an algorithm to select a subset, $\mathcal{Z}^-$, of attributes that can be used to de-correlate the negatively-aligned preference attributes.
\end{itemize}

Importantly, the proposed {\em causal search for skylines (CSS)} approach is agnostic to the underlying candidate enumeration and pruning strategies and, therefore, can be leveraged to improve any popular skyline discovery algorithm, including (but not limited to) nested-loop skylines~\cite{Borzsonyi01theskyline}, sort-first skylines~\cite{Chomicki03skylinewith},  index-supported skylines~\cite{TanEO01}, as well as more advanced skyline algorithms, such as $k$-dominant skylines~\cite{DBLP:conf/sigmod/ChanJTTZ06}.
In Section~\ref{sec:exp}, we experimentally evaluate the proposed {\em causal-skyline} approach on multiple real and synthetic data sets and for different skyline discovery algorithms. Experiment results show that the proposed causally-informed selective de-correlation technique significantly reduces both the number of dominance checks as well as the time needed to obtain the skyline.


\section{Related Works}\label{sec:rel}
\noteA{[R1.O1]}
\replaceA{Skyline queries are  a staple of decision support systems.}{Skyline queries are commonly used in multi-criteria decision making~\cite{ZHANG2026130889, needle_in_haystack, amin2025development, 6916872, 7373349}. In~\cite{needle_in_haystack}, authors use skylines to find the best clustering of data.  \cite{amin2025development} uses skyline queries to generate recommendations based on user preferences.
}%
\maybe{Beyond answering user's preference queries as in the House Hunting problem illustrated in Example \ref{ex:house} skylines have also been in used in multi-objective optimization problems; for example, authors in \cite{needle_in_haystack} use skylines to find the best clustering of data, arguing that best clustering of data may be subjective, authors use multiple cluster validation indices (CVI) as objectives (preferences) and seek preferred clustering schemes based on these objectives.}%
\replaceA{As a result}{Due to their ability to reduce the volume of data within the context of a user provided preference criterion},
the task of processing skyline queries in an efficient manner has attracted considerable attention.
The task of finding the non-dominated set of data points was attempted
by Kung {\em et al.} in 1975 under the name of the
maximum vector problem~\cite{KungLP75}. The algorithm proposed in~\cite{KungLP75} 
is
based on the divide and conquer principle. Kung's algorithm lead to the development of various skyline algorithms designed for specific situations, including those devised
for static environments~\cite{Borzsonyi01theskyline,SFSJ}, high dimensional data sets~\cite{Matousek:1991:CD} and parallel
environments~\cite{StojmenovicM88,Wu06parallelizingskyline,DBLP:conf/sigmod/VlachouDK08,distributedZorder}. It also inspired novel skyline algorithms for  
progressive skyline queries~\cite{TanEO01}, online
processing~\cite{Kossmann02shootingstars,PapadiasTFS05,Zorder},
skyline-join
queries~\cite{SkylineJoin,S2J,SFSJ} and data streams~\cite{3Lin,4Tao,6Zhang}.

Borzsonyi {\em et al.}~\cite{Borzsonyi01theskyline} were the first to
coin and investigate the {\em skyline} computation problem in the
context of databases. The authors extended Kung's divide and conquer
algorithm so that it works well on large databases. 
{They proposed a {\em Block-Nested-Loops (BNL)}
  algorithm, which keeps an in-memory window of incomparable points
and reports a point as a skyline result
only if it is not dominated by any other point in the database.} They also proposed a
{\em divide and conquer} based algorithm, which divides the data space
into several regions, calculates the skyline in each region, and
produces the final skyline from the points in the regional
skylines. 

The {\em Sort-Filter-Skyline (SFS)}
algorithm~\cite{Chomicki03skylinewith}, which is based on the same
principle as the BNL algorithm, improves on 
performance by first sorting
the data according to a monotone function. Bartolini {\em et al.} also developed a sort-based skyline technique called the {\em Sort and Limit Skyline
algorithm (SaLSa)} that uses the idea of presorting input tuples to limit the number of tuples read and compared during skyline query processing~\cite{Salsa}.
 Tan {\em et al.}
proposed progressive skyline algorithms called {\em bitmap} and
{\em index}~\cite{TanEO01}.
The  {\em bitmap} method is
completely non-blocking and exploits a bitmap structure to quickly
identify whether a point is a skyline result or not. The
{\em index} method, on the other hand, exploits a transformation mechanism
and a B$^+$-tree index to return skyline points in batches.
Other
contributions to
skyline query processing include online algorithms, such as~\cite{Kossmann02shootingstars}
and~\cite{PapadiasTFS05},
based on nearest-neighbor
search.
%
Bentley {\em et al.} developed the {\em Fast Linear Expected Time (FLET)} algorithm~\cite{bentley1993fast} which improved upon Kung's divide and conquer based algorithm by probabilistically eliminating points which cannot be part of the skyline. Godfrey {\em et al.}~\cite{godfrey2005maximal} demonstrated that the performance of Kung's divide and conquer based approach is poor with respect to the dimensionality of the problem. They also introduced the {\em Linear Elimination Sort for Skyline (LESS)} algorithm which combines aspects of BNL, SFS, and FLET, without any aspects of divide and conquer approach.

Park {\em et al.} proposed a mechanism for parallel processing of skyline queries to exploit multicore processors~\cite{DBLP:conf/icde/ParkKPKI09}; the authors also proposed a parallel version of the BBS algorithm.
Lee et al.~\cite{lee2007approaching} proposed an index
structure called \textit{ZBtree} to index
and store
data points based
on the Z-order curve. They also
developed several skyline
algorithms that utilize the \textit{ZBtree} index to efficiently process skyline queries.
Huang {\em et al.}~\cite{distributedZorder} also proposed a
parallel skyline algorithm for multi-processor clusters that utilizes
Z-order clustering to reduce dominance checks.
Shang {\em et al.} examined the skyline operator
in the context of
anti-correlated distributions~\cite{SLanti2013}. The authors proposed a probabilistic cardinality model for anti-correlated distributions and analyzed the upper and lower bounds of the expected value of skyline cardinality. They also developed an algorithm called \emph{SOAD (Skyline Operator on Anti-correlated Distributions)} that effectively eliminates non-promising points based on a \emph{determination} and \emph{elimination} framework.

\replaceA{}{
There has also been significant work on estimating the cardinality of skyline sets.
\cite{barndorff1966distribution} analyzes\noteA{[R1.O2]} the distribution of admissible pareto-optimal points and
\cite{bentley1978average} investigates the expected number of maximal vectors in a set, providing insights into the average-case of skyline cardinality. 
\cite{1129924,10.1145/1559845.1559899} proposed skyline cardinality estimation methods using log-sampling (LS).
\cite{10.1145/1559845.1559899} introduced a non-parametric  kernel-based (KB) skyline cardinality estimation technique which improves upon the log-sampling (LS) method
and does not depend on the assumption that the preference attributes have to be independent of each other. 
%
\cite{luo2012sampling} introduces a so-called
purely-sampling based approach (PS),
that does not involve complex mathematical computations and, hence is more efficient than KB. One of the more recent works in skyline cardinality estimation is \cite{10.1145/3588958}, which  outperforms the previous state-of-the-art including~\cite{10.1145/1559845.1559899, 1129924,luo2012sampling}.
}

\section{Preliminaries}

This section introduces the key concepts and provides preliminary
formalisms we will utilize in the rest of the paper. We assume that the reader is familiar with {\em basic} concepts in skylines (such as tuple dominance) and causal graphs (such as a causal path and a confounder). 
 Appendix~\ref{sec:appendix_pre} provides a quick refresher.

\subsection{Causal Graphs and Skylines}\label{sec:prel}
Causality is the relationship between the effect (outcome) and cause (treatment) that
gives rise to it~\cite{causal_inference} and is a topic
of increasing interest in big data analysis~\cite{causal_bigdata,DBLP:journals/tsas/AzadCKLLMSACSMAS24},
machine learning~\cite{causal_regularization}~\cite{causal_image}, and
reinforcement learning~\cite{causal_rl}.
Among others~\cite{10825754,DBLP:conf/bigdataconf/LiCS23,DBLP:journals/tkdd/ShethG00C23}, Pearl
has shown that a priori knowledge, in the form of  \textit{causal
  graphs}, is critical in analyzing and managing data~\cite{Pearl2009}.
%
\begin{figure*}[t]
\centering{
\begin{tabular}{ccc}
	\setlength{\fboxsep}{1.4pt}
	\fbox{
		\includegraphics[width=1.4in]
		{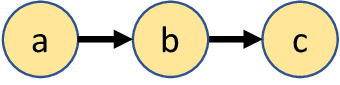}
	}
	&
	\setlength{\fboxsep}{1.4pt}
	\fbox{
		\includegraphics[width=1.4in]
		{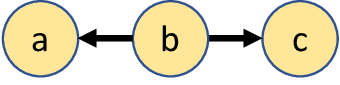}
	}
    &
    \setlength{\fboxsep}{1.4pt}
	\fbox{
		\includegraphics[width=1.4in]
		{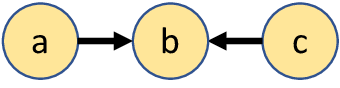}
	}
	\\
    (a) chain & (b) fork  & (c) collider\\
	\end{tabular}
    }
	\caption{Three basic causal structures}
	\label{fig:causal_struct}
\end{figure*}

\replaceB{}{\begin{definition}[Causal Graph]
Let $D$ be a data set defined\noteB{[R2.M2]} in an attribute space, $A$. A causal graph, $G_A = (V_A, E_A, \lambda)$, corresponding to the attribute set $A$ is an edge-labeled directed acylic graph (DAG), where for each attribute $a_i \in A$, there is a vertex $v_i \in V_A$ and each edge $e_h = \langle a_i, a_j \rangle$  (or $a_i \rightarrow a_j$) indicates a known direct causal relationship between the attributes $a_i$ and $a_j$ and  the edge label $\lambda(e_h)$  indicates the nature of the causal relationship between the two attributes.
\end{definition}
Figure~\ref{fig:causal_struct} presents three basic causal
structures introduced in~\cite{Pearl2009}:
(a) a  \textit{chain} structure where $a$ causally affects $c$ through its influence on $b$ (here $b$ is referred to as a mediator);
(b) a   \textit{fork} structure where $b$ is the common cause of both $a$ and $c$ (note that, in this case, $a$ and $c$ are likely dependent but there  is no causation between them);
(c) a \textit{collider} structure where both $a$ and $c$ independently
  cause $b$.
  }
In this paper, we note that, in the context of skylines,
%
%
\replace{}{these}
causal structures underlying the data may introduce statistical dependence among preference attributes: 

\begin{example}[Weakening or Strengthening of Relationships among Preference Attributes]
Let us reconsider our house hunting running example, with preference attributes $\tt Commute$ and $\tt Price$. Let us further assume that the  causal graph underlying the data is as depicted in Figure~\ref{fig:cg_houses}(b). 
As we see in this example, there is a direct (negative) causal relationship between the preference attributes, $\tt Commute$ and $\tt Price$ -- more commute generally translates into lower house prices. In addition, there are also  several indirect contributors to the relationship between these two preference attributes:
\begin{itemize}[leftmargin=*]
\item %
The $\tt Distance\_to\_city\_center$ attribute has a direct causal impact on both $\tt Price$ and $\tt Commute$ attributes (and therefore is a confounder): if we consider a city (like London), where house prices in the city center are higher due to housing density and that most of the commutable jobs are also located in the city center, the $\tt Distance\_to\_city\_center$ attribute would contribute additional negative correlation, strengthening the existing negative correlation between the $\tt Price$ and $\tt Commute$ attributes. 
\item In contrast, 
the $\tt Scenic\_quality$ attribute  impacts both $\tt Commute$ and $\tt Price$  positively as houses with scenic views generally have longer commutes and they also tend to be more expensive than the other houses. Note that this attribute weakens the negative correlation between the two preference attributes.\circend
\end{itemize}

\end{example}

\begin{figure}[t]
\noteC{[R3.O5]}
\adjustbox{width=0.8\columnwidth}{
\centering{%
   \begin{tabular}{ccc}
		\includegraphics[width=1.3in]
		{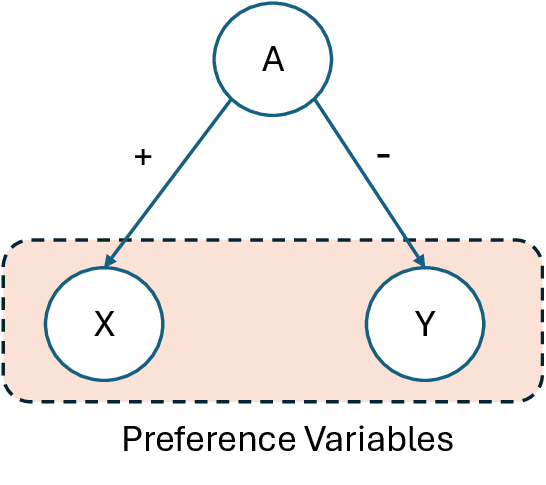}
	 &&
		\includegraphics[width=1.3in]
		{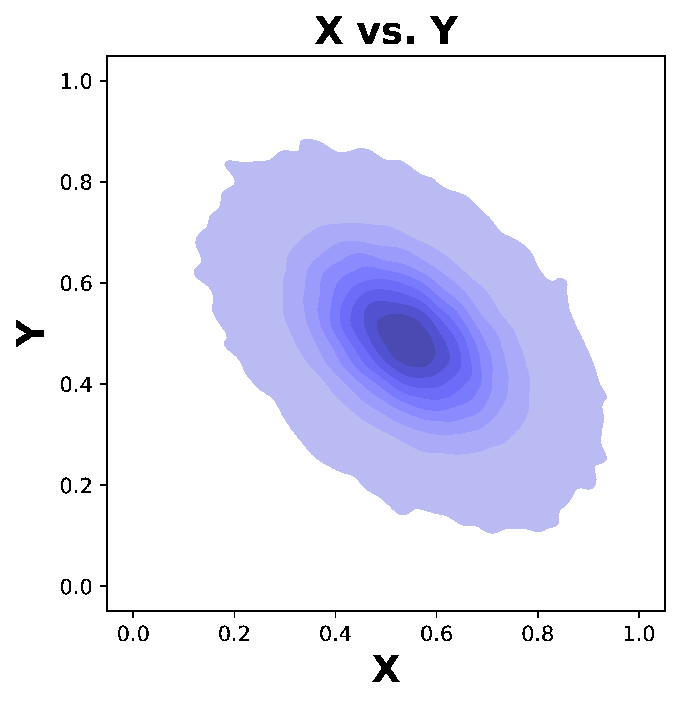}

	\\
 (a) a fork && (b) X vs. Y\\    &\\
		\includegraphics[width=1.3in]
		{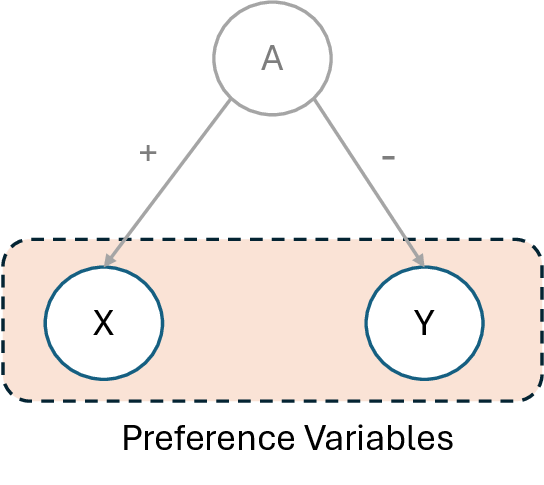}
	 &
		\includegraphics[width=1.3in]
		{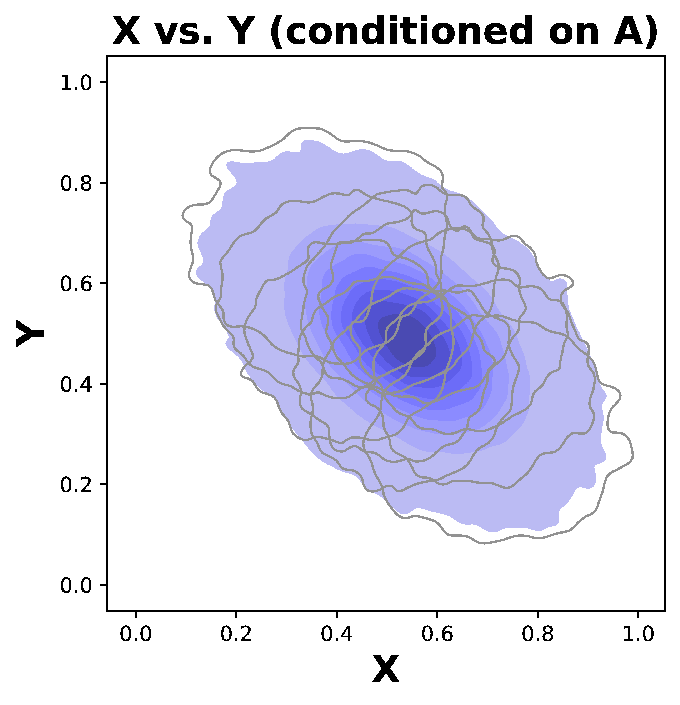}
     &
		\includegraphics[width=1.3in]
		{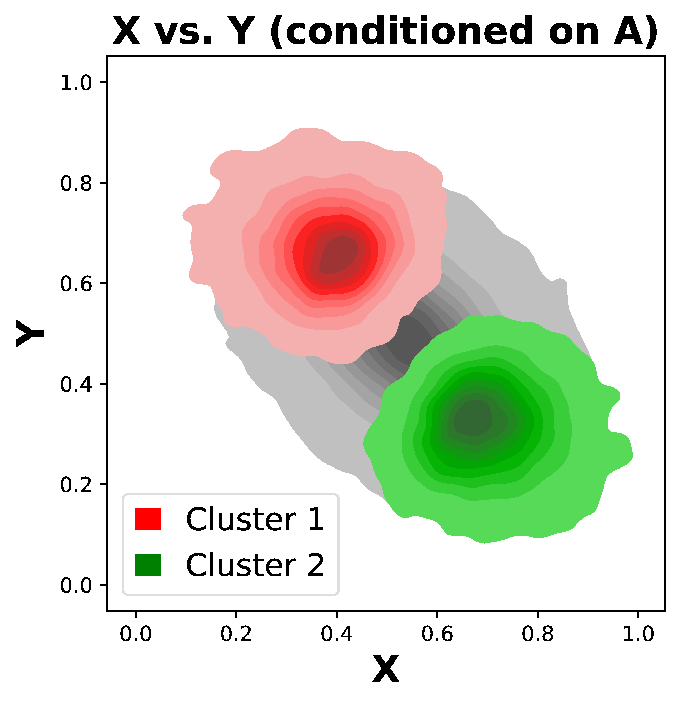}
	\\
 (c) cond. on the fork & \replaceC{}{(d) cluster outlines} & \replaceC{(d)}{(e)} two sample cond. subsets 
   \end{tabular}
}}
\caption{
\replaceC{In this example, t}{(a,b) T}he fork attribute $A$ imposes negative correlation on attributes $X$ and $Y$; \replaceC{}{(c,d)} conditioning of the attribute $A$ \replaceC{, however,}{} creates multiple clusters of data points, each cluster lacking any negative correlation\replaceC{}{; in (e), we highlight two of these resulting clusters in red and green}}\label{fig:fork_condition}
\end{figure}


\subsection{Conditioning and De-Correlation}\label{sec:cond}

{\em Conditioning} is the act of clustering the data entries that share the same value for a given (set of) attribute(s).

\begin{definition}[Conditioning]
Let $D$ be a data set and let  $a \in A$ be an attribute. Let $Dom(a)$ be the domain of this attribute. 
\begin{eqnarray*}
cond(D, a) = 
\{ 
(v,D_v)
&\;|\;&
(v \in Dom(a)) \wedge\\
&&
(t \in D) \wedge
(D_v = \{t.a = v\})
\}.\;\;\;\;\;\;\;\diamond
\end{eqnarray*}
%

\end{definition}
\noindent In SQL, this corresponds to a {\tt GROUP BY} operation.

\begin{figure}[t]
\adjustbox{width=\columnwidth}{
\centering{%
   \begin{tabular}{cccc}
		\includegraphics[width=0.25\columnwidth]
		{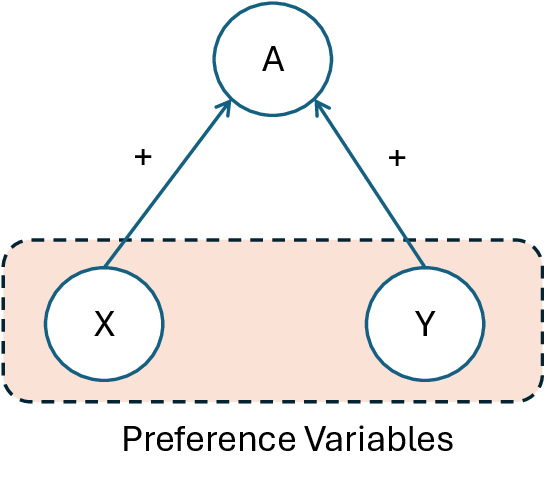}
	 &
		\includegraphics[width=0.25\columnwidth]
		{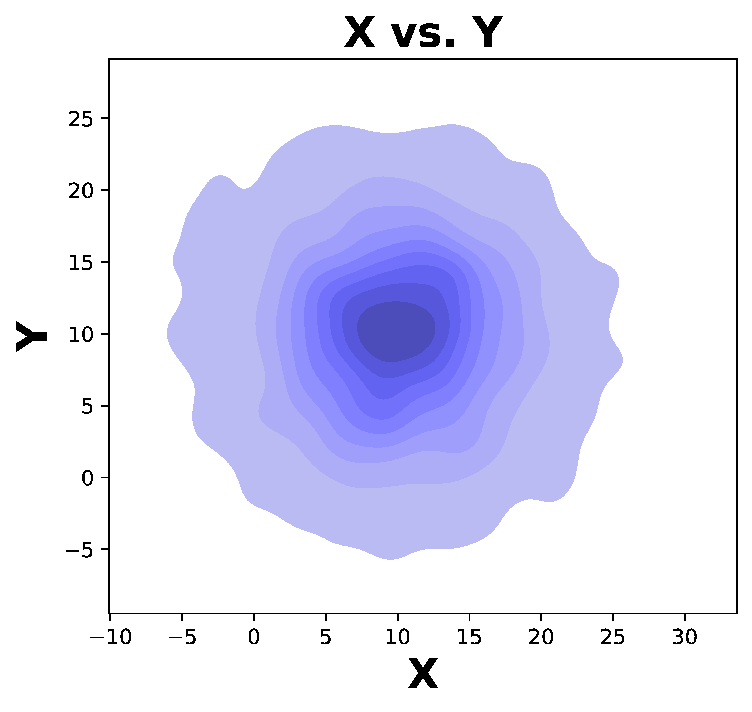}

&
		\includegraphics[width=0.25\columnwidth]
		{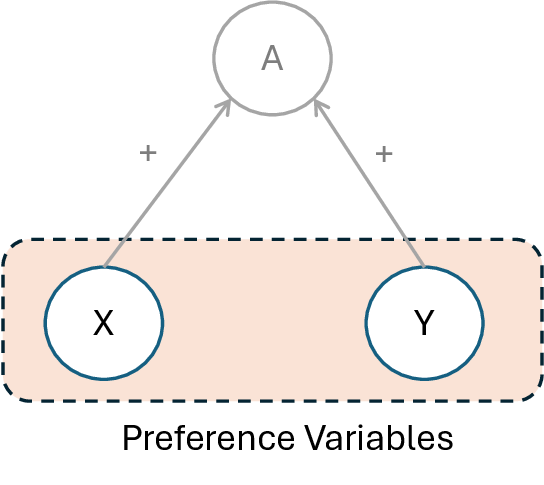}
	 &
		\includegraphics[width=0.25\columnwidth]
        {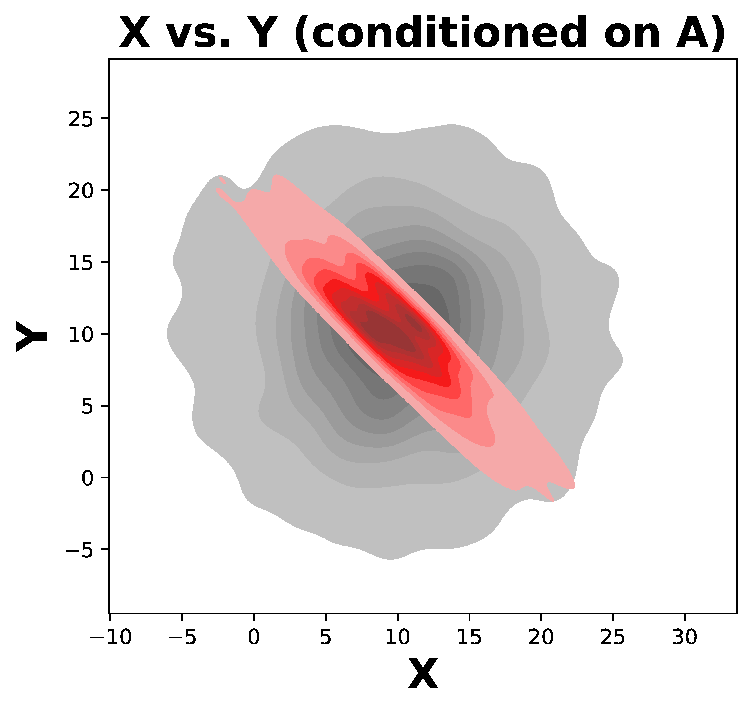}
	\\
 (a) pos. collider & (b) X vs. Y & (c) cond. of A & (d) sample cluster\\
		\includegraphics[width=0.25\columnwidth]
		{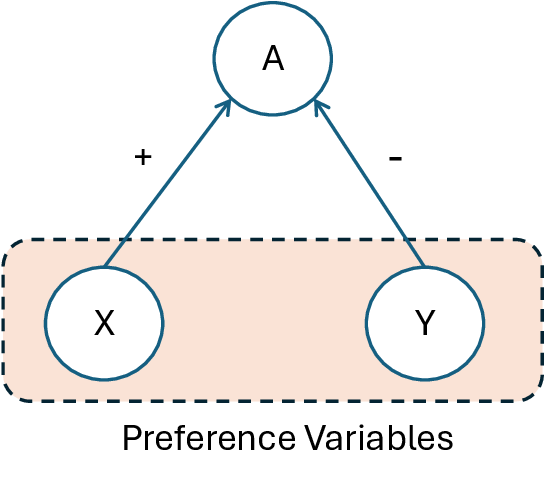}
	 &
		\includegraphics[width=0.25\columnwidth]
        {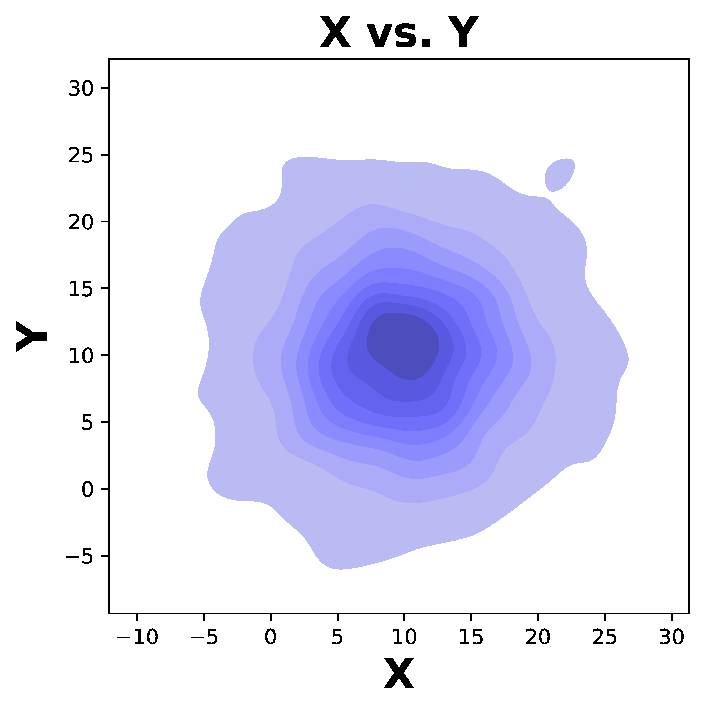}
&
		\includegraphics[width=0.25\columnwidth]
		{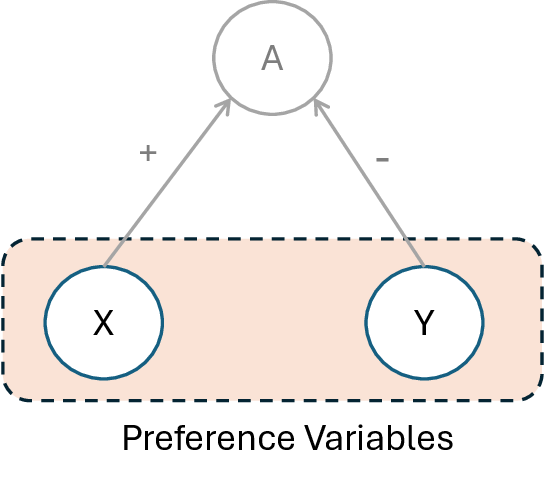}
	 &
		\includegraphics[width=0.25\columnwidth]
        		{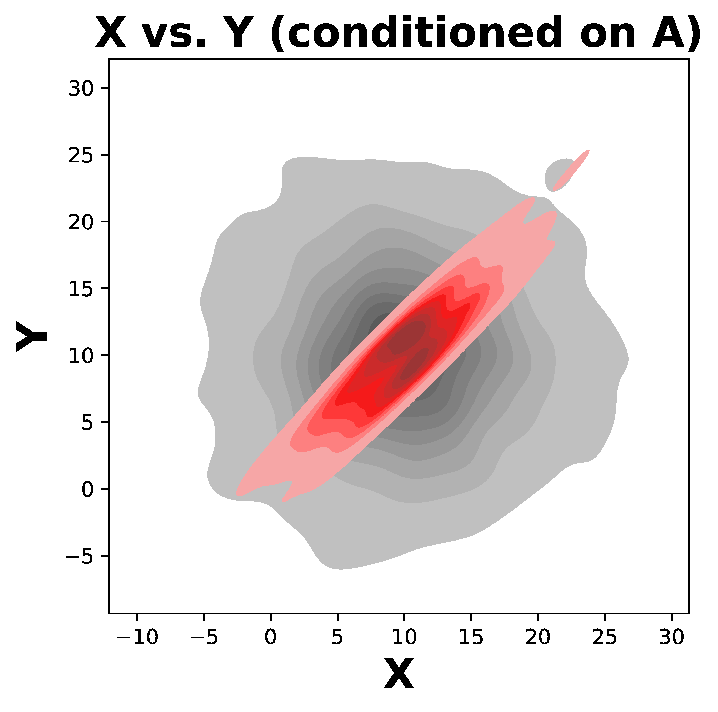}
	\\
 (e) neg. collider & (f) X vs. Y & (g) cond. of  A & (h) sample cluster\\ 
   \end{tabular}
}
}
\caption{The effect of the conditioning of a collider attribute on the distribution of the data: (a,b,e,f) $X$ and $Y$ are originally independent from each other; (c,d,g,h) conditioning of the collider attribute $A$, however, creates clusters of data points, each cluster having correlation with a sign opposite of the overall sign of the collision path  }\label{fig:collider_condition}
\end{figure}

In this paper, we note that the process of  conditioning
can  have impact on the statistical distribution  of the values for  the remaining attributes.  For instance in the \textit{chain}  and \textit{fork} structures 
conditioning on $A$ can eliminate any statistical
dependence between $X$ and $Y$ (this is visualized in Figure~\ref{fig:fork_condition}).
In contrast, in the \textit{collider}
structure 
conditioning on $A$ can introduce a (negative)
correlation
between unrelated variables, $X$ and $Y$ (this is visualized in Figure~\ref{fig:collider_condition}).
%


\begin{observation}[Selective De-Correlation]
As a corollary to the above, we note that negative correlations can be eliminated by conditioning on the mediator/fork attributes with originally negative impact; similarly, positive correlations can be boosted by conditioning on the colliders which have originally negative impact.
\end{observation}

\noindent Consider the following example:

\begin{example}[Conditioning to Weaken Negative Correlations among the Preference Attributes]
In Example~\ref{ex:house}, if we condition on the  $\tt Distance\_to\_city\_center$ by clustering all houses based on their distance to the city center and on whether they have scenic views or not, then within each such cluster, the only remaining causal relationship between  $\tt Price$ and $\tt Commute$ attributes would be the direct causal relationship of $\tt Commute$ on $\tt Price$. 
\circend
\end{example}

\begin{figure}[t]
\adjustbox{width=\columnwidth}{
\centering{
\begin{tabular}{cccc}
	{
		\includegraphics[width=0.25\columnwidth]
		{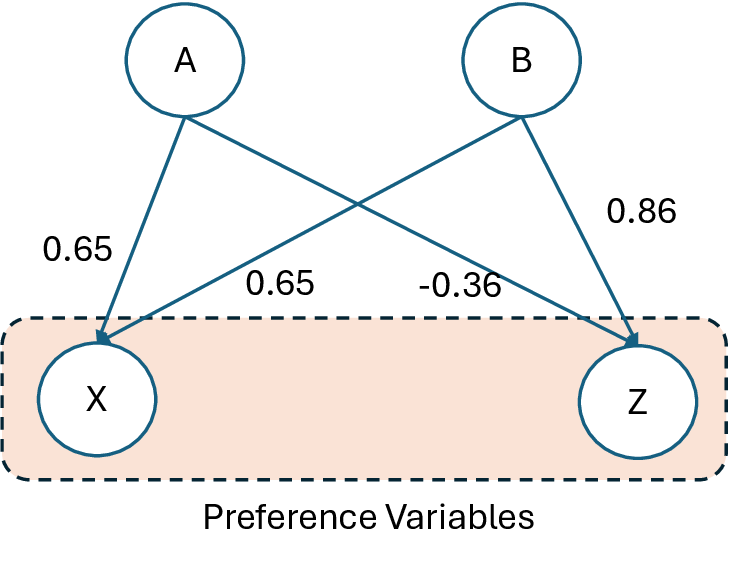}
	} &
 	{
		\includegraphics[width=0.25\columnwidth]
		{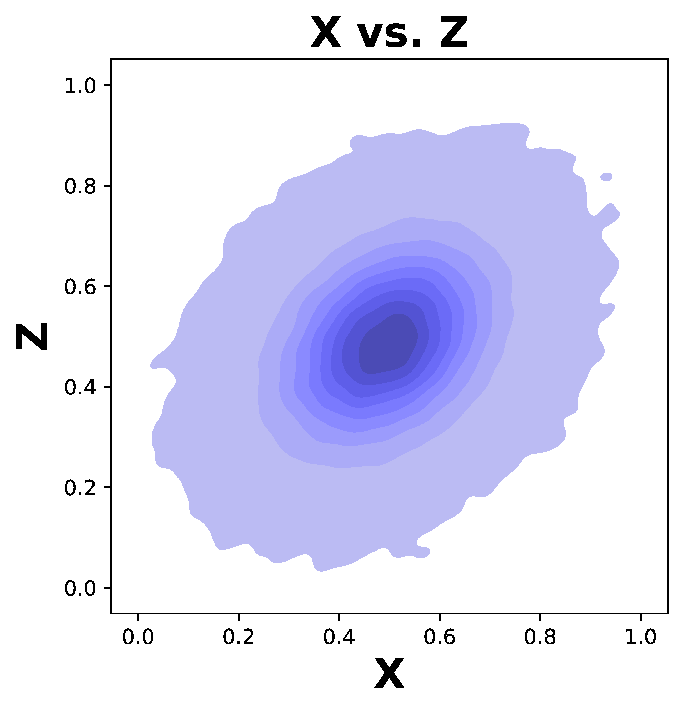}
	} &
    {
		\includegraphics[width=0.25\columnwidth]
		{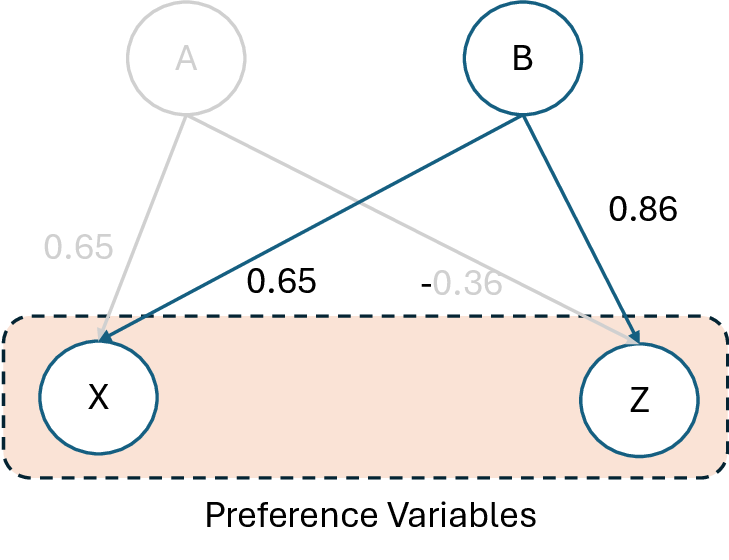}
	} &
    {
		\includegraphics[width=0.25\columnwidth]
		{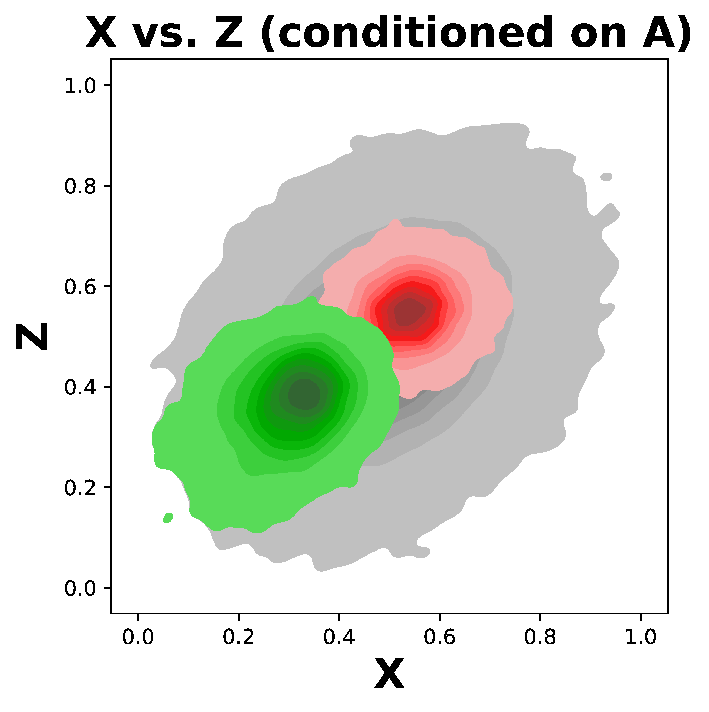}
	} 
	\\
  (a) sample causal graph & (b) scatter plot X vs. Z & (c) conditioning of "A"& (d) corr. preserved \\
    &
    {
        \includegraphics[width=0.25\columnwidth]
        {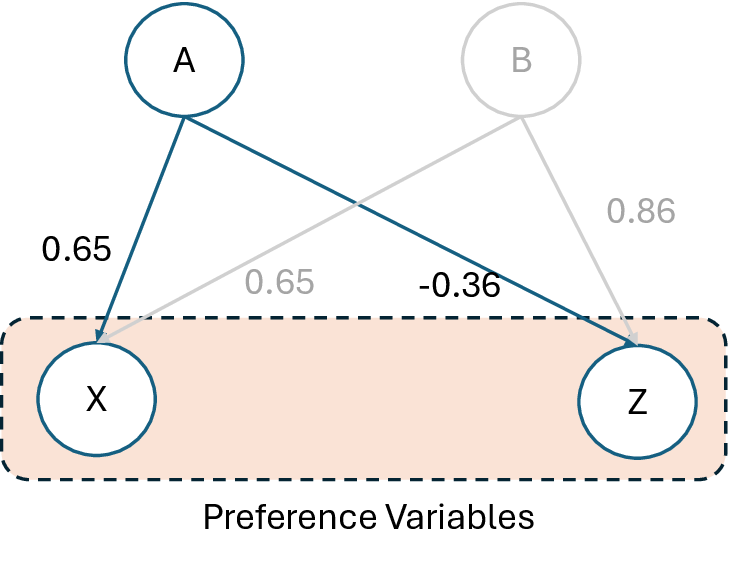}
    } &
    {
        \includegraphics[width=0.25\columnwidth]
        {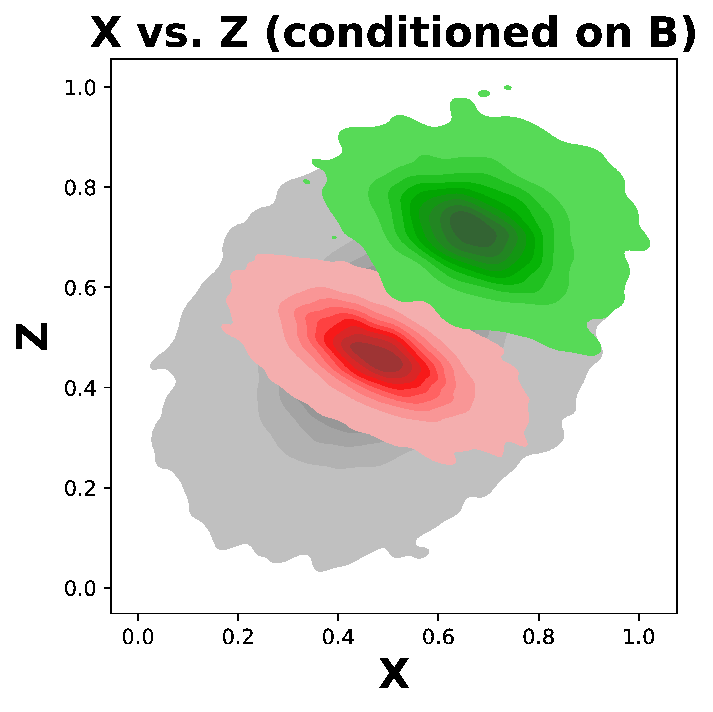}
    } &
    \\
    & (e) conditioning of "B"& (f) corr. reversed & \\
\end{tabular} 
}
}
	\caption{(a,b) A sample causal graph and the corresponding preference data, (c,d) conditioning of "A" preserves  the correlations among the preference variables, while (e,f) conditioning of "B" reverses the correlations among the preference variables
}
	\label{fig:reversing}
\end{figure}

\section{Causal Search for Skylines (CSS) with Selective De-Correlation}
\label{sec:css}
Let us reconsider our house hunting example and the corresponding causal graph visualized in Figure~\ref{fig:cg_houses}. In this example, our preference attributes are $\tt Commute$ and $\tt Price$, and our preference criteria are {\tt min-min}.
As we have discussed in the previous section, in this graph, we have two confounders $\tt Distance\_to\_city\_center$ and $\tt Scenic\_view$, the former introducing a negative correlation between the preference attributes, whereas the latter introducing positive correlation between them. 
While the negative correlation between $\tt Commute$ and $\tt Price$ due to the "$-$" labeled direct causal edge between them cannot be eliminated,
the additional correlations due to the existence of the confounders can potentially be eliminated by conditioning them:
\begin{itemize}[leftmargin=*]
\item The negative correlation imposed on the preference attributes by the $\tt Distance\_to\_city\_center$ confounder can be eliminated by conditioning this attribute. Our intuition is that, since negative correlations are mis-aligned with the {\tt min-min} preference criteria for the $\tt Commute$ and $\tt Price$ skyline attributes, this should help reduce the number of dominance checks and improve the efficiency of the skyline search process. 
\item
In contrast, however, conditioning of the $\tt Scenic\_view$ attribute would eliminate a degree of positive correlation between  our preference attributes. This clearly is not desirable when our preference criteria is {\tt min-min} as the elimination of this positive correlation would effectively emphasize the negative correlation between the preference attributes due to the "$-$" labeled direct causal edge between them.
\end{itemize}
Therefore, while conditioning of the $\tt Distance\_to\_city\_center$ confounder is desirable, conditioning of the $\tt Scenic\_view$ attribute is not.
Note that this indicates a potential approach to improve the efficiency of skyline discovery, when there are only two preference attributes:
(1) identify the confounders; (2) identify if the statistical correlation implied by the confounder is positively or negatively aligned with the preference criteria; and (3) condition those confounders that have negative alignment with the preferences to selectively de-correlate the preference attributes. 
%
While this approach promises an effective way to reduce the number of dominance checks, there are several challenges:
\begin{itemize}[leftmargin=*]
\item Firstly, we do not always have only two preference attributes. We therefore need 
an efficient and effective way to account for attributes that are negatively aligned with the preference attributes.
\item Secondly, data attributes may have partial causal impact on the set of preference attributes and they may be positively-aligned with some of the criteria, while being negatively aligned with the rest. In Figure~\ref{fig:reversing}(e,f), we see a scenario where conditioning on one of the attributes reverses the positive correlation between the preference variables, potentially increasing the cost of the skyline search. We therefore need an effective way to account for such mis-alignments to promote the efficiency of the skyline discovery process, without negatively effecting the pruning of non-promising skyline candidates.  
\end{itemize}
Therefore, we present a {\em causal search for skylines} (CSS) 
which, given a causal graph that describes the underlying causal structure of the data,  identifies a subset, $\mathcal{Z}^-$, of the data attributes that can be used to de-correlate the negatively-aligned preference attributes (while preserving the correlations among the positively-aligned ones). 

\subsection{Problem Formulation}\label{sec:problem}
 \replaceB{}{Table~\ref{tab:notation} in the Appendix \noteB{[R2.M3]}provides the key notations used throughout the paper.}
Let $R(A)$ be a relation with the attribute set $A$ and let $P \subseteq A$ denote the set of preference/skyline attributes. 
Let us consider a skyline setting where all preferences are {\em max} or all of them are {\em min} (other scenarios can easily be converted to these with simple data transformations).
%
Let $G_R$ (or $G$ for short) be an $\pm$-edge-labeled causal graph for the attributes in R.
Let further 
$C_P$ denote a correlation matrix for the attributes in $P$. 

Our  goal is to find a subset of conditioning attributes, $\mathcal{Z} \subseteq A$, such that for all $a_i,a_j \in P$, $C_P[i,j] \leq C^*_P[i,j]$, where $C^*_P$ is the correlation matrix obtained after the conditioning on attributes in $\mathcal{Z}$.
It is, however, especially important that 
\begin{itemize}[leftmargin=*]
\item if $C_P[i,j] \ll 0$,  then we have $C^*_P[i,j] \gtrsim 0$ and
\item if $C_P[i,j] > 0$, then we do not have $C^*_P[i,j] \ll C_P[i,j]$.  
\end{itemize}

\noindent \replaceC{}{
More specifically,\noteC{[R3.O4]} when the correlation between two preference attributes, $i, j$, is significantly negatively aligned with the preference criteria ($C_P[i, j] << 0$),  we would like the correlation between preference variates $i, j$ to be positive, but close to zero ($C_P[i, j] \gtrsim 0$).
%
If, on the other hand, the correlation between a pair of preference attributes, $i, j$, is positively aligned with the preference criteria ($C_P[i,j] > 0$), then we do not want the correlation to have a significantly drop (i.e., $C_{P}^*[i,j] \ll C_P[i,j]$) after we perform conditioning.
}


\subsection{Baseline Algorithm - Algorithm \texorpdfstring{$\#0$}{\#0}: Data Driven Conditioning Set Selection}
\label{sec:ddCSS}
Before discussing algorithms that rely on causal graphs for identifying the conditioning set, we first present a purely data\replace{ }{-}driven approach to conditioning set selection.
\begin{enumerate}[leftmargin=*]
\item{\bf Conditioning set enumeration.}
Let \maybe{$\mathcal{Z} \subseteq A\backslash P$} \default{$\mathcal{Z} \subseteq A$} be a subset of the attribute set \maybe{(excluding the preference attributes)} of the given relation. \maybe{When we condition on the preference variates, we end up decorrelating the data completely, since our aim is to selectively decorrelate, we want $\mathcal{Z} \subseteq A \backslash P$ and not $\mathcal{Z}\subseteq A$.} 
\begin{enumerate}
\item \underline{\em Data conditioning.} Using a {\tt GROUP BY} operation, the input data set, $D$, is grouped
into multiple subsets, $D_1(\mathcal{Z}),\ldots, D_m(\mathcal{Z})$ (where $m$ is the number of unique combinations for the conditioning set of attributes).

    

\item \underline{\em Pairwise correlation computation.} We consider all pairs of $a_i, a_j \in P$ for group $D_k(\mathcal{Z})$ to compute the corresponding  correlation $C_{\mathcal{Z},P}[i,j,k]$ under the given conditioning set.
    

\item \default{\underline{\em Pairwise correlation gain computation.} We then compute the corresponding correlation gain for the given conditioning set, where $C_{P}[i,j]$ is the correlation without any conditioning.
\[
C_{gain}(\mathcal{Z})=C_{avg}(\mathcal{Z}) - \sum_{i\neq j} C_{P}[i,j],
\]
\noindent where 
}
\[
C_{avg}(\mathcal{Z})=\frac{1}{k} \sum_{k} \sum_{i\neq j} C_{\mathcal{Z},P}[i,j,k].
\]
\end{enumerate}
\item {\bf Conditioning set selection.} Given all subsets of the
attribute set, we select the subset, $\mathcal{Z}$, that provides the
largest value of \maybe{$C_{sum}(\mathcal{Z})$} \default{$C_{gain}(\mathcal{Z})$} as the conditioning set.
\end{enumerate}

Once the above process ({\em Step 0}) is completed , we proceed with identifying the skylines for the input data set $D$ as follows:
\begin{enumerate}[left=0.35in]
    \item[\underline{\em Step 1}] using a {\tt GROUP BY} operation, the input data set, $D$, is grouped into multiple subsets, $D_1,\ldots, D_m$ (where $m$ is the number of unique combinations for the conditioning set of attributes),
    \item[\underline{\em Step 2}] candidates skylines are identified for each subset separately, 
    \item[\underline{\em Step 3}] these candidate skylines are merged into $M$, and 
    \item[\underline{\em Step 4}] the final skyline set is identified by running a skyline algorithm on the merged candidate skyline data set, $M$.
\end{enumerate} 

The intuition behind the above process is that the conditioning set we seek should ideally eliminate the negative correlations among the preference variables, while preserving (or even boosting) positive correlations among them.
As we empirically establish in Section~\ref{sec:gainexp}, the above (baseline) algorithm may be effective in identifying conditioning subsets; yet, this baseline algorithm is also quite costly as \replace{}{in Step 0} we need to compute pairwise correlations among $|P|$ attributes for \maybe{$2^{|A|-|P|}$} \default{$2^{|A|}$} conditioning subsets, which can be achieved at \maybe{$O(|P|^2\times 2^{|A|-|P|}\times N)$} \default{$O(|P|^2\times 2^{|A|}\times N)$} time, where $N$ is the size of the input data. In the rest of this paper, we present algorithms to compute conditioning sets relying on causal graphs (when they are available), avoiding the cost of computing pairwise correlations for each potential conditioning subset.

\section{Algorithm \texorpdfstring{$\#1$}{\#1}: Gain-based Negative Path Blocking}
\label{sec:gnblock}
In this section, we  describe an algorithm to address this issue by blocking the negative paths, while preserving the positive ones in the underlying causal graph.
%
Below, we provide the outline of the algorithm and discuss each step in detail:
\begin{enumerate}[leftmargin=*]
\item {\bf Path counting.}
In the first step, we consider all pairs of $a_i, a_j \in P$ and for each such pair, we enumerate the set of paths from $a_i$ to $a_j$ using exhaustive BFS or DFS. Let
\begin{itemize}
\item $Paths_{i,j}$ denote the set of all paths between $a_i$ and $a_j$; 
\item $Open^+_{i,j}$ denote the set of \underline{non-blocked} (i.e., collider free) paths on which the count of "-" labeled edges is an even number (this implies that the overall correlation implied by the path is positive), and
\item $Open^-_{i,j}$ denote the set of \underline{non-blocked} (i.e., collider free) paths on which the count of "-" labeled edges is an odd number (this implies that the  correlation implied by the path is negative). 
\end{itemize}

\item{\bf Conditioning set enumeration.}
Let \maybe{$\mathcal{Z} \subseteq A\backslash P$} \default{$\mathcal{Z} \subseteq A$} be a subset of the attribute set \maybe{(excluding the preference attributes)} of the given relation. \maybe{When we condition on the preference variates, we end up decorrelating the data completely, since our aim is to selectively decorrelate, we want $\mathcal{Z} \subseteq A \backslash P$ and not $\mathcal{Z}\subseteq A$.} 
\begin{enumerate}
    \item \underline{\em Preference pair assessment.} For this subset, consider all pairs of $a_i, a_j \in P$
\begin{enumerate}
\item {\em Identify new paths that provisionally open} between $a_i$ and $a_j$. This can happen if a path that was originally blocked through one or more colliders now gets un-blocked, because each collider on the path (or one of its descendants) gets conditioned. Note that, these paths are only provisionally open, because they may in fact be blocked by other conditioned mediator or fork nodes - we will not know this until they are further confirmed.
\begin{itemize}
\item $Prov^+_{i,j}(\mathcal{Z}) \subseteq Paths_{i,j}$ denotes the set of new paths that (provisionally) open between $a_i$ and $a_j$ by $\mathcal{Z}$, such that the overall correlation implied by the path is positive and
\item $Prov^-_{i,j}(\mathcal{Z}) \subseteq Paths_{i,j}$ denotes the set of new paths that (provisionally) open between $a_i$ and $a_j$  by $\mathcal{Z}$, such that the overall correlation implied by the path is negative. 
\end{itemize}
Note, however, that each un-blocked collider reverses the positive/negative relationship between the end-points. Therefore, whether a provisionally opened path is positive or negative depends on whether the sum of the number of "-" labeled edges and the number of colliders on the path is even or odd.

\item {\em Revise the sets of open paths with the provisional set.} Given the above, we can revise the sets of open paths between $a_i$ and $a_j$ considering the paths that provisionally open thanks to the un-blocked conditioners in $\mathcal{Z}$:
\begin{itemize}
\item ${ O}pen^{+}_{i,j}(\mathcal{Z}) = Open^+_{i,j} \cup Prov^+_{i,j}(\mathcal{Z})$, and
\item ${ O}pen^{-}_{i,j}(\mathcal{Z}) = Open^-_{i,j} \cup Prov^-_{i,j}(\mathcal{Z})$, 
\end{itemize}

\item {\em Identify the paths that get blocked by $\mathcal{Z}$.} Next, we consider the open paths between $a_i$ and $a_j$ and see if any of these get blocked by the conditioning of any mediators or forks:
\begin{itemize}
\item $Blocked^+_{i,j}(\mathcal{Z}) \subseteq { O}pen^{+}_{i,j}(\mathcal{Z})$ denotes the set of positive paths that were (provisionally) open between $a_i$ and $a_j$ that are blocked by conditioning of $\mathcal{Z}$; and 
\item $Blocked^-_{i,j}(\mathcal{Z}) \subseteq { O}pen^{-}_{i,j}(\mathcal{Z})$ denotes the set of negative paths that were (provisionally) open between $a_i$ and $a_j$ that are blocked by conditioning of $\mathcal{Z}$.
\end{itemize}

\item \label{enum:impact} {\em Evaluate the positive and negative path impact of $\mathcal{Z}$,} in terms of the number of positive and negative paths between $a_i$ and $a_j$ that were open at the beginning versus the number of such paths open after the conditioning of $\mathcal{Z}$: 
\begin{itemize}
\maybe{
$
imp^+_{i,j}(\mathcal{Z}) = \frac{(|{ O}pen^{+}_{i,j}(\mathcal{Z}) \backslash Blocked^+_{i,j}(\mathcal{Z})|)
- |Open^+_{i,j}|
}{|Open^+_{i,j}| + |Open^-_{i,j}|} 
$,
}
\maybe{
$
imp^-_{i,j}(\mathcal{Z}) = \frac{(|{ O}pen^{-}_{i,j}(\mathcal{Z}) \backslash  Blocked^-_{i,j}(\mathcal{Z})|)
- |Open^-_{i,j}|
}{|Open^+_{i,j}| + |Open^-_{i,j}|} 
$.
}
\item 
$imp^+_{i,j}(\mathcal{Z}) = |Open^{+}_{i,j}(\mathcal{Z}) \backslash Blocked^+_{i,j}(\mathcal{Z})| - |Open^+_{i,j}|$,
\item 
$imp^-_{i,j}(\mathcal{Z}) = |Open^{-}_{i,j}(\mathcal{Z}) \backslash  Blocked^-_{i,j}(\mathcal{Z})| - |Open^-_{i,j}|$.
\end{itemize}
An impact value $< 0$ indicates a drop in the corresponding open paths, whereas a value $> 0 $ indicates an increase.
\end{enumerate}
\item \underline{\em Compute the overall gain provided by $\mathcal{Z}$.} The overall impact of the conditioning of $\mathcal{Z}$ is a function of the changes  in the numbers of  positive and negative open paths as well as of the (absolute value) of the correlation between $a_i$ and $a_j$.
\maybe{
\[
gain(\mathcal{Z}) = \sum_{a_i,a_j \in P} [imp^+_{i,j}(\mathcal{Z}) - imp^-_{i,j}(\mathcal{Z})]\times |C_P[i,j]|
\]
}
\[
gain(\mathcal{Z}) = \sum_{a_i,a_j \in P} [imp^+_{i,j}(\mathcal{Z}) - imp^-_{i,j}(\mathcal{Z})]
\]
\end{enumerate}

\item {\bf Conditioning set selection.} 
For conditioning, we select $\mathcal{Z} \subseteq A$ that provides the largest value of $gain(\mathcal{Z})$. 
\end{enumerate}
Note that if we are working with graphs where edges are not labeled with $\pm$, we assume that all open paths are negative paths. Note also that, if we have \default{more precise} edge impact values
associated with the edges, we can also accommodate them. More specifically, we revise the computation of the impact values, $imp^+_{i,j}(\mathcal{Z})$ and $imp^-_{i,j}(\mathcal{Z})$, in Step~\ref{enum:impact} of the above algorithm to take into account the causal weights ($cw$) of the paths: 
\begin{itemize}
    \item ${\mathfrak O}^+_{i,j} = \sum_{p \in Open^+_{i,j}} cw(p)$,
    \item ${\mathfrak O}^-_{i,j} = \sum_{p \in Open^-_{i,j}} cw(p)$,
    \item ${\mathfrak OB}^+_{i,j}(\mathcal{Z}) = \sum_{p \in Open^{+}_{i,j}(\mathcal{Z}) \backslash Blocked^+_{i,j}(\mathcal{Z})} cw(p)$,
    \item ${\mathfrak OB}^-_{i,j}(\mathcal{Z}) = \sum_{p \in Open^{-}_{i,j}(\mathcal{Z}) \backslash Blocked^-_{i,j}(\mathcal{Z})} cw(p)$,
\maybe{$imp^+_{i,j}(\mathcal{Z}) = \frac{{\mathfrak OB}^+_{i,j}(\mathcal{Z}) - {\mathfrak O}^+_{i,j}
}{{\mathfrak O}^+_{i,j} + {\mathfrak O}^-_{i,j}} 
$, and}
\maybe{$imp^-_{i,j}(\mathcal{Z}) = \frac{{\mathfrak OB}^-_{i,j}(\mathcal{Z}) - {\mathfrak O}^-_{i,j}
}{{\mathfrak O}^+_{i,j} + {\mathfrak O}^-_{i,j}} 
$,}
\default{
    \item $imp^+_{i,j}(\mathcal{Z}) = {\mathfrak OB}^+_{i,j}(\mathcal{Z}) - {\mathfrak O}^+_{i,j}$, and
    \item $imp^-_{i,j}(\mathcal{Z}) = -({\mathfrak OB}^-_{i,j}(\mathcal{Z}) - {\mathfrak O}^-_{i,j})$,
}
\end{itemize}
where, given a path $p$, the causal weight $cw(p)$ of the path is determined as the multiplication of the causal edges on the causal path; i.e,
$ cw(p) = \prod_{e \in p} cw(e)$.

\begin{algorithm}[t]
\caption{Skylines with Gain-based Negative Path Blocking (\replaceC{gn-Skyline}{\em gnSky})}\label{alg:gnblock}
{\scriptsize
\begin{algorithmic}
\algrenewcommand\algorithmicrequire{\textbf{Input}}
\algrenewcommand\algorithmicensure{\textbf{Output}}
\Require{$R(A)$, $P \subseteq A$, $D$}
\Ensure{Skyline set $S$, on preference attributes, $P$}
\hrule\\
\Procedure{\replaceC{gn-Skyline}{gnSky}}{$R(A)$; $P$, $D$}
\State\Comment{{\em Step 0:} Gain-based Negative Path Blocking}
\State $C_P \gets correlation\_matrix(P,D)$
\For{$a_i,a_j \in P$}
    \State $(Paths_{i,j}; Open^+_{i,j}; Open^-_{i,j}) \gets paths(G,a_i,a_j)$
\EndFor
\For{$\mathcal{Z} \subseteq A\backslash P$}
    \State $I(\mathcal{Z}) \gets \emptyset$
    \For{$a_i,a_j \in P$}
    \State $(Prov^+_{i,j};Prov^-_{i,j}) \gets prov\_open(G,\mathcal{Z},a_i,a_j)$
    \State\Comment{Below, $*$ stands for + or -}
    \State $Open^*_{i,j}(\mathcal{Z}) \gets Open_{i,j} \cup Prov^*_{i,j}(\mathcal{Z})$
    \State $Blocked^*_{i,j}(\mathcal{Z}) \gets get\_blocked(Open^*_{i,j}(\mathcal{Z}), \mathcal{Z})$
    \State 
    $
    \begin{aligned}
imp^*_{i,j}(\mathcal{Z}) \gets &compute\_impact(Open^*_{i,j}(\mathcal{Z}),\\
&Blocked^*_{i,j}(\mathcal{Z}), Open^+_{i,j})
    \end{aligned}
$  
\State \State $I(\mathcal{Z}) \gets I(\mathcal{Z}) \cup imp^*_{i,j}(\mathcal{Z}) $
    \EndFor
    \State $gain(\mathcal{Z}) \gets compute\_gain (I(\mathcal{Z}),C_P)$
\EndFor
\State $\mathcal{Z}^- \gets argmax_\mathcal{Z}\; gain(\mathcal{Z})$
\State\Comment{{\em Step 1:} Conditioning step}
\State $D_1,\ldots, D_m \gets {\tt GROUP\_BY}_{\mathcal{Z}^-}(D)$
\State\Comment{{\em Step 2:} Candidate skylines per conditioning value}
\State $\forall_i S_i \gets skyline(P,D_i)$
\State\Comment{{\em Step 3:} Merging of candidate skylines}
\State $C \gets \bigcup_i S_i$
\State\Comment{{\em Step 4:} Computing of the final skylines}
\State $S \gets skyline(P,C)$
\EndProcedure
\end{algorithmic}
}
\end{algorithm}

This gives us the following {\em gain-based negative path blocking \replaceC{}{causal skylines} (\replaceC{gnBlock}{gnSky})} \replaceC{causal skyline}{} algorithm\noteC{[R3.O2]}: 
Once the conditioning set is selected based on gain as described above (let us refer to the above process as {\em Step 0}), we proceed to identify the skylines for the input data set $D$ \replaceC{as follows:
\begin{enumerate}[left=0.35in]
    \item[\underline{\em Step 1}] using a {\tt GROUP BY} operation, the input data set, $D$, is grouped into multiple subsets, $D_1,\ldots, D_m$ (where $m$ is the number of unique combinations for the conditioning set of attributes),
    \item[\underline{\em Step 2}] candidates skylines are identified for each subset separately, 
    \item[\underline{\em Step 3}] these candidate skylines are merged into $M$, and 
    \item[\underline{\em Step 4}] the final skyline set is identified by running a skyline algorithm on the merged candidate skyline data set, $M$.
\end{enumerate}    
}{using the same 4 steps\noteC{[R2.O3]} (Steps 1 through 4) in the baseline algorithm (Section~\ref{sec:ddCSS}).}
The outline of the overall algorithm is presented in Algorithm~\ref{alg:gnblock}. Note that the effectiveness of the algorithm depends on, among other things,  the cost of identifying candidate skylines for the subsets in {\em Step 2} (which is a function of the number and size of the groups and the correlation within each group) and the size of the merged candidate skyline set obtained in {\em Step 3}. Therefore, as we discuss in the next section, the grouping operation in {\em Step 1} is critical in the overall performance of  causal skyline search.

\section{Algorithm \texorpdfstring{$\#2$}{\#2}: Leaky Negative Path Blocking}
\label{sec:lnblock}
While the algorithm presented in the previous section helps eliminate negative correlations by blocking paths on the causal graph that lead to negative correlations among the preference attributes, this algorithm suffers from several shortcomings as outlined below.


\begin{figure}[t]
\adjustbox{width=\columnwidth}{
\centering{
\begin{tabular}{cccc}
	{
		\includegraphics[width=1.4in]
		{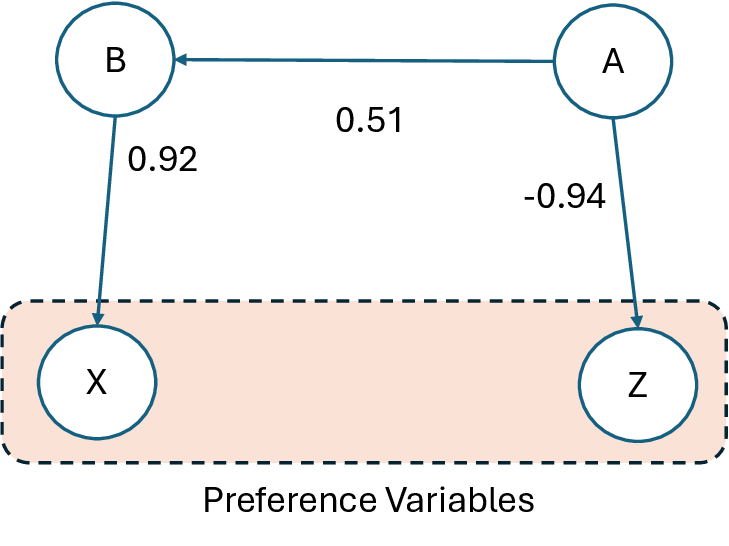}
	} &
 	{
		\includegraphics[width=1.4in]
		{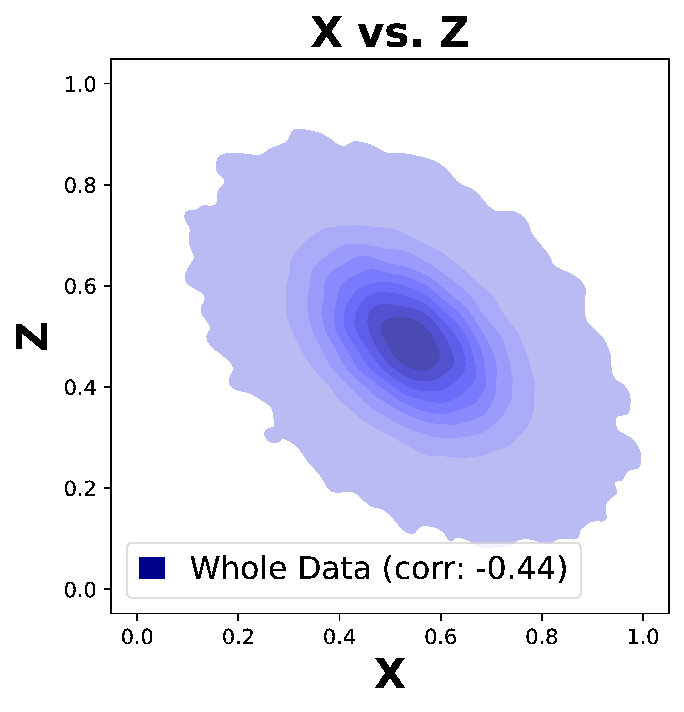}
	} &
 	{
		\includegraphics[width=1.4in]
		{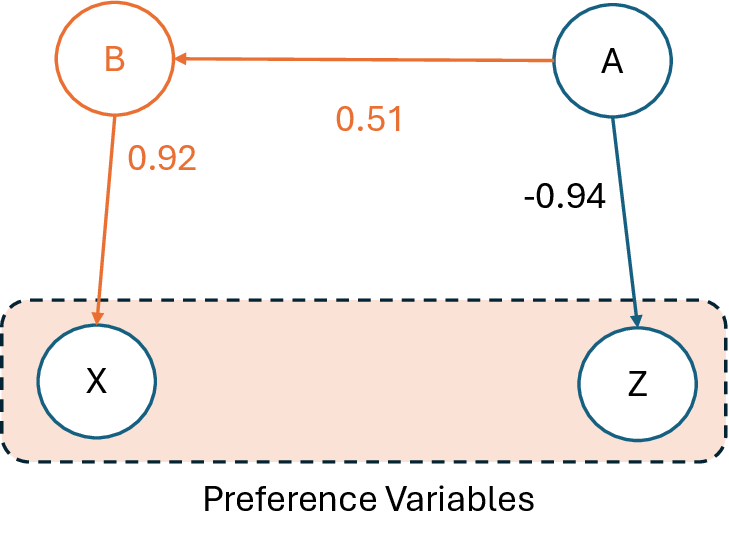}
	} &
 	{
		\includegraphics[width=1.4in]
		{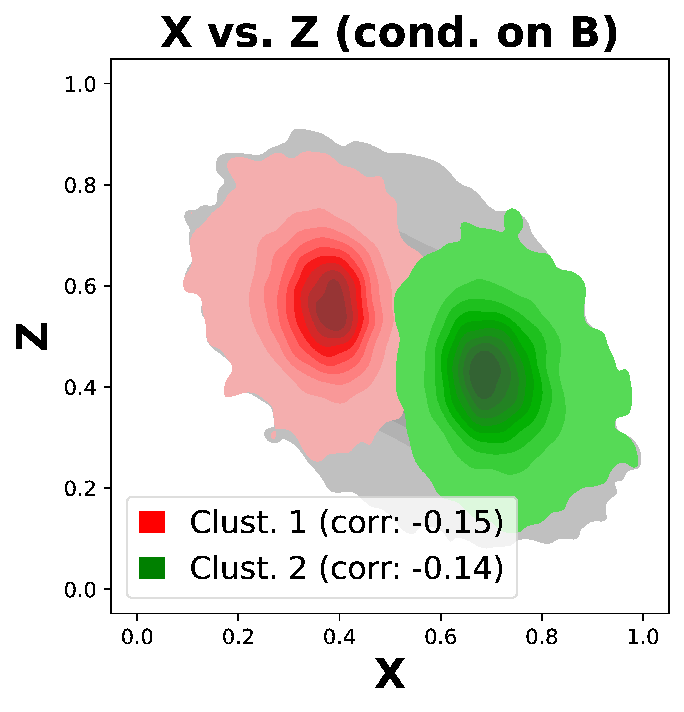}
	}
	\\
    (a) sample causal graph & (b) scatter plot of $X$ vs. $Z$ & (c) leaky cond. of $\{B\}$ & (d) high leakage \\
    &
 	{
		\includegraphics[width=1.4in]
		{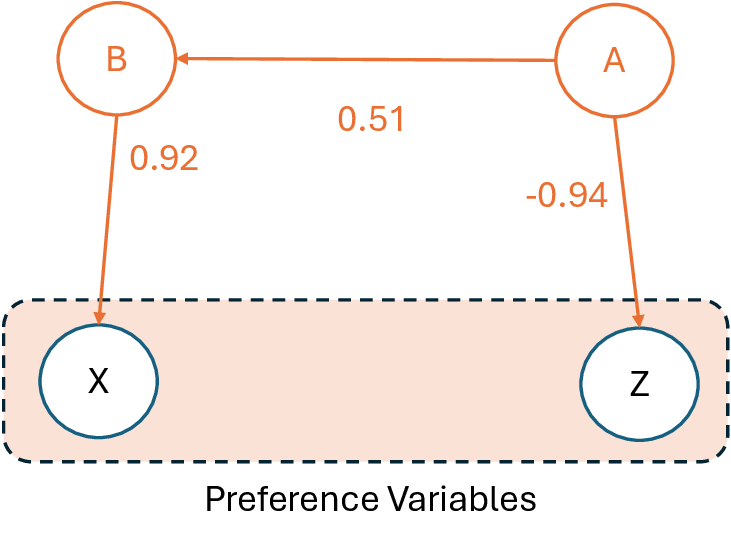}
	} &
 	{
		\includegraphics[width=1.5in]
		{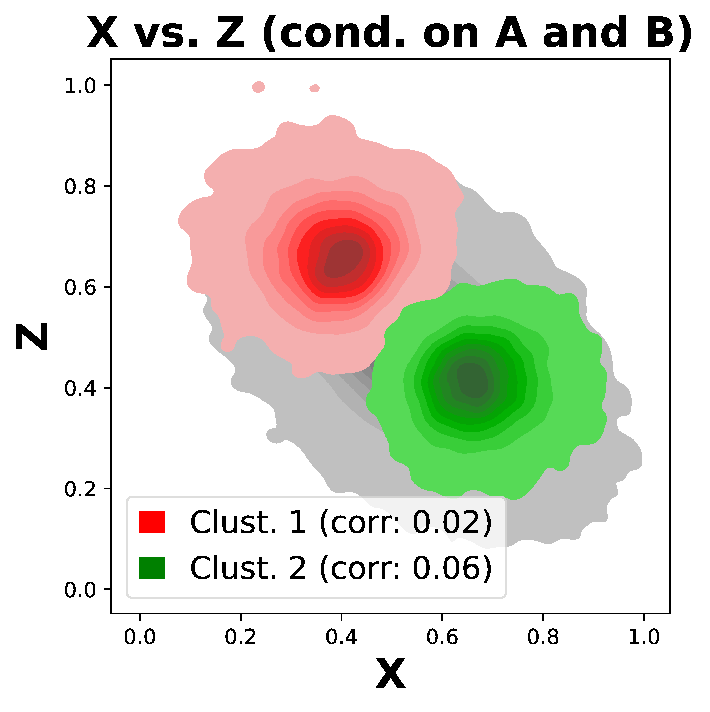}
	}
    &
	\\
    & (e) leaky cond. of $\{A,B\}$ & (f) low leakage &
\end{tabular}
}
}
\caption{(a) A sample causal graph, (b) scatter plot of two attributes $X$ and $Z$, (c,d) two sample clusters obtained through high leakage conditioning set $\{B\}$, and (e,f)  two sample clusters obtained through low leakage conditioning set $\{A,B\}$ -- in the causal graphs, orange colored nodes and edges indicate leaky conditioning/blocking; as we see in this figure, the larger the number of conditioned attributes on a path, the lesser the information leak on that path}
\label{fig:leaking}
\end{figure}

\subsection{Issue \texorpdfstring{$\#1$}{\#1}: Leaky Blocking}

As we have discussed in Section~\ref{sec:cond}, the {\tt GROUP BY} operation is analogous to conditioning and  the algorithm presented in the previous subsection leverages this to block (for mediator and fork structures) or open (for colliders) paths on a causal graph to eliminate negative correlations among the preference attributes, while maintaining the positive correlations. The problem, of course, is that in practice grouping the input tuples by a single value combination of the conditioning attributes may be overly selective -- it is possible that, the {\tt GROUP BY} operation returns a single tuple per each unique value combination of the conditioning set of attributes; in such a case, we would have $M=D$ and the cost of the {\em Step 4} of the algorithm above would be equal to the cost of running the skyline on the original data, rendering the total cost of {\em Steps 1 through 3} a redundant overhead.
In order to avoid this outcome, we can revise the {\em Step 1} of the algorithm above as follows:
\begin{enumerate}[left=0.35in]
    \item[\underline{\em Step 1*}] given the target number, $m$, of groups, use a clustering algorithm to group the input  data set, $D$, into multiple subsets, $D_1,\ldots, D_m$.
\end{enumerate}
On the positive side, this revision enables us to control the number of groups that are created in {\em Step 1}, enabling us to deal with the situations where the combined selectivity of the conditioning set of attributes is high. On the negative side, however, the replacement of {\em single-value grouping} in {\em Step 1} with {\em similar value grouping} in {\em Step 1*} potentially reduces the effect of conditioning and path blocking. We next illustrate this problem, which we refer to as {\em leaky conditioning} or {\em leaky blocking}, with an example:

\begin{example}[Leaky Conditioning/Blocking]
This is visualized in Figure~\ref{fig:leaking}, here we see a sample causal graph, which has two alternative conditioning sets, $\{B\}$ and $\{A,B\}$ with the same gain. In this example, the data has been grouped into 10 subsets using K-means clustering and in Figures~\ref{fig:leaking}(d) and (f), we are showing two of the resulting clusters for each of the conditioning set scenarios. 

As we see in the figure, when we use the groups formed by the conditioning set, $\{B\}$, the correlation between preference attributes $X$ and $Z$ is brought down from $-0.44$ for the original data set to \default{$\sim -0.15$} for the resulting clusters; 
while this conditioning set blocks the negative path between  $X$ and $Z$, the blockage is imperfect or {\em leaky}. 

In contrast, when we use the conditioning set, $\{A,B\}$, the correlation between preference attributes $X$ and $Z$ is brought down from \default{$-0.44$} to \default{$\sim 0.04$} for the resulting clusters; in other words,  this conditioning set blocks the negative path between  $X$ and $Z$ very effectively, almost completely eliminating the negative correlation between the preference attributes in the resulting clusters.

\end{example}
Therefore, when we are relying on cluster-based {\em similar-value grouping} in {\em Step 1} of the algorithm, rather than {\em single-value grouping}, the value of $gain$ defined in Section~\ref{sec:gnblock} is not sufficient to differentiate among the path blocking alternatives. Instead, as we discuss in the next section, we need to expand the definition of $gain$ in a way that accounts also for the leakages of the conditioning attributes.

\subsection{Issue \texorpdfstring{$\#2$}{\#2}: Noisy Causal Information Passing}
\default{
Imperfect causal information passing is not limited to only the blocked nodes. Even open nodes may be imperfect conduits of causality.
As discussed in Appendix~\ref{sec:cg}, a
 common causal model~\cite{Wright1921,
Pearl2000}, for example, represents the causal relationship $X \rightarrow Y$ as $Y = c_{xy} X + \epsilon_y$, 
where $c_{xy}$ denotes the causal impact of $X$ on $Y$, whereas $\epsilon_y$, the noise that acts on attribute $Y$ along this causal edge. When the noise, $\epsilon_y \sim 0$, the causal impact of $X$ and $Y$ is clear. 
When  $|\epsilon_y| \gg 0$, the causal relationship between $X$ and $Y$ will be noisy and the impact of $X$ on $Y$ may be hidden, even for large values of $c_{xy}$.
}

\subsection{Leaky Negative Path Blocking}
\label{sec:lnblock-algo}

\noteC{[R3.O2]}We address leaky blocking and noisy information passing problems by blocking the negative paths, while preserving the positive ones in the underlying causal graph, in a way that takes into account leaky conditioning due to cluster-based {\em similar-value grouping} in {\em Step 1} of the algorithm.
The algorithm shares the same outline as the {\em gain-based negative path blocking \replaceC{(gnBlock)}{skyline (gnSky)}} algorithm in Section~\ref{sec:gnblock}, but instead of treating each node in the graph as open or closed, it considers a degree of information passing, $\pi$, on a conditioned attribute node: 
\default{A conventionally open/non-blocked node passes complete information ($\pi$ = 1), whereas a conventionally (i.e., {\em single-value grouping} based) blocked node passes no information at  all  ($\pi$ = 0). In contrast, a {\em noisy open} node passes $\pi =\lambda_o$ information, whereas a {\em similar-value grouping} based blocked node passes $\pi =\lambda_b$ information. Here $0 <\lambda_b \ll \lambda_o < 1 $ are the corresponding degrees of information passage for imperfectly open and blocked nodes, respectively. }
%
%
The outline of the {\em leaky negative path blocking \replaceC{(lnBlock)}{skyline (lnSky)}} algorithm is as follows:
\begin{enumerate}[leftmargin=*]
\item {\bf Path counting.}
In the first step, we consider all pairs of $a_i, a_j \in P$ and for each such pair, we enumerate the set of paths from $a_i$ to $a_j$ using exhaustive BFS or DFS. Let $Paths_{i,j}$, $Open^+_{i,j}$, and $Open^-_{i,j}$ be defined as before.

\item{\bf Conditioning set enumeration.}
Let \maybe{$\mathcal{Z} \subseteq A\backslash P$} \default{$\mathcal{Z} \subseteq A$} be a subset of the attribute set \maybe{(excluding the preference attributes)} of the given relation.
\begin{enumerate}
    \item \underline{\em Preference pair assessment.} For this subset, consider all pairs of $a_i, a_j \in P$
\begin{enumerate}
\item {\em Identify new paths that provisionally open} between $a_i$ and $a_j$. As before, this can happen if a path that was originally blocked through one or more colliders now gets un-blocked, because each collider on the path (or one of its descendants) gets conditioned. Let $Prov^+_{i,j}(\mathcal{Z})$ and $Prov^-_{i,j}(\mathcal{Z})$ be defined as before.

Note that, in this case, how much information passes on a path $p \in Prov^+_{i,j}(\mathcal{Z}) \cup Prov^-_{i,j}(\mathcal{Z})$ depends on how much information passes on the conditioned colliders.

\item {\em Revise the sets of open paths with the provisional set.} Given the above, we can revise the sets of open paths between $a_i$ and $a_j$ considering the paths that provisionally open thanks to the un-blocked conditioners in $\mathcal{Z}$:
\begin{itemize}
\item ${ O}pen^{+}_{i,j}(\mathcal{Z}) = Open^+_{i,j} \cup Prov^+_{i,j}(\mathcal{Z})$, and
\item ${ O}pen^{-}_{i,j}(\mathcal{Z}) = Open^-_{i,j} \cup Prov^-_{i,j}(\mathcal{Z})$, 
\end{itemize}

\item {\em Identify the paths that get blocked by $\mathcal{Z}$.} Next, we consider the open paths between $a_i$ and $a_j$ and see if any of these get blocked by the conditioning of any mediators or forks:
\begin{itemize}
\item $Blocked^+_{i,j}(\mathcal{Z}) \subseteq { O}pen^{+}_{i,j}(\mathcal{Z})$ denotes the set of positive paths that were (provisionally) open between $a_i$ and $a_j$ that are blocked by conditioning of $\mathcal{Z}$; and 
\item $Blocked^-_{i,j}(\mathcal{Z}) \subseteq { O}pen^{-}_{i,j}(\mathcal{Z})$ denotes the set of negative paths that were (provisionally) open between $a_i$ and $a_j$ that are blocked by conditioning of $\mathcal{Z}$.
\end{itemize}
In this case, conditioning does not entirely eliminate information passage on the paths, but reduces it due to leakages. 
\item {Compute noise and leakage adjusted causal weights of the paths.}
Given an open or closed path $p$, let the causal weight $cw(p)$ of the path be determined as the multiplication of the weights of the causal edges on the causal path; i.e,
$cw(p) = \prod_{e \in p} cw(e).$
Let $o_p(\mathcal{Z})$ be the number of open nodes on $p$ and let $b_p(\mathcal{Z})$ be the number of blocked nodes on $p$, given $\mathcal{Z}$.
The  noise and leakage adjusted causal weight of $p$ is computed as 
\[
\hat{cw}(p) = cw(p) \times \lambda_o^{o_p(\mathcal{Z})} \times \lambda_b^{b_p(\mathcal{Z})}.
\]


\item {\em Evaluate the leaky impact of $\mathcal{Z}$ on positive and negative paths} between $a_i$ and $a_j$, in terms of the amount of information flow on them:
\begin{itemize}

\maybe{$initial\_flow^+_{i,j} = \sum_{p \in { O}pen^+_{i,j}} \hat{cw}(p)$,}
\maybe{$initial\_flow^-_{i,j} = \sum_{p \in { O}pen^-_{i,j}} \hat{cw}(p)$,}
\maybe{$final\_flow^+(\mathcal{Z})_{i,j} = \sum_{p \in { O}pen^+_{i,j}(\mathcal{Z})\cup { B}locked^+_{i,j}(\mathcal{Z})} \hat{cw}(p)$,}
\maybe{$final\_flow^-(\mathcal{Z})_{i,j} = \sum_{p \in { O}pen^-_{i,j}(\mathcal{Z}) \cup { B}locked^-_{i,j}(\mathcal{Z})} \hat{cw}(p)$,}
\maybe{
$
l\_imp^+_{i,j}(\mathcal{Z}) = \frac{final\_flow^{+}_{i,j}(\mathcal{Z})
- initial\_flow^+_{i,j}
}{initial\_flow^+_{i,j} + initial\_flow^-_{i,j}} 
$,}
\maybe{
$
l\_imp^-_{i,j}(\mathcal{Z}) = \frac{final\_flow^{-}_{i,j}(\mathcal{Z})
- initial\_flow^-_{i,j}
}{initial\_flow^+_{i,j} + initial\_flow^-_{i,j}}.$ 
}
%
    \item ${\mathfrak O}^+_{i,j} = \sum_{p \in Open^+_{i,j}} \hat{cw}(p)$,
    \item ${\mathfrak O}^-_{i,j} = \sum_{p \in Open^-_{i,j}} \hat{cw}(p)$,
    \item ${\mathfrak OB}^+_{i,j}(\mathcal{Z}) = \sum_{p \in Open^{+}_{i,j}(\mathcal{Z}) \cup Blocked^+_{i,j}(\mathcal{Z})} \hat{cw}(p)$,
    \item ${\mathfrak OB}^-_{i,j}(\mathcal{Z}) = \sum_{p \in Open^{-}_{i,j}(\mathcal{Z}) \cup Blocked^-_{i,j}(\mathcal{Z})} \hat{cw}(p)$,
    \item $l\_imp^+_{i,j}(\mathcal{Z}) = {\mathfrak OB}^+_{i,j}(\mathcal{Z}) - {\mathfrak O}^+_{i,j}$, and
    \item $l\_imp^-_{i,j}(\mathcal{Z}) = -({\mathfrak OB}^-_{i,j}(\mathcal{Z}) - {\mathfrak O}^-_{i,j})$.
\end{itemize}
Note that, a (leaky)  impact value $< 0$ indicates a drop of information flow on the corresponding open paths, whereas a value $> 0 $ indicates an increase.
\end{enumerate}
\item \underline{\em Compute the overall leaky gain provided by $\mathcal{Z}$.} The overall (leaky) impact of the conditioning of $\mathcal{Z}$ is a function of the changes of the amount of positive and negative information flow as well as of the (absolute value) of the correlation between $a_i$ and $a_j$.
\maybe{
\[
l\_gain(\mathcal{Z}) = \sum_{a_i,a_j \in P} &[l\_imp^+_{i,j}(\mathcal{Z}) - l\_imp^-_{i,j}(\mathcal{Z})]\\&\times
|C_P[i,j]|.
\]
}
\[
l\_gain(\mathcal{Z}) = \sum_{a_i,a_j \in P} [l\_imp^+_{i,j}(\mathcal{Z}) - l\_imp^-_{i,j}(\mathcal{Z})]
\]

\end{enumerate}

\item {\bf Conditioning set selection.} Given all subsets of the attribute set, we select the one, $\mathcal{Z}$, that provides the largest value of the leaky gain, $l\_gain(\mathcal{Z})$. 
\end{enumerate}
Note that, as before, if we are working with graphs where edges are not labeled with $\pm$, we can assume that all open paths are negative paths. Note also that, if we have edge impact values associated with the edges, we can also accommodate them (multiplicatively). 

This gives us the following {\em leaky negative path blocking \replaceC{(lnBlock)}{causal skyline (lnSky)}} \replaceC{causal skyline}{} algorithm: 
Once the conditioning set is selected based on the leaky gain as described above (we refer to the above process as {\em Step 0} of the {\em \replaceC{lnBlock}{lnSky}} algorithm), we proceed to identify the skylines for the input data set $D$ \noteC{[R3.O3]}\replaceC{as follows:
\begin{enumerate}[left=0.35in]
    \item[\underline{\em Step 1}] given the target number, $m$, of groups, use a clustering algorithm to group the input  data set, $D$, into multiple subsets, $D_1,\ldots, D_m$,
    \item[\underline{\em Step 2}] candidates skylines are identified for each subsets separately, 
    \item[\underline{\em Step 3}] these candidate skylines are merged into $M$, and 
    \item[\underline{\em Step 4}] the final skyline set is identified by running a skyline algorithm on the merged candidate skyline data set, $M$.
\end{enumerate}
}{using the same 4 steps in the baseline algorithm (Section~\ref{sec:ddCSS}) and the gain-based negative path blocking skyline ({\em gnSky}) algorithm.}

\begin{algorithm}[t]
\caption{Skylines with Leaky Negative Path Blocking (\replaceC{ln-Skyline}{\em lnSky})}\label{alg:lnblock}
{\scriptsize
\begin{algorithmic}
\algrenewcommand\algorithmicrequire{\textbf{Input}}
\algrenewcommand\algorithmicensure{\textbf{Output}}
\Require{$R(A)$; $P \subseteq A$, $D$,$m$}
\Ensure{Skyline set $S$, on preference attributes, $P$}
\hrule\\
\Procedure{\replaceC{ln-Skyline}{lnSky}}{$R(A)$, $P$, $D$, $m$}
\State\Comment{{\em Step 0:} Leaky Gain-based Negative Path Blocking}
\State $C_P \gets correlation\_matrix(P,D)$
\For{$a_i,a_j \in P$}
    \State $(Paths_{i,j}; Open^+_{i,j}; Open^-_{i,j}) \gets paths(G,a_i,a_j)$
\EndFor
\For{$\mathcal{Z} \subseteq A\backslash P$}
    \State $I(\mathcal{Z}) \gets \emptyset$
    \For{$a_i,a_j \in P$}
    \State $
    \begin{aligned}
        (Prov^+_{i,j};Prov^-_{i,j}) &\gets\\ &l\_prov\_open(G,\mathcal{Z},a_i,a_j)
    \end{aligned}$
    \State $Open^*_{i,j}(\mathcal{Z}) \gets Open_{i,j} \cup Prov^*_{i,j}(\mathcal{Z})$
    \State $
    \begin{aligned}
Blocked^*_{i,j}(\mathcal{Z}) \gets get\_l\_blocked(&\mathcal{Z},\\
&Open^*_{i,j}(\mathcal{Z}))    
    \end{aligned}$
    
    \State $
    \begin{aligned}
    l\_imp^*_{i,j}(\mathcal{Z}) \gets& compute\_l\_impact(Open^*_{i,j}(\mathcal{Z}),\\&Blocked^*_{i,j}(\mathcal{Z}), Open^+_{i,j})
    \end{aligned}$
    \State $I(\mathcal{Z}) \gets I(\mathcal{Z})\cup l\_imp^*_{i,j}(\mathcal{Z})$     
    \EndFor
    \State $l\_gain(\mathcal{Z}) \gets compute\_l\_gain (I(\mathcal{Z}),C_P)$
\EndFor
\State ${\mathcal{Z}^-} \gets argmax_\mathcal{Z}\; gain(\mathcal{Z})$
\State\Comment{{\em Step 1:} Conditioning step}
\State $D_1,\ldots, D_m \gets {\tt cluster}(D,{\mathcal{Z}^-},m)$
\State\Comment{{\em Step 2:} Candidate skylines per conditioning value}
\State $\forall_i S_i \gets skyline(P,D_i)$
\State\Comment{{\em Step 3:} Merging of candidate skylines}
\State $C \gets \bigcup_i S_i$
\State\Comment{{\em Step 4:} Computing of the final skylines}
\State $S \gets skyline(P,C)$
\EndProcedure
\end{algorithmic}
}
\end{algorithm}

The outline of the algorithm is presented in Algorithm~\ref{alg:lnblock}.

\subsection{Complexity Analysis}\label{sec:complexity} \noteB{[R2.O3]}\replaceB{The algorithm consists}{Both {\em gnSky} and {\em lnSky} algorithms consist} of 5 steps (Steps 0 through 4). We analyze the complexity of each step separately:
\begin{itemize}[leftmargin=*]
\item {\em Step 0 - Gain computation and conditioning set selection:} 
%
%
\maybe{Once the pairwise correlations are available, the remaining} \default{The} cost of this step depends on the number of paths considered which is a function of the number of preference variables and the number of paths in the causal graph between pairs of these variables. More specifically, in the worst case, the algorithm would enumerate $((|P|\times |P|-1)/2)\times 2^{|A|}$ 
number of paths. Each of these paths would then be evaluated for each possible conditioning set. 
Given the above,
the overall worst case cost of the algorithm is \default{$O(|P|^2 \times 2^{2|A|})$ or equally $O(|P|^2 \times 4^{|A|})$.} \maybe{$O(|P|^2 \times (N + 2^{|A|} \times 2^{|A|-|P|}))$ or equally $O(|P|^2 \times (N + 2^{2|A|-|P|}))$.} 

\item {\em Step 1 - Data conditioning:}
\replaceA{Group by based implementation}{GROUP BY}  would require \replaceA{sorting of the data at $O(Nlog(N))$ time}{$O(N)$ time for hashing\noteA{[R1.O4]} if the values are categorical or $O(Nlog(N))$ time if the values are ordinal and we need to sort the data for similarity based grouping.}
%
The cost of the clustering based implementation would depend on $m$, $N$, and the clustering algorithm used. 
\item {\em Step 2 - Candidate skylines:} Assuming that the data is uniformly divided into $m$ subsets, the cost of this step is $m\times cost\_skyline(N/m)$; i.e., $O(m \times (N/m)^2)$. Note that, in the worst case (where one of the subsets is almost as large as the original data), of course, the cost would be $\sim cost\_skyline(N)$, which is, generally, $O(N^2)$. 
\item {\em Step 3 - Merging of candidate skylines:} Let $s_i$ be the number of skylines returned for the $i^{th}$ subset. The cost of this step is $O(\sum_{i} s_i)$\replace{}{.}
\item {\em Step 4 - Obtaining final skylines:} The cost of this step is $cost\_skyline( \sum_{i} s_i)$, which is $O((\sum_{i} s_i)^2)$\replace{}{.}
\end{itemize}
Note that the cost of Step $0$ depends on the size and complexity of the causal graph and, while the worst case size is exponential in the number of attributes, in practice, the number of paths that need to be enumerated is much smaller 
{and, as we see in Section~\ref{sec:exp},} the cost of Step 0, (which is \default{independent of} the data size, $N$) is generally negligible relative to the cost of skyline computations, which is $O(N^2)$. 
%
\maybe{Moreover, once the impact values for a causal graph is computed, the computation of the gain value for a new data set with the same causal structure can be done at
$O(|P|^2 \times N) + 2^{|A|-|P|}$ time, reusing the impact values that are structural and do not change from one data set to another.} \default{Moreover, once the impact values for a causal graph is computed, the gain values can be re-used without any recomputation for a new data set with the same causal structure.} 
%
Consequently, the overall cost of Steps 1 through 4 depends on (a) how uniformly the data conditioning step splits the data into $m$ subsets and (b) how well the conditioning reduces the number of skyline objects (through elimination of negative correlations among the preference attributes).

\subsection{
Computation of the 
Scaling Parameters}\label{sec:opt}
The Algorithm $\#2$ (\replaceC{ln-Skyline}{\em lnSky}) alleviates the shortcomings of Algorithm $\#1$ (\replaceC{gn-Skyline}{\em gnSky}) by accounting for the leaky blocking and noisy causal information passage by leveraging two scaling factors, $\lambda_b$ and $\lambda_o$, respectively.
We therefore need a mechanism to 
quantify these degrees of leakage and noise as a function of the structure of the causal graph. 
Let $p$ be a path between two preference variables $X$ and $Y$ and $\mathcal{Z}$ be the set of conditioning attributes; to compute the conditioning gain for the path, $p$, we need to compute two terms:
\begin{itemize}[leftmargin=*]
\item $corr_p(X,Y)$; i.e., the correlation between $X$ and $Y$, implied by the causal path $p$,  before the conditioning step, and
\item $corr_{p,\mathcal{Z}}(X,Y)$; i.e., the correlation between $X$ and $Y$, implied by the causal path $p$,  after the conditioning of attributes in $\mathcal{Z}$.    
\end{itemize}
Given these, the conditioning gain for the path given the conditioning set $\mathcal{Z}$ can be computed as $c\_gain(p,{\mathcal{Z}}) = corr_{p,\mathcal{Z}}(X,Y) - corr_p(X,Y).$
\default{
We discuss how to compute these terms, under the assumption that the 
 sets of variables have multi-variate Gaussian distributions, in Appendix \ref{sec:appendix_a}}.


\section{Experiments and Analysis}\label{sec:exp}

Next, we present experiment results evaluating the effectiveness of the 
proposed
causally-informed skyline (CSS) search\footnote{The source code, along with all causal graphs used in our experiments and their corresponding preference sets and dominance criteria are available at 
\url{https://github.com/EmitLab/Causal-Search-for-Skylines}.}.

\begin{table}[t]
\caption{Experiment parameters (bold  indicate default values)
}\label{exp:param}
\begin{adjustbox}{width=0.65\columnwidth}
\begin{tabular}{|l|c|}\hline
{\bf Parameter}&{\bf Values}\\ \hline\hline
Number of data elements & 50K, 100K, {\bf 200K}, 400K\\ \hline
Number of preference attributes & 2, 3, 4\\ \hline
Number of de-correlation clusters & 2 - 9, {\bf 10}, 11 - 20\\ \hline
$\langle\lambda_o,\lambda_b\rangle$ pairs & \makecell{$\langle1.0,0.0\rangle$, $\langle0.9,0.1\rangle$, $\langle0.8,0.2\rangle$,\\ $\langle0.7,0.3\rangle$, $\bm{\langle0.6,0.4\rangle}$}\\ \hline
\end{tabular}
\end{adjustbox}
\end{table}

\begin{table}[t]
\caption{Real Causal Graphs}\label{exp:real}
\begin{adjustbox}{width=0.7\columnwidth}
\begin{tabular}{|l|c|c|}\hline
{\bf Name}&{\bf \# Variates}&{\bf \# Tuples}\\ \hline\hline
Abalone \cite{abalone_1} & 8 & 4177\\ \hline
Adult \cite{adult_2} & 14 & 48842\\ \hline
Auto MPG \cite{auto_mpg_9} & 7 & 398\\ \hline
Concrete Compressive Strength \cite{concrete_compressive_strength_165} & 9 & 1030\\ \hline
Dry Bean \cite{dry_bean_602} & 16 & 13611\\ \hline
Hong Kong Weather \cite{Hindy_2021} & 10 & 18262\\ \hline
Individual Household Power Consumption \cite{individual_household_electric_power_consumption_235} & 9 & 2075259\\ \hline
Seoul Bike Demand \cite{seoul_bike_sharing_demand_560} & 13 & 8760\\ \hline
Wine Quality (White) \cite{wine_quality_186} & 11 & 4898\\ \hline
\end{tabular}
\end{adjustbox}
\end{table}

\subsection{Experiment Setup}

\subsubsection{Causal Graphs and Data Sets}\label{sec:cg-datasets}
We consider data sets with both real and synthetic causal graphs for evaluating CSS. For each causal graph, we consider difference preference sets, with varying numbers of preference attributes.
\begin{itemize}[leftmargin=*]
\item{\em Real Causal Graphs:} Table~\ref{exp:real} presents the real causal graphs 
considered in these experiments. Since in this section we also want to investigate the impact of the data size, we varied the data set size for these graphs by leveraging data augmentation using perturbed oversampling with Gaussian noise \cite{beinecke2021gaussian, 10.1145/2907070}.

\item{\em Synthetic Causal Graphs:} In addition to these real causal graphs, we also considered a number of synthetic causal graphs, 
with varying complexities (number of nodes, edge densities, and different causal primitives). The causal edge weights for these causal graphs were set to $+1$ or \default{$+0.5$ for strongly and weakly positive causal edges, respectively. Similarly, $-1.0$ and $-0.5$ were used for strongly and weakly negative causal edges, respectively.} 
The data for these causal graphs were also generated synthetically.

\item \replaceB{}{
{\em Inferred Causal Graphs:} \noteB{[R2.O2]} We also experimented with causal graphs inferred from the data, rather than being provided as  input by the user.
While there are many causal discovery algorithms~\cite{chickering2002optimal, spirtes2000causation, shimizu2006linear}, in the experiments, we used the NOTEARS~\cite{zheng2018dags}  causal inference algorithm, which assumes a simple linear structural equation model, to discover the causal graph and the  edge weights.
}

\end{itemize}

\subsubsection{Skyline Algorithms}\label{sec:sky-algos}
As competitors to CSS, we consider the following skyline algorithms:
\begin{itemize}[leftmargin=*]
\item{\em Block Nested Loop Skylines (BNL\maybe{and BNL-HM})~\cite{Borzsonyi01theskyline}} BNL scans the dataset once, keeping a window of currently non-dominated tuples -- no specific ordering of tuples is assumed. Every new tuple is compared against this window: the tuple is discarded if it is dominated or inserted into the window, evicting any tuples it dominates.  \maybe{In addition, we also consider a {\em hierarchical merge } version of the BNL algorithm (BNL-HM), which}
\item{\em Sort-First Skylines (SFS)~\cite{Chomicki03skylinewith}} SFS sorts the data on a monotone ranking criterion ({\em $\langle sum \rangle$} in the experiments), then performs a scan in that order, maintaining a window of candidates\replaceA{}{\footnote{\label{fn:LESS}\replaceA{}{Although LESS~\cite{godfrey2005maximal} improves upon SFS\noteA{[R1.O5]}, we opted for SFS instead due to its lesser complexity and fewer hyper-parameters. In fact, the primary optimization in LESS, the elimination filter (EF), is applied during the external sorting stage, which does not conflict with our proposal.}}}. The presort reduces the number of comparisons. 
%
%

\item{\em Sort-and-Limit Skyline (SaLSa)~\cite{Salsa}} SaLSa builds on SFS by introducing "lazy" sorting: it interleaves (partial) sorting with skyline testing, in such a way that any removed tuple cannot re-enter the skyline. For SaLSa, we have used the {\em $\langle$max, sum$\rangle$} monotonic function as recommended by the authors.
\item {\em Branch and Bound Skyline (BBS)~\cite{PapadiasTFS05}} BBS operates on an R-tree index, exploring nodes in best-first order using a priority queue \replace{-- The key advantage of this approach is that}{pruning} whole sub-trees \replace{can be pruned}{} if their bounding rectangles are dominated by skyline tuples already found. 
\item {\em Divide and Conquer Skylines (D\&C)~\cite{Borzsonyi01theskyline, KungLP75}} D\&C partitions the space along the median values into sub-regions, computes the skyline of each partition independently\replace{}{,}
and \replace{then}{} merges the partial skylines\replace{,}{} discarding any tuples dominated by another from any partition. \replace{We note that, t}{T}he divide-and-conquer strategy of the D\&C algorithm can be seen as pure de-correlation of the preference attributes.
\end{itemize}

\subsubsection{CSS Implementation}
For each of the above skyline algorithms, we consider the causal version as follows: (a) given a causal graph and the set of preference attributes, we compute the expected gains for each possible conditioning set; (b) given the target number of de-correlation clusters, we apply Euclidean-distance based $k$-means on the conditioning attributes, normalized to zero mean and unit variance; (c) we compute the causal skylines for each cluster\footnote{While both D\&C and CSS allow for parallelization, in this section we assume sequential executions of these algorithms.},  
and (d) obtained skylines from the clusters are unioned \replace{}{and}
\default{the skyline is searched among these tuples}.
%

Note that the leaky negative path blocking (\replaceC{lnBlock}{\em lnSky}) algorithm  presented in Section~\ref{sec:lnblock} uses two parameters $\lambda_b \ll \lambda_o$, denoting the degrees of information passage on imperfectly open and blocked nodes. Table~\ref{exp:param} lists the values considered in the experiments.

\default{In our implementation, all algorithms use the main memory to compute the skylines, and no temporary file was ever needed (i.e., BNL, SFS, and SaLSa always completed in a single pass).}
\default{The skyline algorithms, their corresponding causal equivalents, and K-means clustering have been implemented in Python 3.12.7 with Numpy 2.1.2. 
The experiments were executed using Chameleon~\cite{keahey2020lessons} on\replace{systems with}{} AMD EPYC 7763 64-Core Processor, 256GB RAM, and 500GB SSD.}

\subsubsection{Baselines}
In addition, for each of the above skyline algorithms, we consider the following baselines:
\begin{itemize}[leftmargin=*]
\item{\em Non-causal:} This is the original, non-causal\replace{version of the}{} algorithm.
\item \default{{\em Data-driven \replace{}{(ddSky)}:} Here, we utilize the data-driven strategy to compute the expected gains for each possible conditioning set and subsequently follow the clustering based approach for computing skylines as outlined in the leaky negative path blocking skyline (\replaceC{lnBlock}{\em lnSky}) algorithm.}
\item{\em
Preference de-correlation:} In this baseline, we consider a non-selective,
\default{preference attribute based} de-correlation strategy: we apply K-means clustering directly on the set of preference attributes, thereby de-correlating the data without paying attention to whether this would eliminate positive or negative correlations.
\end{itemize}

\subsubsection{Evaluation Criteria}
As evaluation criteria, we consider both (a) {\em the number of dominance checks}
 and (b) {\em execution time} of the skyline algorithm (wall clock), which also includes the time to de-correlate the data based on the selected de-correlation attribute set.
We separately report the time to compute the de-correlation set.
Since the dominance checks show similar patterns, we report them in  Appendix~\ref{sec:app_c}.

\begin{figure}[t]
\centerline{
\begin{tabular}{cc}
\includegraphics[width=0.47\columnwidth]{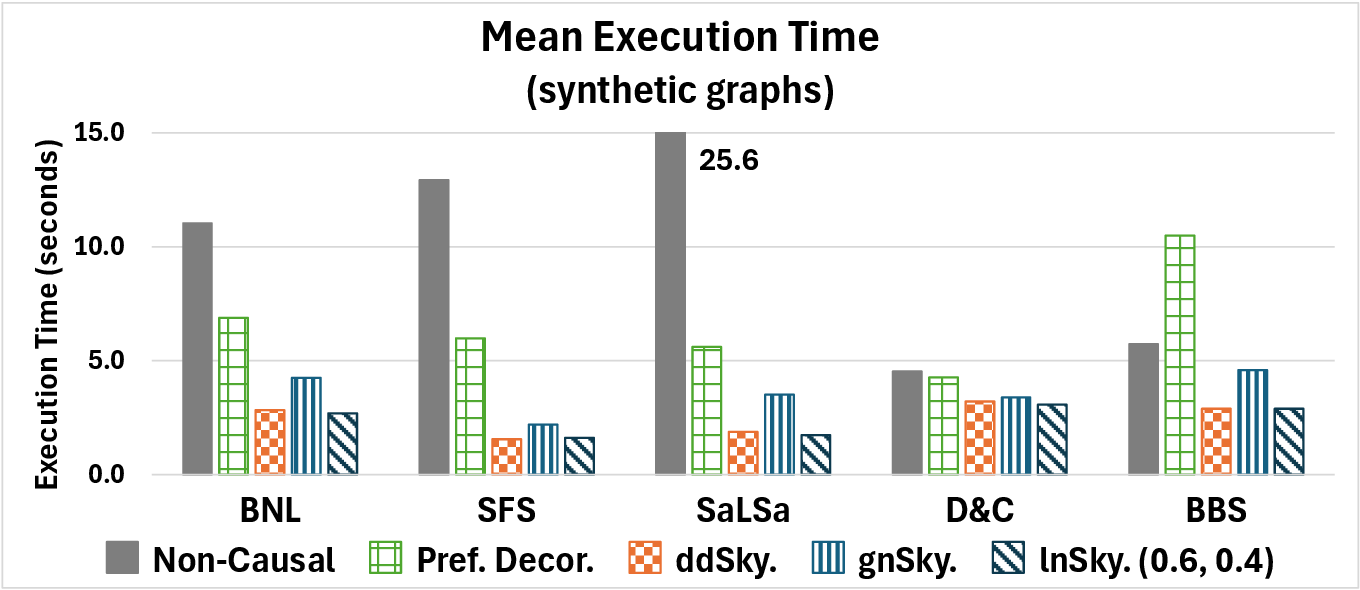} &
\includegraphics[width=0.47\columnwidth]{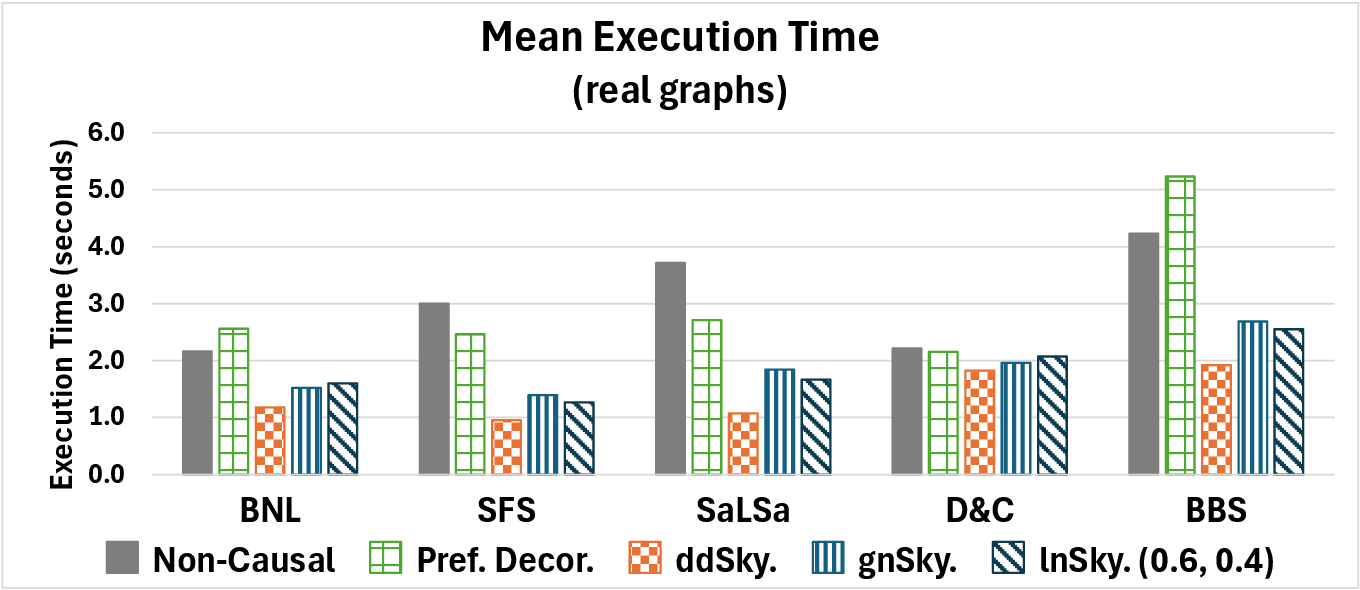}\\
\multicolumn{1}{c}{(a) Synthetic causal graphs} & \multicolumn{1}{c}{(b) Real-world causal graphs}\\
\end{tabular}
}
\caption{Cost of skyline\replace{ computation}{s} with (a) synthetic and (b) real-world causal graphs \underline{(the lower, the better)} -- 200K data}\label{fig:absolute}
\end{figure}

\begin{figure}[t]
\centerline{
\begin{tabular}{cc}
\includegraphics[width=0.47\columnwidth]{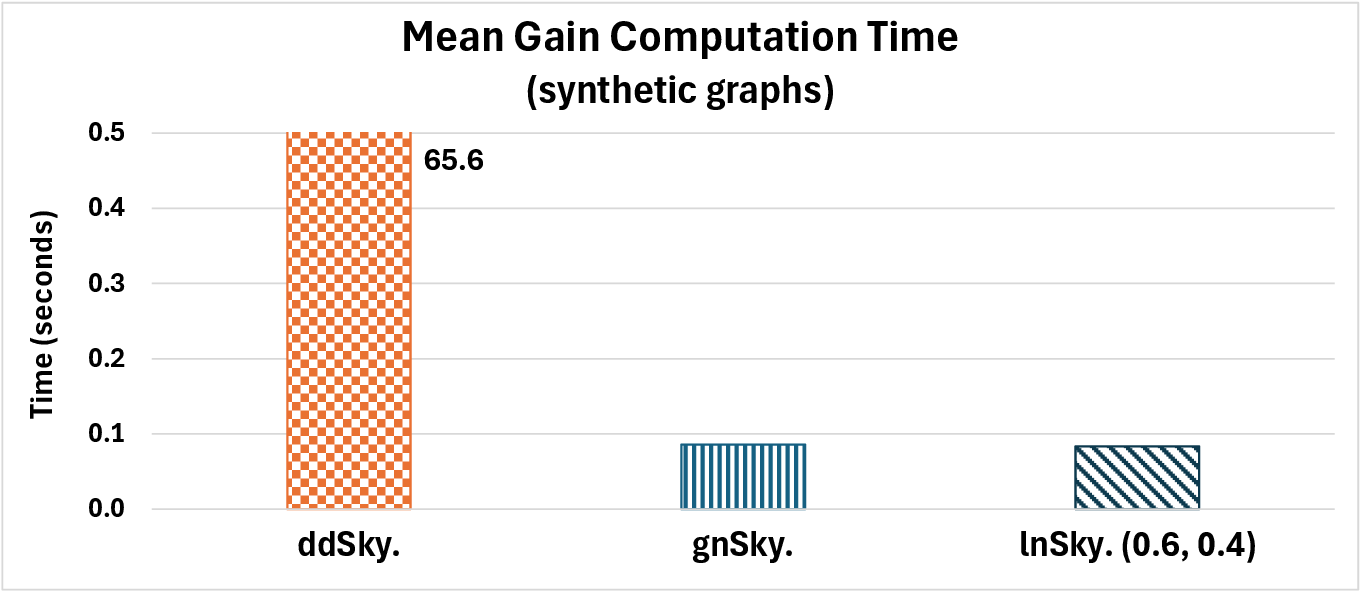} & 
\includegraphics[width=0.47\columnwidth]{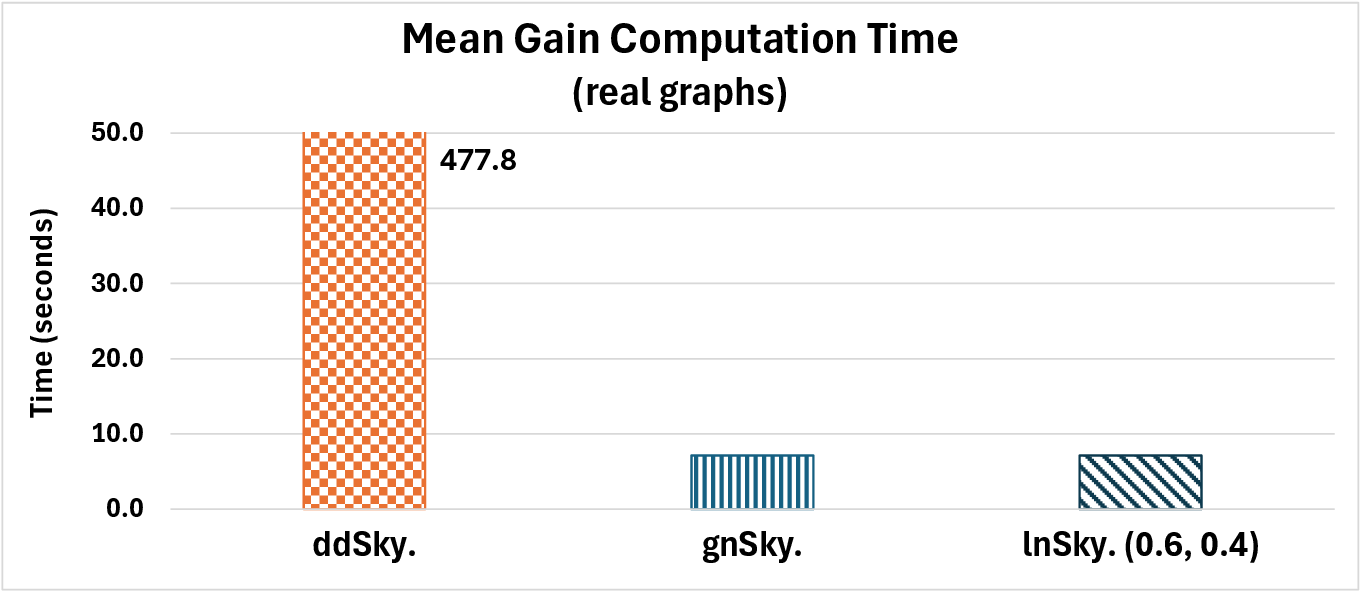}\\
(a) Synthetic causal graphs &(b) Real-world causal graphs
\end{tabular}
}
\caption{Cost of the expected gain computation step
\underline{(the lower, the better)} -- 200K data
}
\label{fig:gain_time}
\end{figure}

\begin{figure}[t]
\centerline{
\begin{tabular}{cc}
\includegraphics[width=0.47\columnwidth]{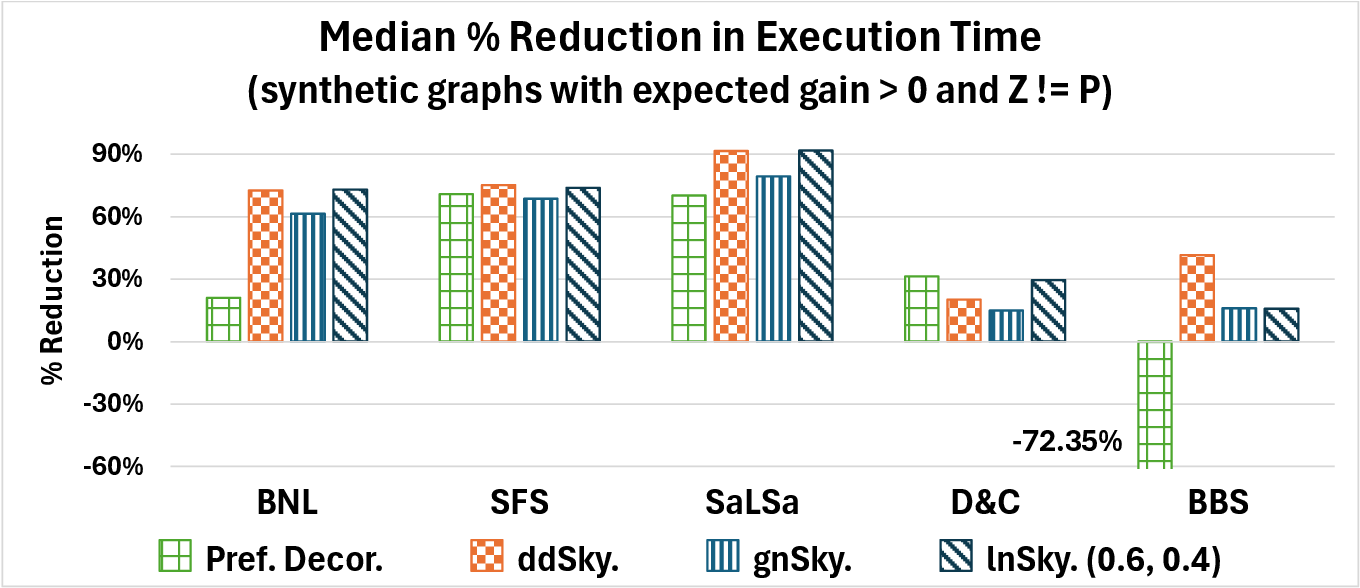}&
\includegraphics[width=0.47\columnwidth]{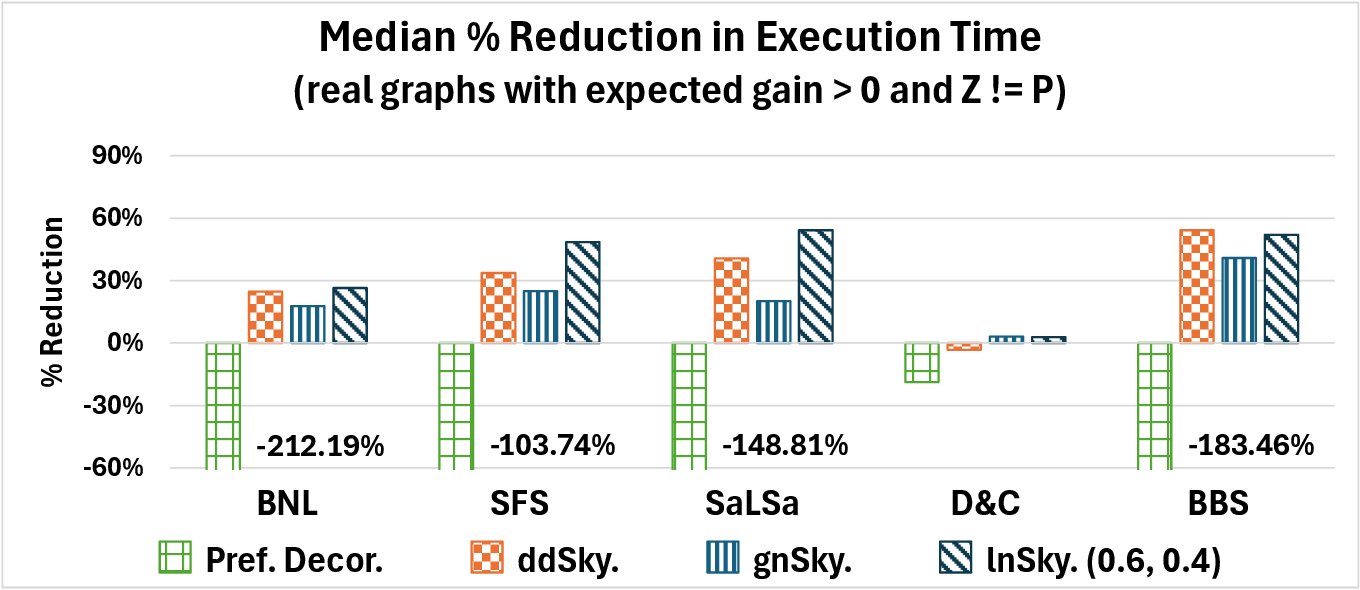}\\
\multicolumn{1}{c}{ (a) Synthetic causal graphs}& \multicolumn{1}{c}{ (b) Real-world causal graphs}\\
\end{tabular}
}
\caption{Median cost reduction \underline{(the higher, the better)}; for these experiments,  expected gain is positive and larger than the gain predicted for the preference attribute set  (i.e., {\em we \underline{expect to see benefits} over both vanilla skylines and preference attribute de-correlation}) -- 200K data}\label{fig:relative-good}
\end{figure}


\begin{figure}[t]
\centerline{
\begin{tabular}{cc}
\includegraphics[width=0.47\columnwidth]{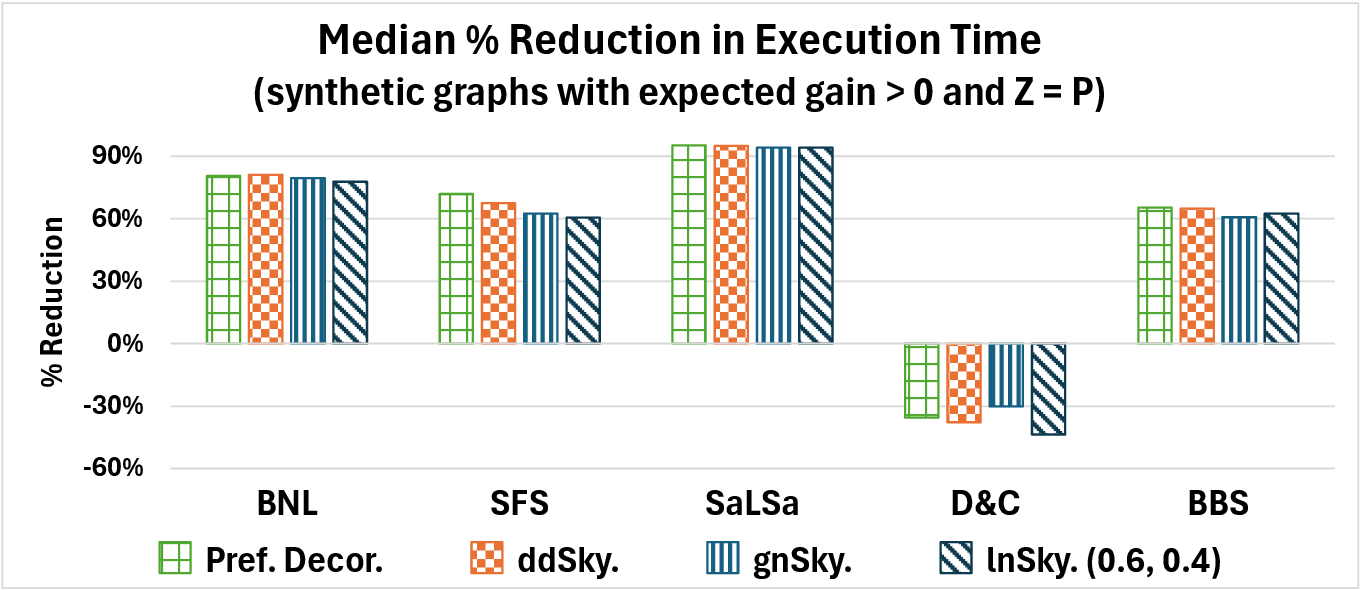} & \includegraphics[width=0.47\columnwidth]{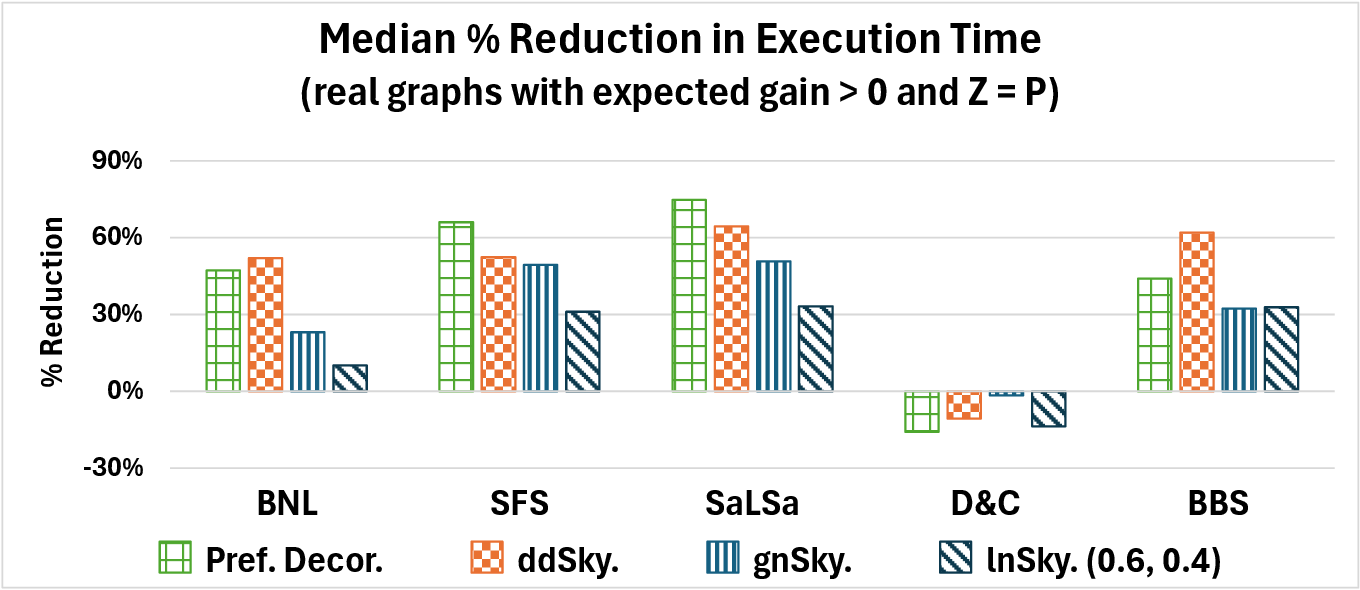}\\
(a) Synthetic causal graphs & (b) Real-world causal graphs\\
\end{tabular}
}
\caption{Median cost reduction \underline{(the higher, the better)}; for these experiments,   expected gain is positive and equal to the gain predicted for the preference attribute set (i.e., we expect to see benefits over \replace{both}{} vanilla skylines, but we do not expect to beat preference attribute de-correlation) -- 200K data}\label{fig:relative-pref}
\end{figure}


\begin{figure}[t]
\centerline{
\begin{tabular}{cc}
\includegraphics[width=0.47\columnwidth]{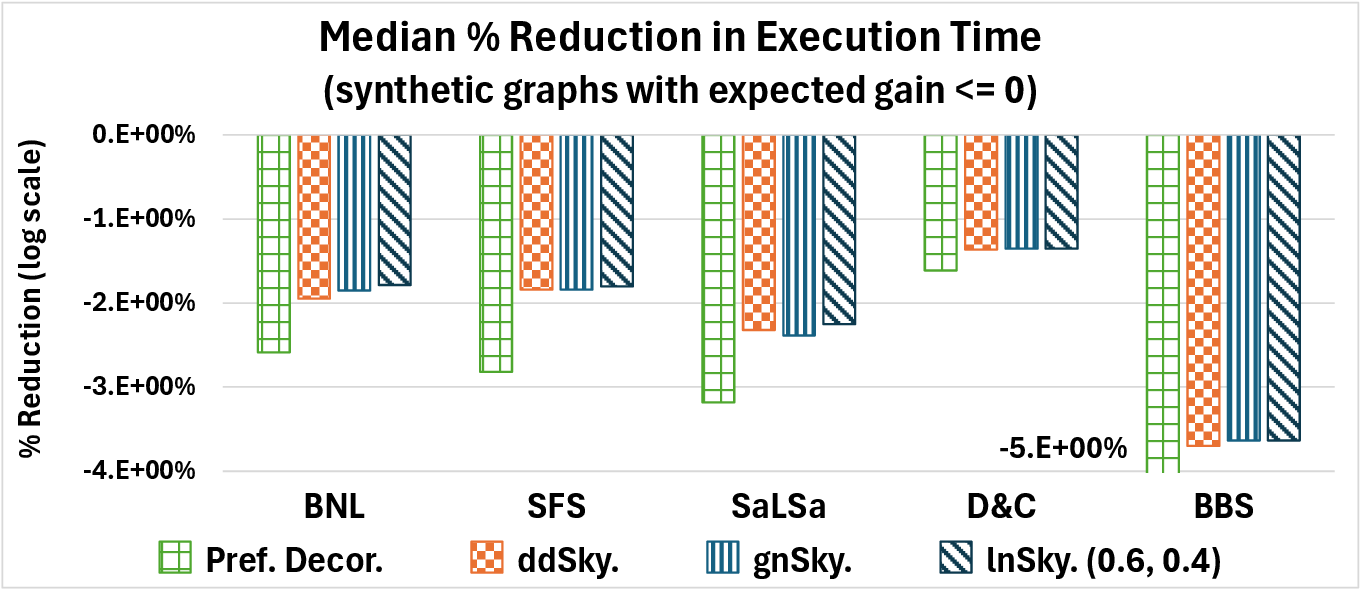}&
\includegraphics[width=0.47\columnwidth]{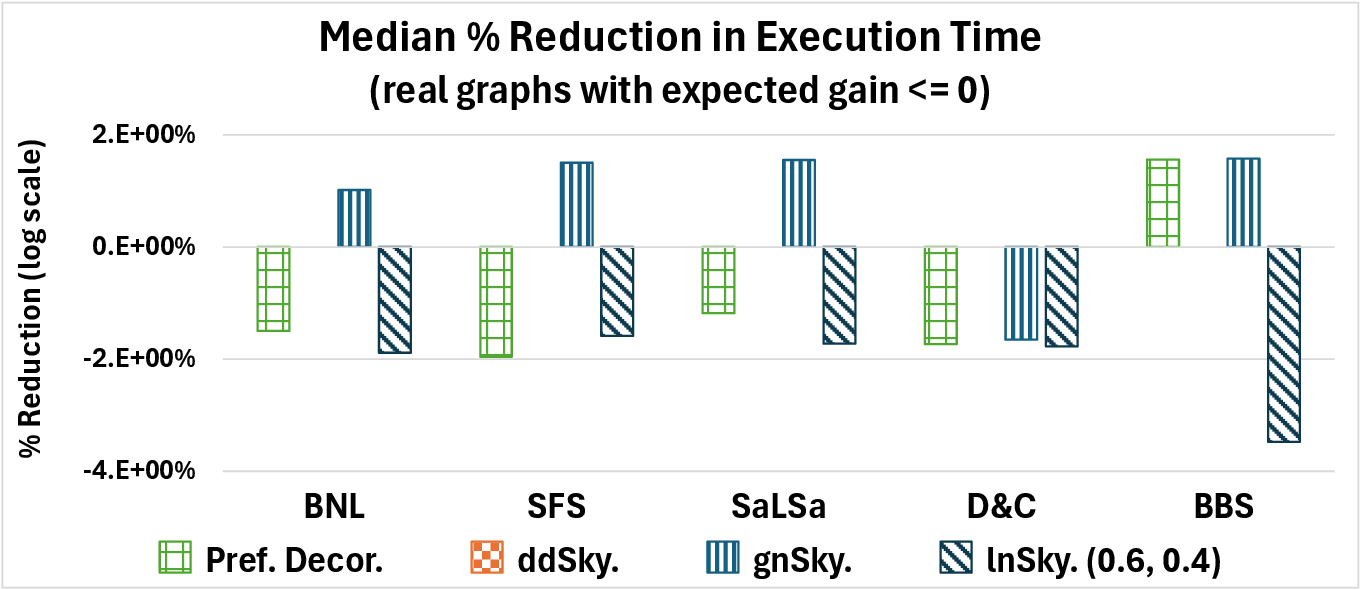}\\

(a) Synthetic causal graphs& (b) Real-world causal graphs\\
\end{tabular}
}
\caption{Median cost reduction \underline{(the higher, the better)}; for these experiments,  expected gain is negative (i.e., {\em we do not expect to see benefits from de-correlation}) -- 200K data}\label{fig:relative-negative}
\end{figure}

\begin{figure}[t]
\centerline{
\begin{tabular}{cc}
\includegraphics[width=0.47\columnwidth]{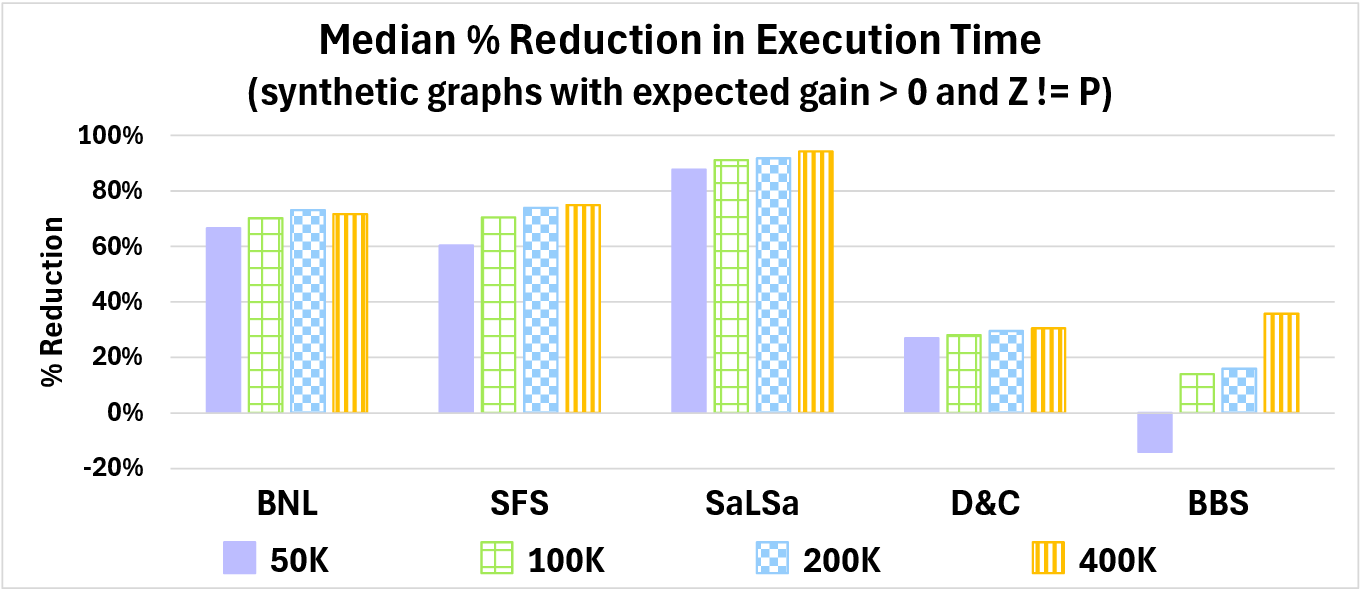}&\includegraphics[width=0.47\columnwidth]{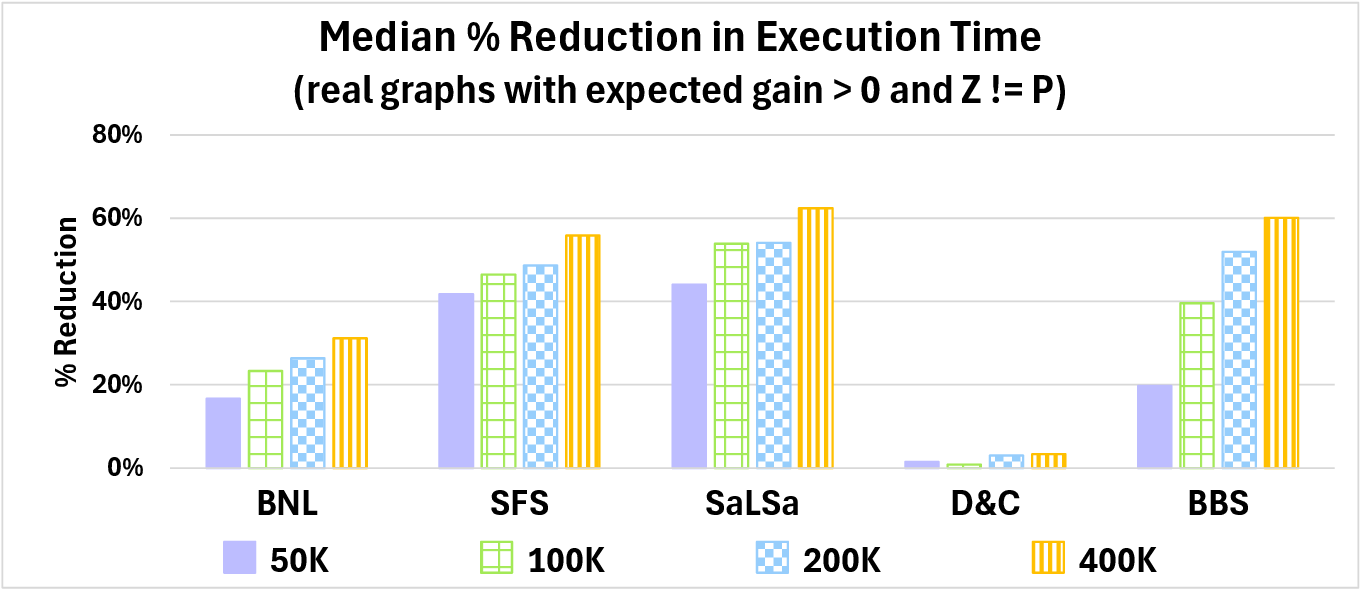}\\
(a) Synthetic causal graphs & (b) Real-world causal graphs\\
\end{tabular}
}
\caption{Median cost reduction \underline{(the higher, the better)} for different numbers of tuples when using \default{{\em \replaceC{lnSkyline}{lnSky}(0.6,\nsbp0.4)}}; for these experiments, the expected gain is positive and larger than the gain predicted for the preference attribute set (i.e., {\em we expect to see benefits over both vanilla skylines and preference attribute de-correlation})}\label{fig:tuples-var}
\end{figure}


\begin{figure}[t]
\centerline{
\begin{tabular}{cc}
\includegraphics[width=0.47\columnwidth]{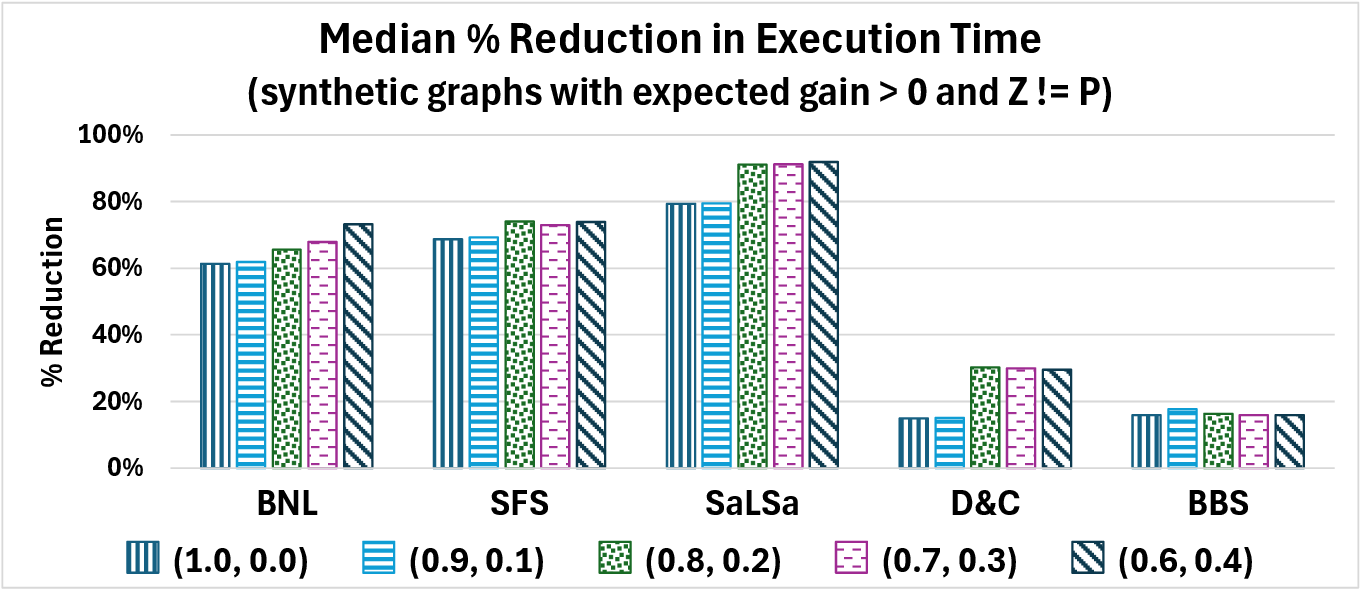} & \includegraphics[width=0.47\columnwidth]{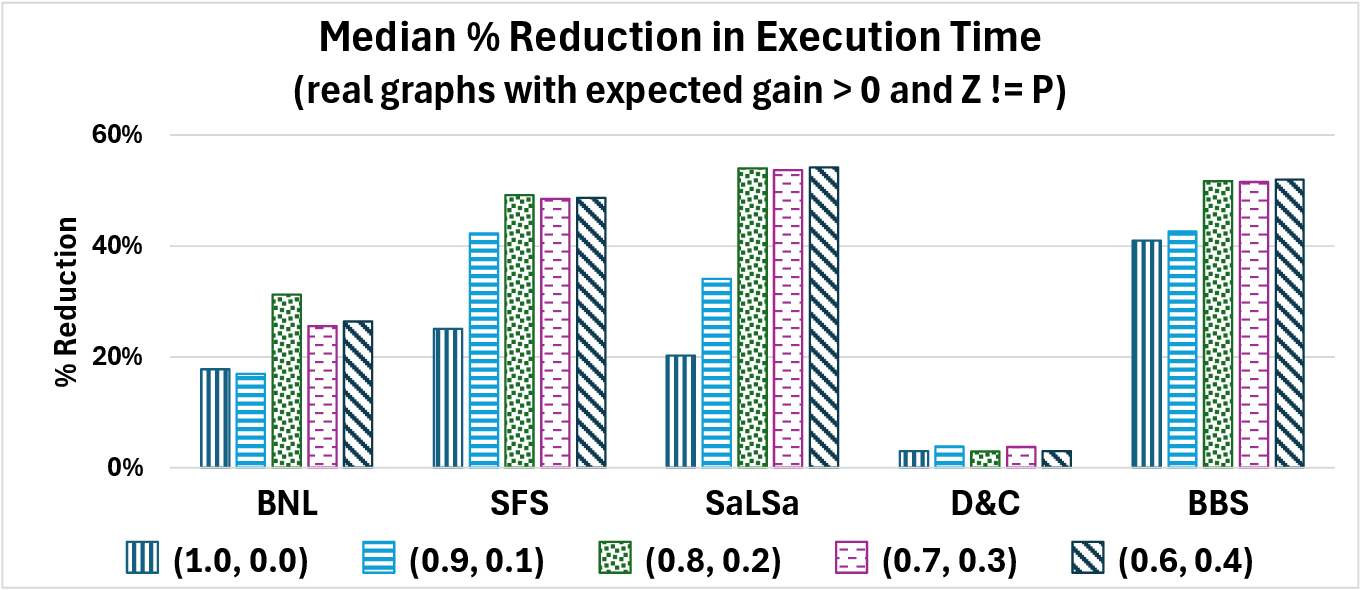}\\
(a) Synthetic causal graphs & (b) Real-world causal graphs\\
\end{tabular}
}
\caption{
Median cost reduction \underline{(the higher, the better)} for \replace{different}{} $\langle\lambda_o,\lambda_b\rangle$ pairs; for these experiments,  expected gain is positive and larger than the gain predicted for the preference attribute set (i.e., {\em we expect to see benefits over both vanilla skylines and preference attribute de-correlation}) -- 200K data}\label{fig:leakage-var}
\end{figure}

\subsection{Experiment Results}

In this section, we present and interpret the experiment results on both synthetic and real-world causal graphs. Importantly, since the divide-and-conquer (D\&C) algorithm can be seen as the pure de-correlation of the preference attributes, we distinguish between scenarios where there is a conditioning attribute set predicted to perform better than conditioning on the preference attribute set and the cases where the causal analysis predicts that the preference attribute set itself is the best conditioning attribute set.
%

\subsubsection{Effectiveness of Causal Search for Skylines (CSS)}\label{sec:results_effective}\label{sec:gainexp}

\paragraph{\bf Overview.}
The way CSS would be used in practice is to leverage the causal version of the algorithm if there exists a conditioning attribute set with positive predicted gain and revert to the vanilla version otherwise. The benefits of this approach is visualized in Figure~\ref{fig:absolute}: 
%
As we see in the figure, 
assuming that the expected gains have already been computed,
both {\em \replace{ddSkyline}{ddSky}} and {\em \replaceC{lnSkyline}{lnSky}} substantially outperform the vanilla versions, indicating that this adaptive strategy effectively leverages CSS when it is predicted to be advantageous, while defaulting to the vanilla approach in scenarios where CSS is not expected to be ideal.

\default{
Interestingly, when \replace{}{the} {\em \replaceC{lnSkyline}{lnSky}} \replace{algorithms are}{algorithm is} considered, SFS and SaLSa become the fastest algorithms due to the significant reduction of dominance checks from which they benefit, even beating more recent algorithms, such as D$\&$C and BBS.
}

\paragraph{\bf Cost of the Expected Gain Computation Step} 
Figure \ref{fig:gain_time} shows the time required to identify the best conditioning attribute set based on different strategies (Step 0).
%
As we expected (see Section~\ref{sec:ddCSS}), the gain computation phase of {\em \replace{ddSkyline}{ddSky}} is prohibitively expensive. This prevents it from being applicable in practice, even though it can provide significant gains in terms of 
execution time as we have already seen in Figure~\ref{fig:absolute}.
%
%
Importantly, the cost of gain computation for {\em \replaceC{lnSkyline,}{lnSky}} (which is  \default{independent of} the data size, $N$) is  negligible relative to the cost of skyline computations, making \replace{the causal
graph  based}{}  {\em \replaceC{lnSkyline}{lnSky}} to be the more practical de-correlation strategy.

\paragraph{\bf Effectiveness of the Gain Prediction}
In Figure~\ref{fig:absolute}, we have seen that gain prediction can be used \replace{as tool}{} to decide whether and how to de-correlate the data to obtain significant reductions in the cost of skyline search.
%
Figure~\ref{fig:relative-good} through~\ref{fig:relative-negative} deep dives in the investigation of the effectiveness of the gain prediction by focusing on the cases where (a) the expected gain is positive and better than that of the preference attributes, (b) gain is positive but not better than that of the preference attributes, and (c) expected gain is negative: 

{\bf (a)} Figure~\ref{fig:relative-good} 
presents the results for scenarios in which there exists a conditioning attribute set with a positive expected gain, {\em predicted to outperform conditioning on the preference attribute set}.
%
As we see here, in these scenarios,
%
both {\em \replace{ddSkyline}{ddSky}} and {\em \replaceC{lnSkyline}{lnSky}} significantly outperform vanilla (non-causal) baseline algorithms as well as preference-attribute set based de-correlation. 
%
Interestingly, in the case of real-world causal graphs, the preference-attribute based de-correlation performs even worse than the vanilla skyline algorithms, while the proposed data- and causality-driven extensions provide significant gains. 

{\bf (b)}
Figure~\ref{fig:relative-pref} shows the results for 200K data samples in scenarios where a positive expected gain exists, but the best conditioning attribute set is {\em predicted to coincide with the preference attribute set itself.}
%
As expected, in the synthetic case, all strategies perform similarly with the preference-attribute based de-correlation or {\em \replace{ddSkyline}{ddSky}}.
%
In the real data case, we see that {\em \replaceC{lnSkyline}{lnSky}} has a somewhat lower reduction in 
execution time, indicating that the causal gain prediction does not perfectly identify these scenarios.

{\bf (c)}
Figure~\ref{fig:relative-negative} shows the results for 200K data samples in scenarios where no conditioning attribute set with positive expected gain exists\footnote{ 
Note that for real causal graphs, we did not observe any scenario where {\em \replace{ddSkyline}{ddSky}} finds a conditioning attribute set with negative gain.}.
As expected, conditioning on an attribute set with negative expected gain significantly hurts performance.
%
Interestingly, {\em \replaceC{gnSkyline}{gnSky}} shows improvements in certain cases with real causal graphs. This likely occurs because {\em \replaceC{gnSkyline}{gnSky}} misclassifies conditioning attribute sets: it sometimes assigns negative gain to sets that actually have positive gain. Conversely, this also explains why {\em \replaceC{gnSkyline}{gnSky}} performs worse than {\em \replaceC{lnSkyline}{lnSky}} in cases with true positive gain, as it incorrectly predicts negative conditioning attribute sets to yield positive gain.
%
In contrast, if {\em \replaceC{lnSkyline}{lnSky}} predicts that a conditioning attribute set has negative gain, it should be avoided. In such cases, the non-causal version is preferable.

\subsubsection{Impact of the Number of Tuples}
As we see in Figure \ref{fig:tuples-var}, for both synthetic and real-world graphs, performance gains  improve as the size of the data gets larger, indicating that the proposed {\em \replaceC{lnSkyline}{lnSky}} de-correlation strategy enables the skyline algorithms to scale better with the data size.

\begin{figure}[t]
\centerline{
\begin{tabular}{cc}
\includegraphics[width=0.47\columnwidth]{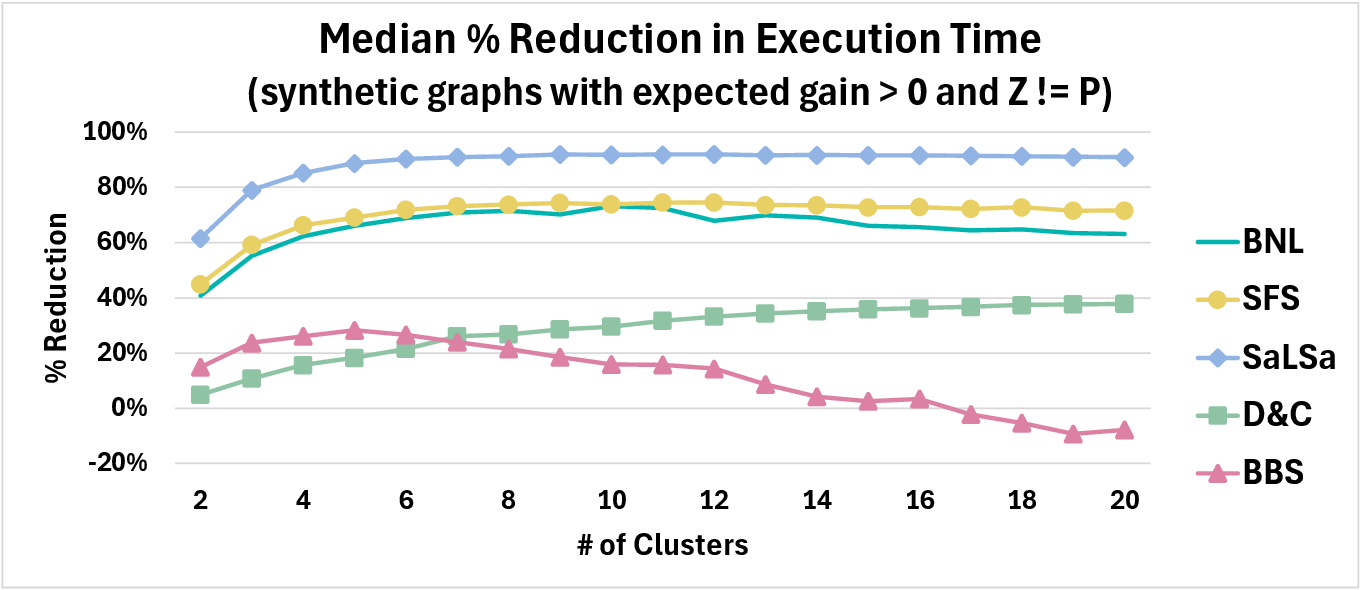} &
\includegraphics[width=0.47\columnwidth]{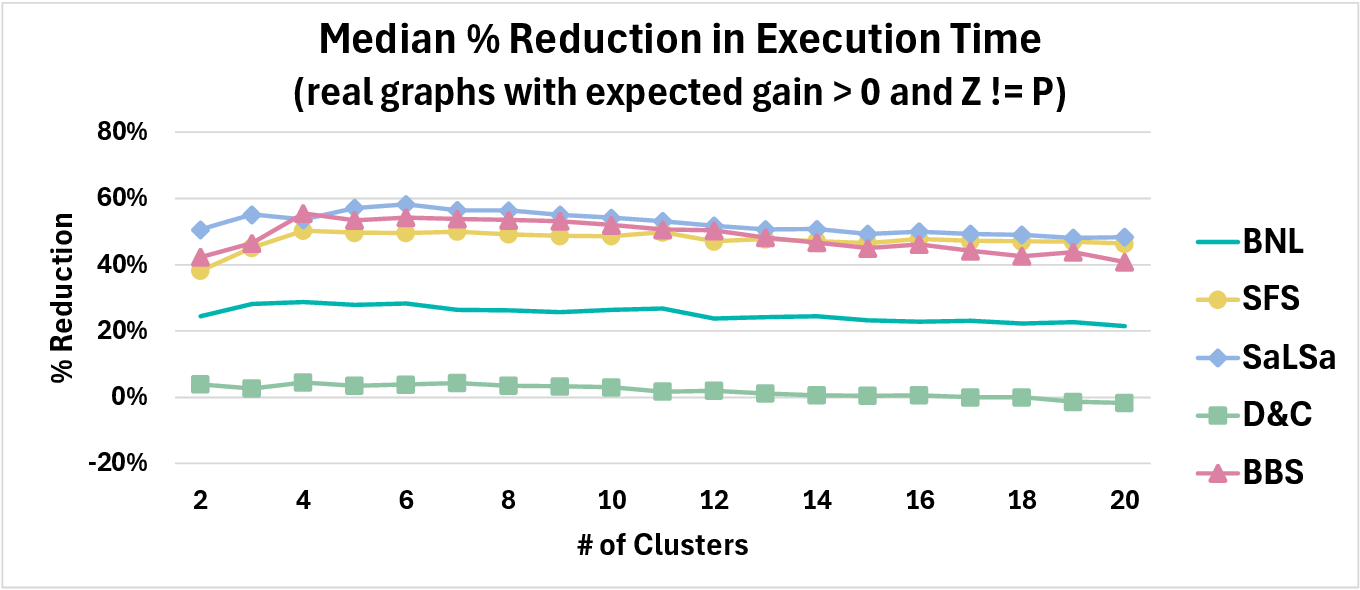}\\
(a) Synthetic causal graphs & (b) Real-world causal graphs\\
\end{tabular}
}
\caption{
Median cost reduction \underline{(the higher, the better)} for different numbers of de-correlation clusters when using \default{{\em \replaceC{lnSkyline}{lnSky}(0.6,\nsbp0.4)}}; for these experiments, the expected gain is positive and larger than the gain predicted for the preference attribute set (i.e., {\em we expect to see benefits over both vanilla skylines and preference attribute de-correlation}) -- 200K data}\label{fig:cluster-var}
\end{figure}

\subsubsection{Effect of the \replaceC{lnSkyline}{lnSky} Parameters $\langle\lambda_o,\lambda_b\rangle$}
Figure \ref{fig:leakage-var} shows the effect of leakage parameters on \replaceC{lnSkyline}{\em lnSky} search time. 
For synthetic graphs, {\em \replaceC{lnSkyline}{lnSky}(0.6, 0.4)} performs the best.
For real-world graphs,  the best option is {\em \replaceC{lnSkyline}{lnSky}(0.8, 0.2)}.
\replace{Note that i}{I}n the rest of this section, we \replace{have used}{use} the leakage parameter pair {\em \replace{(0.6, 0.4)}{$\langle 0.6, 0.4 \rangle$}} for both synthetic and real-world graphs for overall experimental setup consistency, even though that pair is not the best option for real-world graphs.

\subsubsection{Effect of Number of De-correlation Clusters}
As outlined in Section~\ref{sec:css}, the proposed CSS approach relies on K-means clusters of (suitably normalized) data to achieve selective de-correlation.
Figure~\ref{fig:cluster-var} illustrates how the number of de-correlation clusters affects the skyline search efficiency.
As we see in the figure, a small number of clusters (on the order of 10-15 clusters for 200K data) is sufficient to obtain the necessary de-correlation -- a larger number of clusters either do not provide additional benefits or (in the case of causal BNL) actually starts hurting the performance due to the increasing merge overhead.

\begin{figure}[t]
\centerline{
\begin{tabular}{c}
\includegraphics[width=0.6\columnwidth]{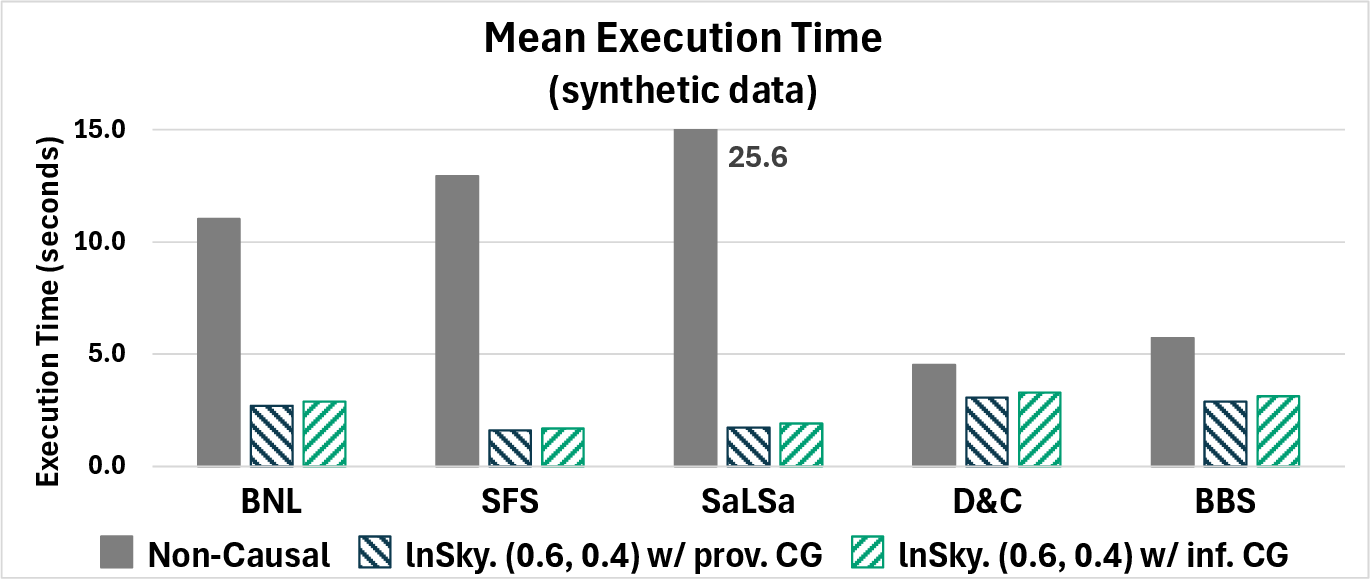}\\ \replaceB{}{(a) causal graphs inferred from synthetic data}\\
\includegraphics[width=0.6\columnwidth]{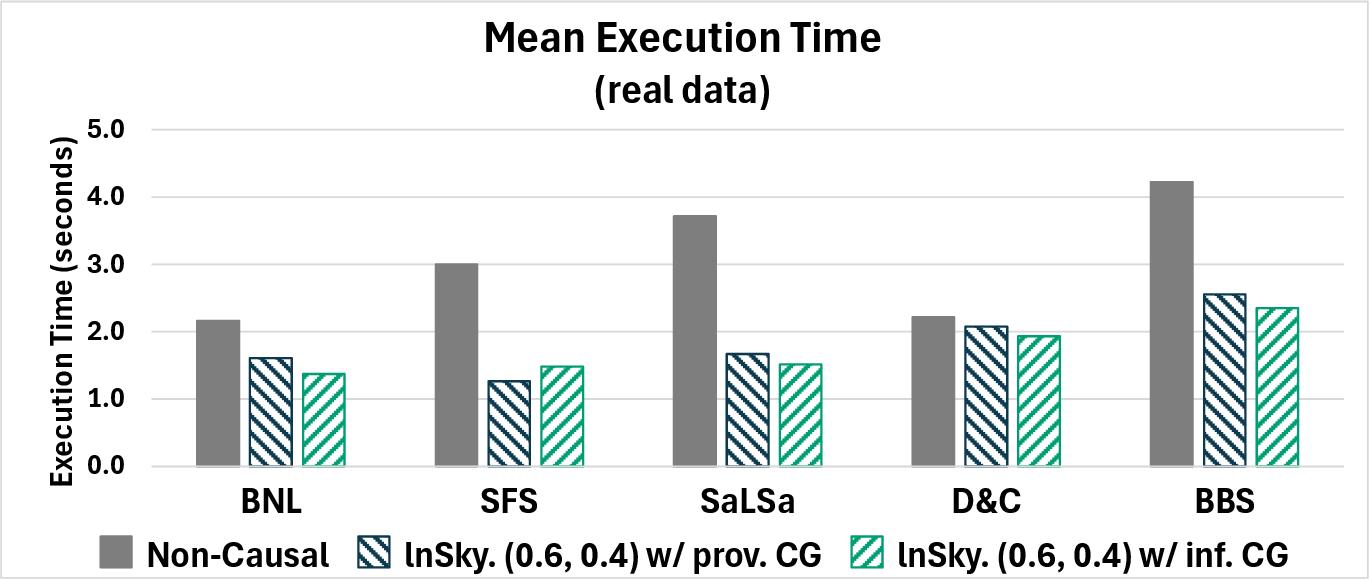}\\
 \replaceB{}{(b) causal graphs inferred from real data}
\end{tabular}
}
\caption{
\replaceB{}{Cost of skyline computation with causal graphs inferred from (a) synthetic data (i.e., data generated from synthetic causal graphs) and (b) real data \underline{(the lower, the better)}, when using {\em lnSky(0.6,\nsbp0.4)} -- 200K data}}\label{fig:inferred-graphs}
\end{figure}

\replaceB{}{
\subsubsection{Impact of Inferred Causal Graphs}\label{sec:inferred}
In the preceding experiments,\noteB{[R2.O2]} we assumed that both the causal graph structure and the corresponding edge strengths were user provided. In practice, however, such causal knowledge may not be available.
In such scenarios causal inference and causal effect estimation techniques  can help 
infer the underlying causal graph
and causal strengths. 
In this section, instead of using the user provided graphs, we use the causal graphs inferred by the 
NOTEARS~\cite{zheng2018dags}  causal inference algorithm from (a) synthetic data (i.e., data generated from synthetic causal graphs) and (b) real-world data.
%
Figure~\ref{fig:inferred-graphs} compares the performance of {\em lnSky(0.6, 0.4)} when using causal graphs inferred by the NOTEARS algorithm with its performance when using the user provided causal graphs, as well as with the baseline variants of the skyline algorithms. As we see in the figure, whether the causal graphs are user provided or inferred, lnSky provides significant gains. For data obtained through synthetic causal graphs, using inferred graphs performs slightly worse than using the true causal graphs. Interestingly, for real causal graphs, {\em lnSky} using the inferred causal graphs outperforms {\em lnSky} using domain-expert graphs, further demonstrating that a perfect causal graph is not required for CSS to provide significant benefits. 
}

\section{Conclusions}

In this paper, we introduced Causal Search for Skylines (CSS), a novel approach that leverages pruning opportunities arising from positively aligned data distributions in skyline queries. CSS exploits the causal structure of the data to selectively de-correlate attributes, preserving positive correlations while removing negative ones. Our experiments showed that CSS achieves greater efficiency than existing methods, and we discussed its potential limitations. We also outlined a strategy for applying CSS in cases where it offers clear benefits while avoiding it otherwise. 

\replaceA{}{
One line\noteA{[R1.O3]} of work we will explore in the future is to see how our divide-and-conquer style approach can be further improved by methods, such as parallelizing the skyline computation of individual clusters. Also, in CSS,
we use the same skyline algorithm for computing both the skyline of each cluster and also to compute the final skyline after combining the skyline output of all the clusters; however, the skyline merge step can potentially be implemented more efficiently.  In addition to these, we plan to explore other ways of "selective de-correlation", such as  arranging the order of the variates in a causally informed manner and, generating new attributes through combinations of the preference attributes to produce causally derived conditioning sets.
}

\begin{acks}
We sincerely thank the anonymous reviewers for their constructive feedback. This work is supported by NSF grant 2311716, "CausalBench: A Cyberinfrastructure for Causal-Learning Benchmarking for Efficacy, Reproducibility, and Scientific Collaboration", 
along with NSF grants 2309030, 2412115, 2435886 and a USACE grant GR40695.

Github Co-Pilot was used to assist in the code development. ChatGPT was used for minor editorial support to improve spelling, grammar, punctuation, and clarity, as specified in ACM guidelines~\cite{acmgenai}.
\end{acks}

\bibliographystyle{ACM-Reference-Format}
\bibliography{datastream.bib}


\clearpage

\appendix
\setcounter{page}{1}

\section{Appendix - A: Preliminaries}\label{sec:appendix_pre}

\begin{table*}[t]
\centering
\caption{\replaceB{}{Summary of the key symbols and notations}}
\label{tab:notation}
{\small
\replaceB{}{
\begin{tabular}{ll}
\hline
\textbf{Symbol} & \textbf{Description} \\
\hline
$D$ & Input dataset consisting of tuples/points \\
$A$ & Set of all attributes (feature space) \\
$P \subseteq A$ & Set of preference (skyline) attributes \\
$t$ & A tuple (data point) in dataset $D$ \\
$\Theta$ & Set of preference criteria for attributes in $P$ \\
$\theta_a$ & Preference criterion for attribute $a$ (e.g., min or max) \\
$S$ & Skyline set (non-dominated tuples) \\
$R(A)$ & Relation defined over attribute set $A$ \\
$G$ or $G_A$ & Causal graph over attributes in $A$ \\
$(V_A, E_A)$ & Vertices and edges of the causal graph \\
$\lambda(e)$ & Sign/label of a causal edge $e$ ($+$ or $-$) \\
$C_P$ & Correlation matrix for preference attributes $P$ \\
$C^*_P$ & Correlation matrix for preference attributes $P$ after conditioning on attributes in $\mathcal{Z}$\\
$C_P[i,j]$ and $C_P^*[i,j]$& Correlation between preference attributes $a_i$ and $a_j$ \\
$\mathcal{Z}$ & Conditioning (grouping) attribute set \\
$\mathcal{Z}^-$ & Selected conditioning set for de-correlation \\
$D_k(\mathcal{Z})$ & Subset of data conditioned on $\mathcal{Z}$ \\
$Paths_{i,j}$ & All causal paths between $a_i$ and $a_j$ \\
$Open^+_{i,j}$ & Open causal paths with positive overall effect \\
$Open^-_{i,j}$ & Open causal paths with negative overall effect \\
$gain(\mathcal{Z})$ & Gain from conditioning on $\mathcal{Z}$ ({\em gnSky}) \\
$l\_gain(\mathcal{Z})$ & Leaky gain from conditioning on $\mathcal{Z}$ ({\em lnSky}) \\
\hline
$Prov^+_{i,j}(\mathcal{Z})$ & Provisionally opened positive paths due to conditioning \\
$Prov^-_{i,j}(\mathcal{Z})$ & Provisionally opened negative paths due to conditioning \\
$Blocked^+_{i,j}(\mathcal{Z})$ & Positive paths blocked by conditioning \\
$Blocked^-_{i,j}(\mathcal{Z})$ & Negative paths blocked by conditioning \\
$imp^+_{i,j}(\mathcal{Z})$ & Impact of conditioning on positive paths ({\em gnSky}) \\
$imp^-_{i,j}(\mathcal{Z})$ & Impact of conditioning on negative paths ({\em gnSky}) \\
$l\_imp^+_{i,j}(\mathcal{Z})$ & Leaky impact on positive paths ({\em lnSky}) \\
$l\_imp^-_{i,j}(\mathcal{Z})$ & Leaky impact on negative paths ({\em lnSky}) \\
$cw(p)$ & Causal weight of path $p$ \\
$\lambda_o, \lambda_b$ & Information-passing factors for open and blocked nodes \\
\hline
\end{tabular}
}}
\end{table*}

\replaceB{}{Table~\ref{tab:notation} provides\noteB{[R2.M3]} the key notations used in the paper.}

\subsection{Tuple Dominance and Skylines}\label{sec:prelims:domDefns}

Let $D$ be a set of data points in a feature/attribute space, $A$, and the set, $P = {a_1, \ldots, a_p} \subseteq A$, denote the set of preference attributes.
Furthermore, let  $\Theta= \{\theta_{a_1},\ldots, \theta_{a_p}\}$, denote the corresponding preference criteria, such that $\theta_* \in \{min, max\}$.
According to the conventional definition~\cite{Borzsonyi01theskyline}
of dominance,
a tuple $t$ \textit{dominates} another tuple $q$
if and only if tuple $t$ is better than or equal to ($\succeq$) tuple $q$ in
all preference attributes and better than ($\succ$) $q$ in at least one attribute of
the preference attribute set. 

\begin{definition} [Tuple dominance ($dom$)]\label{define:domT}
Let $u$ and $t$ be two tuples in
$D$
and $t.a_h$ be the value of attribute, $a_h$, of tuple $t$.
Let $P$ be the
user provided
preference attribute set.
Tuple $u$ \textit{dominates}
$t$  ($u~dom_P~t$) in the preference
attribute set $P$
iff
\[(\forall_{a_i \in {P}}\;\; u.a_i \succeq_{a_i} t.a_i) \wedge (\exists_{ a_k \in {P}}
\;\;
u.a_k \succ_{a_k} t.a_k),\]
where $\succ_{a_*}$ and $\succeq_{a_*}$ are the value dominance relationships according to the preference function $\theta_{a_*}$.
\diaend
\end{definition}

Given this, the skyline of the  dataset, $D$, then
consists of the subset of tuples that are not dominated by any other
tuple in $D$:

\begin{definition}[Skyline]\label{define:sky}
Let $D$ be a data set. The skyline, $S\subseteq D$, with respect to the preference attribute set $P$ and preference criteria $\Theta$ is the \underline{maximal} subset of $D$, where
\[S = \{t \in D \; |\; \nexists_{u \in D} \;\; u\;\; dom_P\;\; t\}.\]
\diaend
\end{definition}
Note that, from these, it is possible to infer that 
\[(\forall_{u \in D\backslash S}\exists_{t\in S}\;\; t\;\; dom_P\;\; u) \wedge (\nexists_{u,t \in S}\;\; t\;\; dom_P\;\; u).\]

\subsection{Causal Graphs}\label{sec:cg}
Let $D$ be a data set defined in an attribute space, $A$. A causal graph, $G_A = (V_A, E_A, \lambda)$, corresponding to the attribute set $A$ is an edge-labeled directed acylic graph (DAG), where for each attribute $a_i \in A$, there is a vertex $v_i \in V_A$ and each edge $e_h = \langle a_i, a_j \rangle$  (or $a_i \rightarrow a_j$) indicates a known direct causal relationship between the attributes $a_i$ and $a_j$ and  the edge label $\lambda(e_h)$ would indicate the nature of the causal relationship between the two attributes.

In a simple causal model, the label $\lambda(e_h) \in \{+,-\}$ would indicate positive or negative causal impact. Intuitively, this indicates that in the context in which the data set $D$ has been collected, there is a process that governs the relationship between the values of $a_i$ and $a_j$; in other words, the values of $a_i$ and $a_j$ cannot be independently varied -- any change in the value of $a_i$ comes with a corresponding change in the value of $a_j$ as described by the edge label (but not vice versa).

\begin{example}[Causal graph for house hunting]
Let us consider our running example -- house hunting. Figure~\ref{fig:cg_houses}(b) shows a causal graph, $G$, that outlines the causal relationships among the attributes of the data set. In this example, we see that the $\tt Distance\_to\_city\_center$ attribute causally impacts both $\tt Price$ and $\tt Commute$ attributes of the data.  
\circend
\end{example}
In linear structural causal models~\cite{Wright1921,Pearl2000}, the  label $\lambda(e_h)$ associated with a causal edge $e_h = a_i \rightarrow a_j$ would be of the form $a_j = c_{ij} a_i + \epsilon_{ij}$, 
where the constant $c_{ij}$ denotes the causal impact of $a_i$ on $a_j$, whereas $\epsilon_{ij}$ would denote the random noise that acts on attribute $a_j$ along this causal edge.

\subsection{Causal Paths}
A causal path, then, is a sequence of (backward or forward) causal edges that causally relates two given attributes; moreover, the causal path is said to be a directed causal path if all of the edges in the sequence point in the cause-effect direction:

\begin{definition}[Causal Path]
Let $G_A = (V_A, E_A)$ be a causal 
graph on attribute set $A$ and let $a_i$ and $a_j$ be two attributes in $A$. The sequence $[e_1, e_2, \ldots, e_k]$ of edges is said to be a {\em causal path} from $a_i$ to $a_j$ (denoted as $a_i \leftrightsquigarrow a_j$) iff 
there is a sequence $[a_{s_1}, \ldots, a_{s_k}, a_{s_{k+1}}]$ of attributes, such that   
\begin{eqnarray*}
(\forall_{1\leq h \leq k}\;\; 
(e_h = \langle a_{s_h}, a_{s_{h+1}} \rangle ) \vee
(e_h = \langle a_{s_{h+1}}, a_{s_{h}} \rangle))
\wedge\\
(a_{s_1} = a_i) \wedge
(a_{s_{k+1}} = a_j).
\end{eqnarray*}
A causal path is said to be a {\em directed causal path} (denoted as $a_i \dpath a_j$) iff 
\[
(\forall_{1\leq h \leq k}\;\; e_h = \langle a_{s_h}, a_{s_{h+1}} \rangle )\wedge
(e_{s_1} = a_i) \wedge
(e_{s_{k+1}} = a_j).
\]
All causal paths that are not directed are referred to as {\em undirected causal paths}. 
\diaend
\end{definition}

\begin{example}[Sample paths on the House hunting causal graph]
In our running example in Figure~\ref{fig:cg_houses}, the edge sequence $[e_4,e_5]$ is a directed path from the $\tt Distance\_to\_city\_center$ attribute to the $\tt Price$ attribute.
The edge sequence $[e_1,e_2]$, on the other hand, is an undirected causal path from the $\tt Commute$ attribute to  the $\tt Price$ attribute. 
\circend
\end{example}


\begin{figure*}[t]
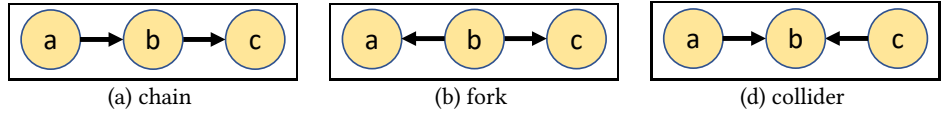

\begin{tabular}{ccc}
	\setlength{\fboxsep}{1.4pt}
	\fbox{
		\includegraphics[width=1.4in]
		{chain.eps}
	}
	&
	\setlength{\fboxsep}{1.4pt}
	\fbox{
		\includegraphics[width=1.4in]
		{fork.eps}
	}
	&
	\setlength{\fboxsep}{1.4pt}
	\fbox{
		\includegraphics[width=1.4in]
		{collider.eps}
	}
	\\
	(a) chain & (b) fork & (d) collider \\
	\end{tabular}
	\caption{Three basic causal structures}
	\label{fig:causal_structures}
\end{figure*}

\subsection{Causal Graph Patterns and Statistical Relationships}

Figure~\ref{fig:causal_structures} presents three basic causal
structures introduced in~\cite{Pearl2009} -
(a): a  \textit{chain} structure where $a$ causally affects $c$ through its influence on $b$ (here $b$ is referred to as a mediator);
(b): a   \textit{fork} structure where $b$ is the common cause of both $a$ and $c$ -- note that, in this case, $a$ and $c$ are likely dependent but there  is no causation between them;
(c): a \textit{collider} structure where both $a$ and $c$ independently
  cause $b$,  but there is no causation between them.

\section{Appendix - B: Dominance Check Results}\label{sec:app_c}

\begin{figure}[t]
\centerline{
\begin{tabular}{cc}
\includegraphics[width=0.47\columnwidth]{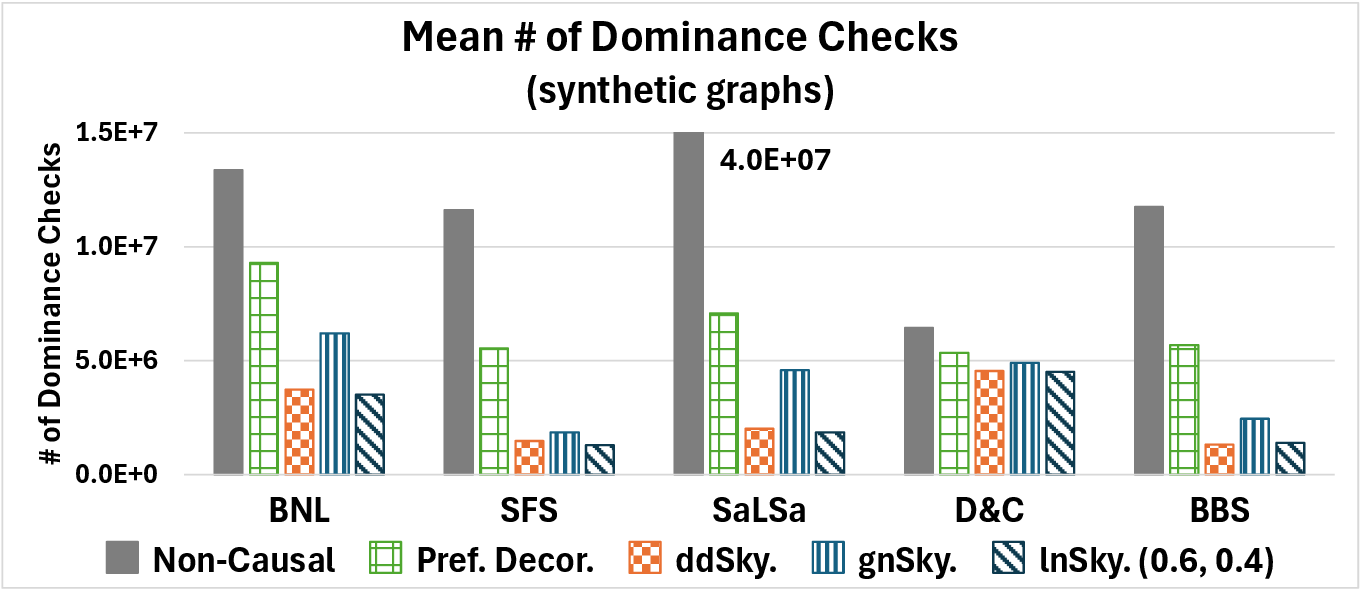}&
\includegraphics[width=0.47\columnwidth]{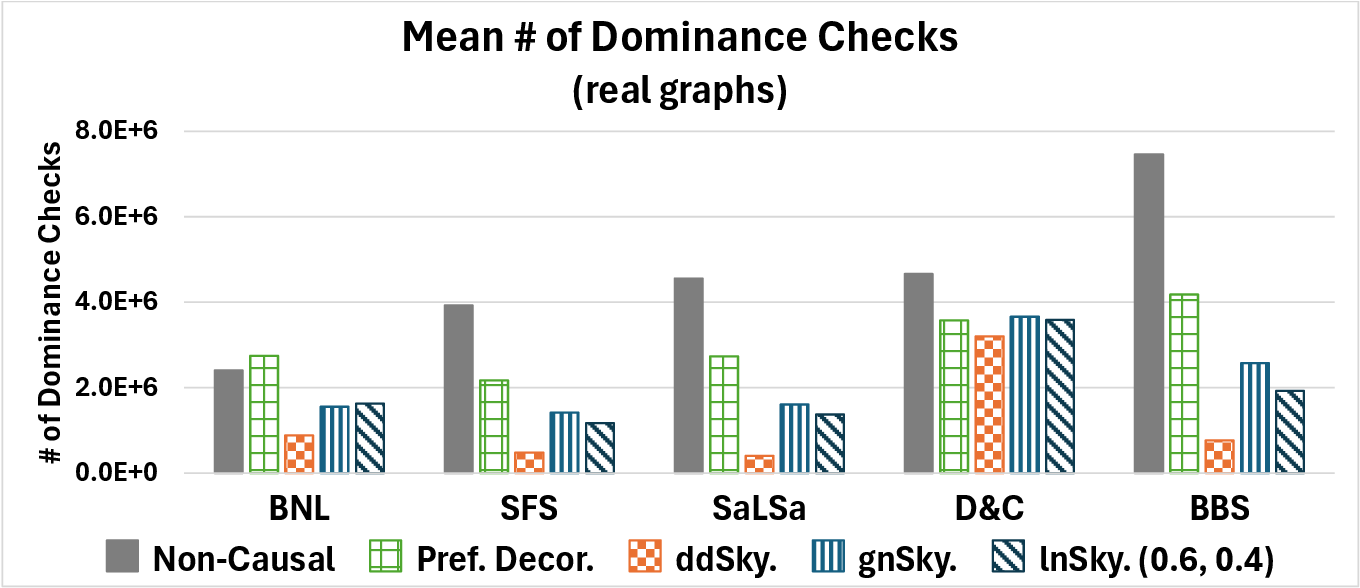}\\
\multicolumn{1}{c}{(a) Synthetic causal graphs} & \multicolumn{1}{c}{(b) Real-world causal graphs}\\
\end{tabular}
}
\caption{Cost of skyline computation with (a) synthetic and (b) real-world causal graphs \underline{(the lower, the better)} -- 200K data}\label{fig:absolute_dom}
\end{figure}

\begin{figure}[t]
\centerline{
\begin{tabular}{cc}
\includegraphics[width=0.47\columnwidth]{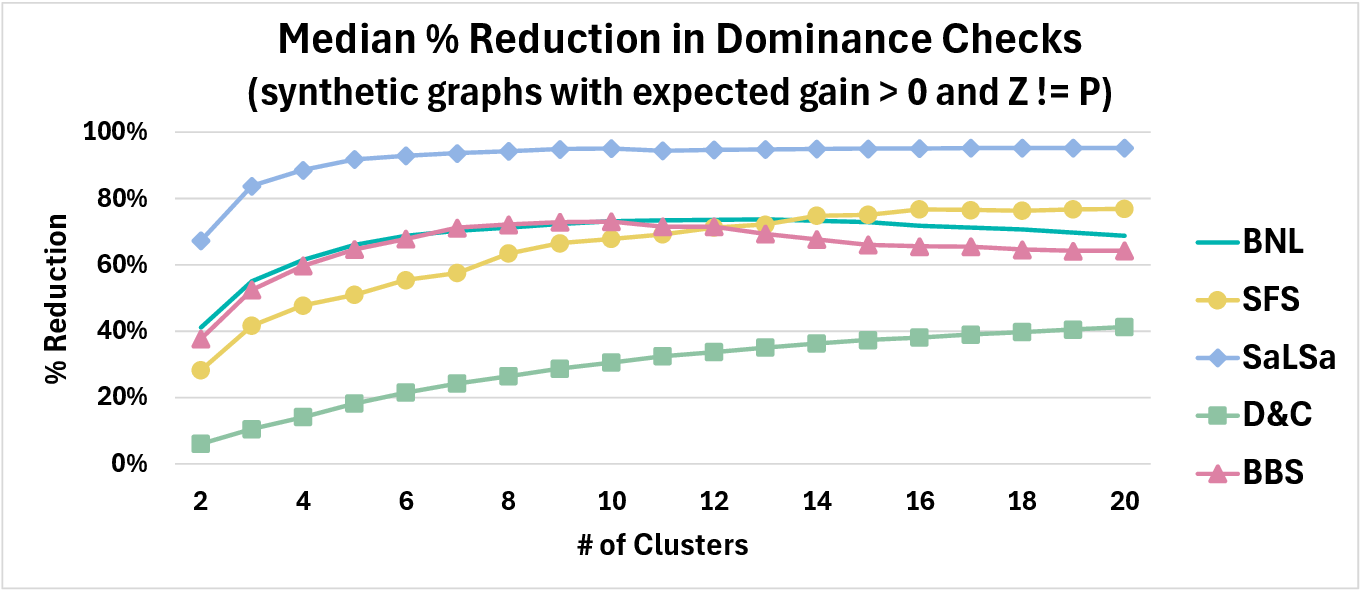} &\includegraphics[width=0.47\columnwidth]{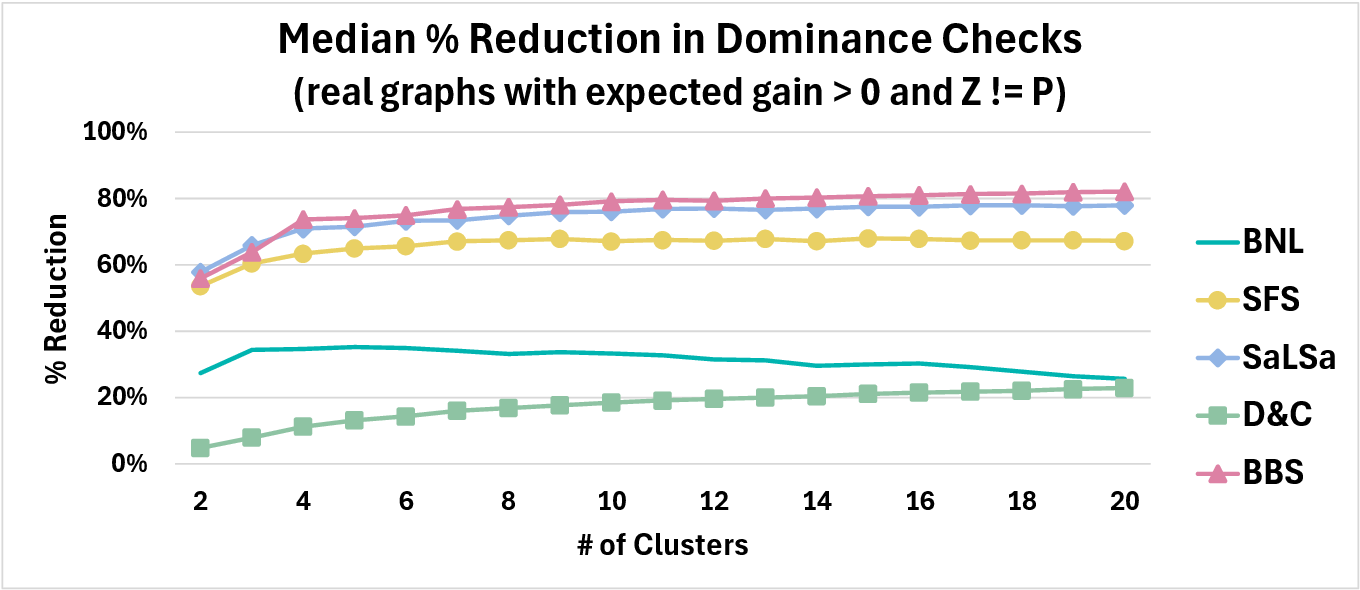}\\
(a) Synthetic causal graphs & (b) Real-world causal graphs\\
\end{tabular}
}
\caption{Median cost reduction \underline{(the higher, the better)} for different numbers of de-correlation clusters when using \default{{\em \replace{lnSkyline}{lnSky}(0.6,\nsbp0.4)}}; for these experiments, the expected gain is positive and larger than the gain predicted for the preference attribute set -- 200K data}\label{fig:cluster-var_dom}
\end{figure}

\begin{figure}[t]
\centerline{
\begin{tabular}{cc}
\includegraphics[width=0.6\columnwidth]{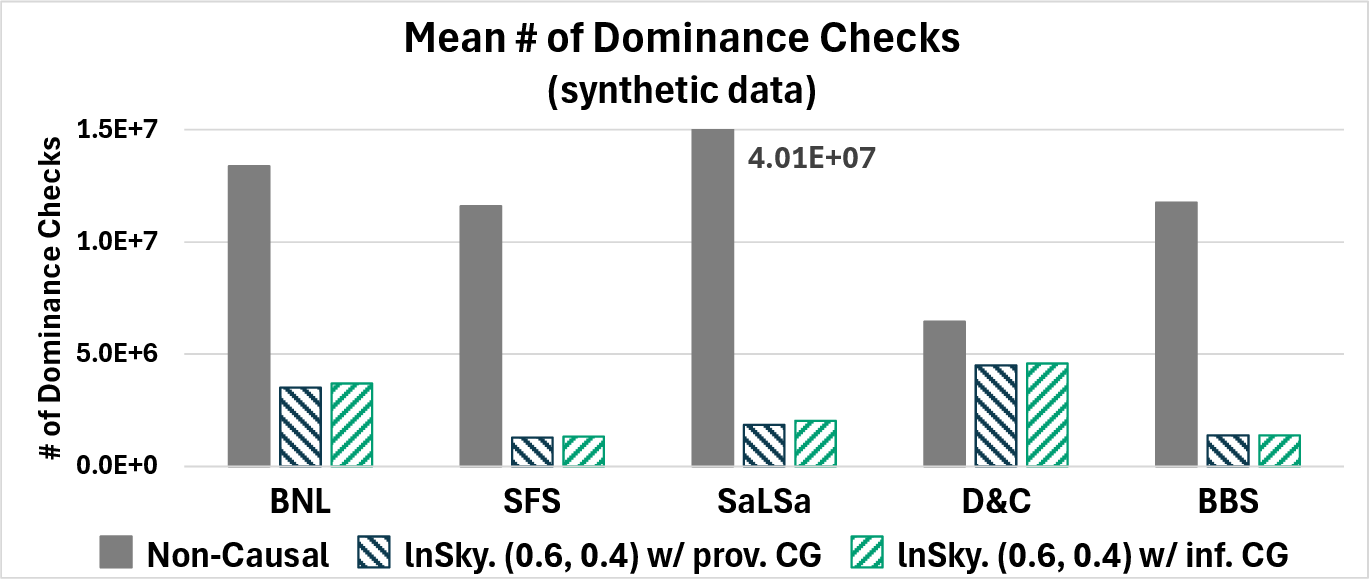}\\ 
\replaceB{}{(a) causal graphs inferred from synthetic data} \\
\includegraphics[width=0.6\columnwidth]{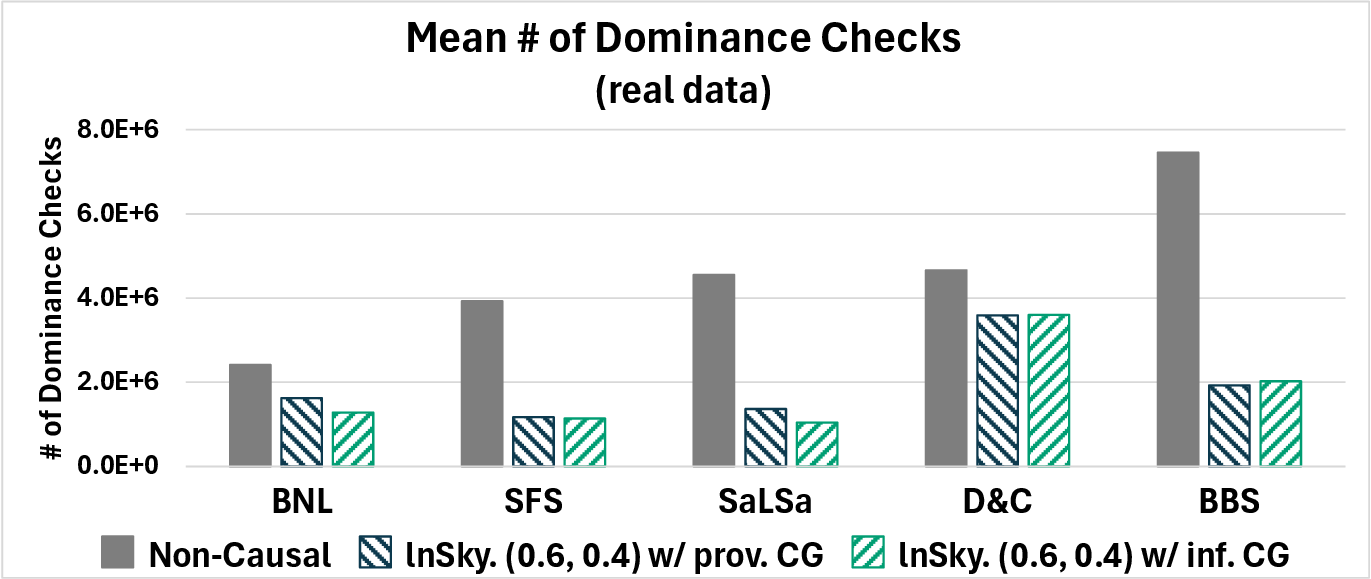}\\
 \replaceB{}{(b) causal graphs inferred from real data}\\
\end{tabular}
}
\noteB{[R2.O2]}
\caption{
\replaceB{}{Cost of skyline computation with causal graphs inferred from (a) synthetic data (i.e., data generated from synthetic causal graphs) and (b) real data \underline{(the lower, the better)}, when using {\em lnSky(0.6,\nsbp0.4)} -- 200K data}}\label{fig:inferred-graphs_doms}
\end{figure}

In Section~\ref{sec:exp} of the paper, we reported the execution times of the skyline algorithms and the execution time gains of CSS. In this section, in Figures~\ref{fig:absolute_dom} through \replaceB{\ref{fig:cluster-var_dom}}{\ref{fig:inferred-graphs_doms}},
we report the corresponding dominance check results, which parallel the performance results reported earlier in Figures~\ref{fig:absolute} through
\replaceB{\ref{fig:cluster-var}}{\ref{fig:inferred-graphs}}.

\begin{figure}[t]
\centerline{
\begin{tabular}{cc}
\includegraphics[width=0.47\columnwidth]{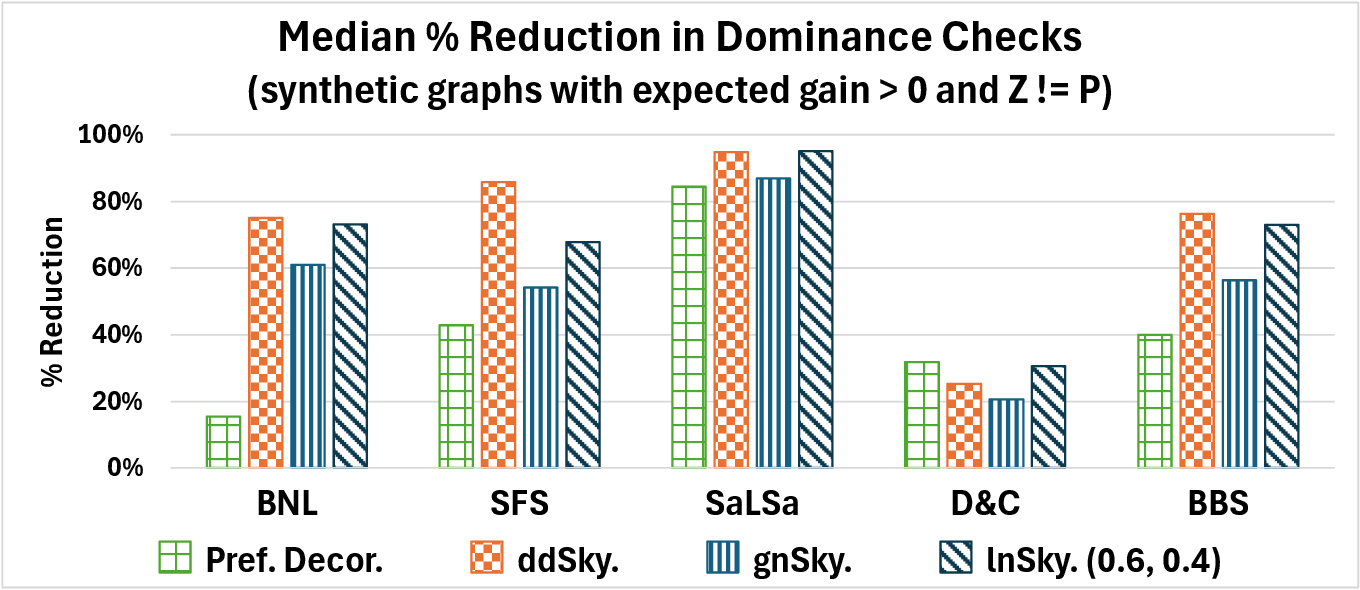}&
%
\includegraphics[width=0.47\columnwidth]{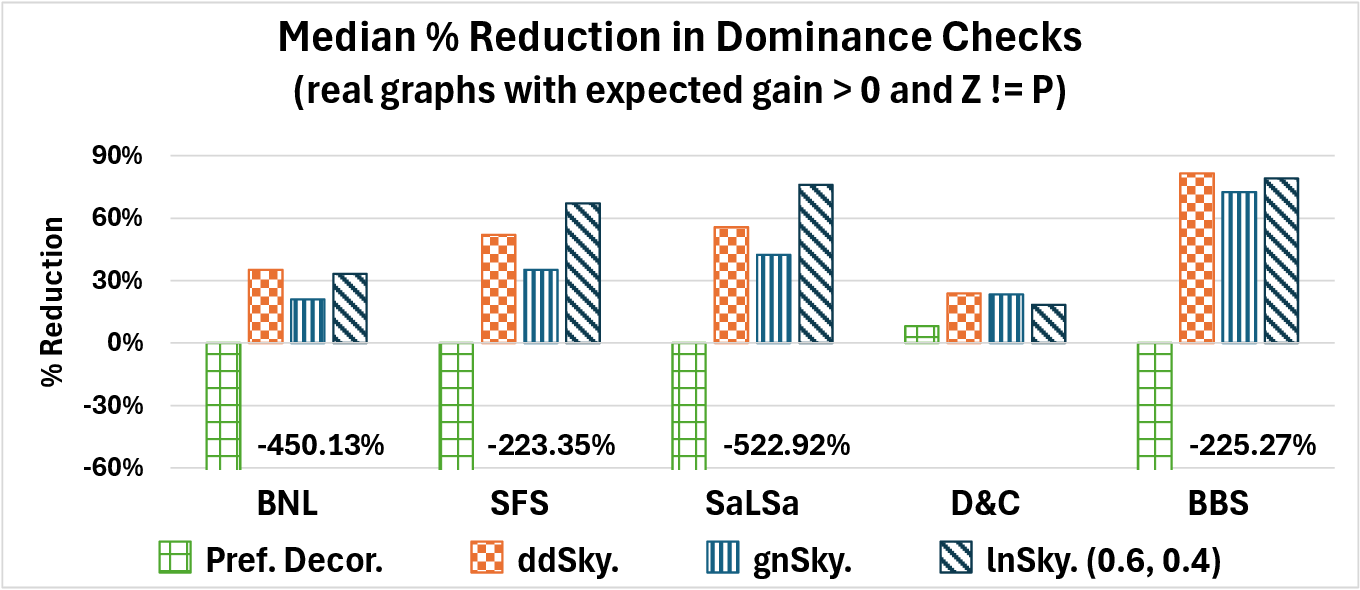}\\
\multicolumn{1}{c}{ (a) Synthetic causal graphs}&\multicolumn{1}{c}{ (b) Real-world causal graphs}\\
\end{tabular}
}
\caption{Median cost reduction \underline{(the higher, the better)}; for these experiments, the expected gain is positive and larger than the gain predicted for the preference attribute set  (i.e., we expect to see benefits over both vanilla skylines and preference attribute de-correlation) -- 200K data}\label{fig:relative-good_dom}
\end{figure}

\begin{figure}[t]
\centerline{
\begin{tabular}{cc}
\includegraphics[width=0.47\columnwidth]{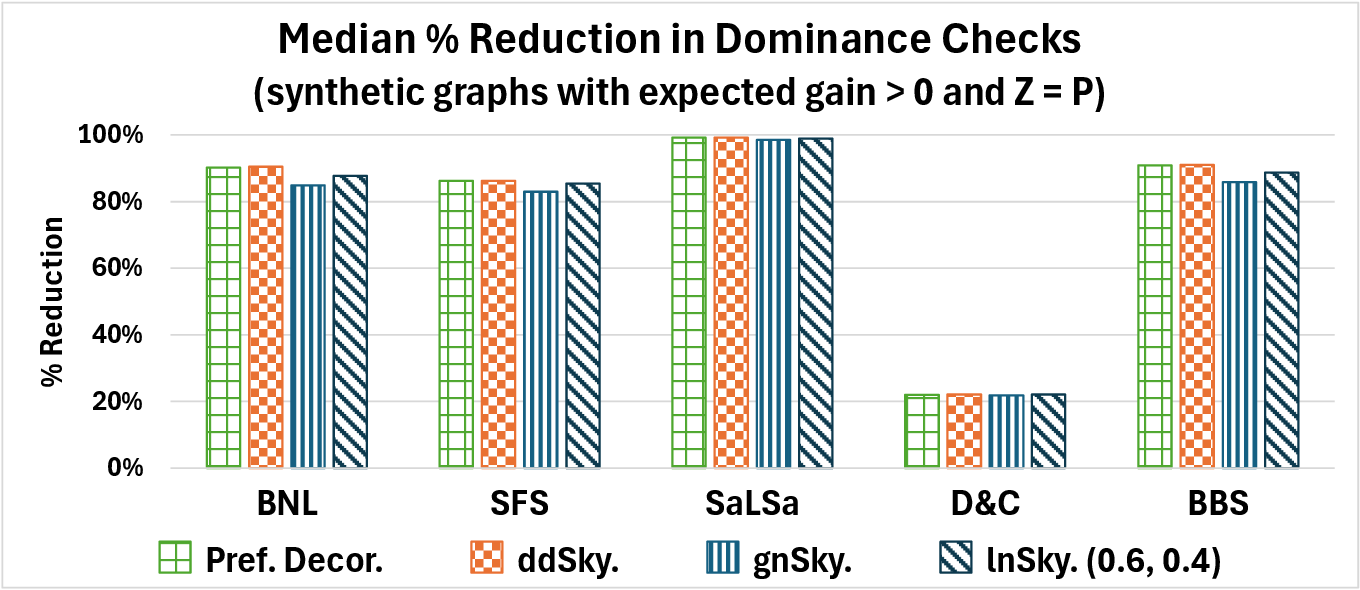}&
%
\includegraphics[width=0.47\columnwidth]{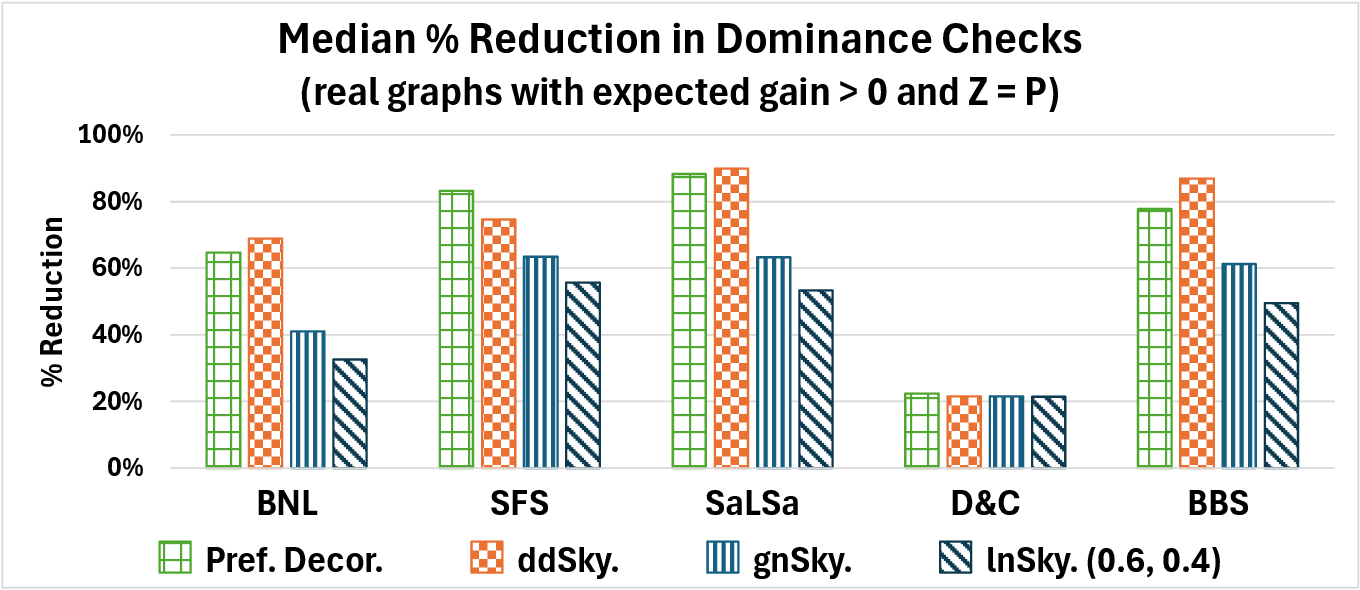}\\
\multicolumn{1}{c}{ (a) Synthetic causal graphs} & \multicolumn{1}{c}{ (b) Real-world causal graphs}\\
\end{tabular}
}
\caption{Median cost reduction \underline{(the higher, the better)}; for these experiments, the  expected gain is positive and equal to the gain predicted for the preference attribute set (i.e., we do not expect to beat preference attribute de-correlation) -- 200K data}\label{fig:relative-pref_dom}
\end{figure}

\begin{figure}[t]
\centerline{
\begin{tabular}{cc}
\includegraphics[width=0.47\columnwidth]{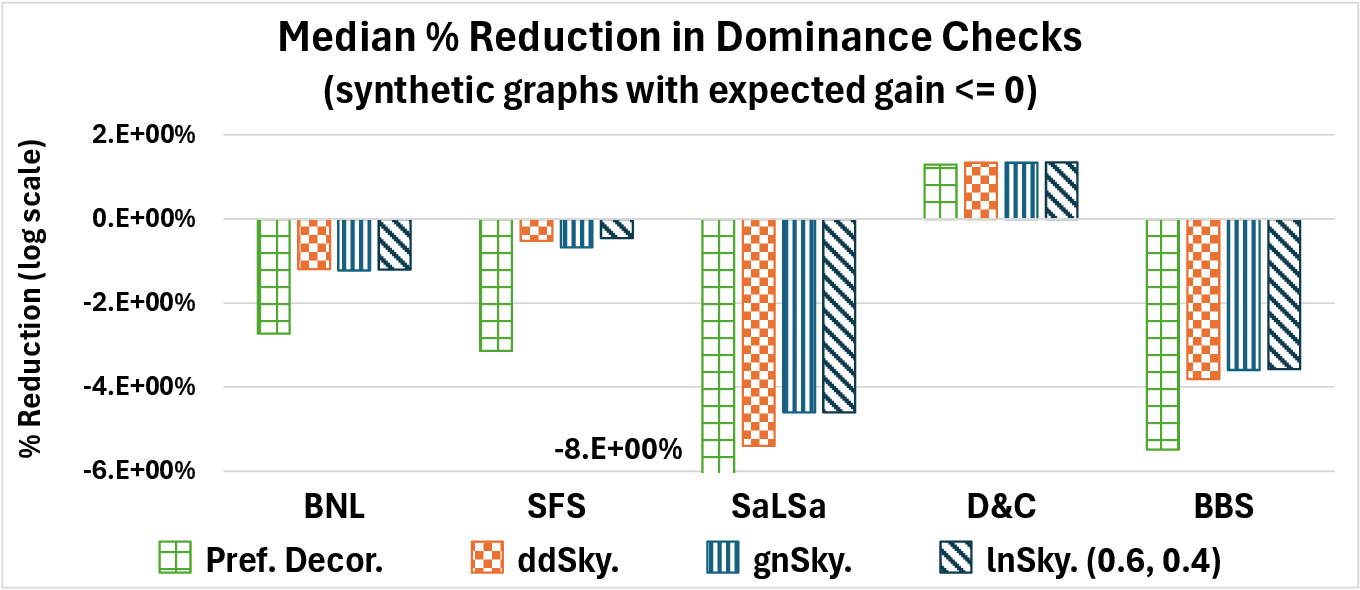}&
\includegraphics[width=0.47\columnwidth]{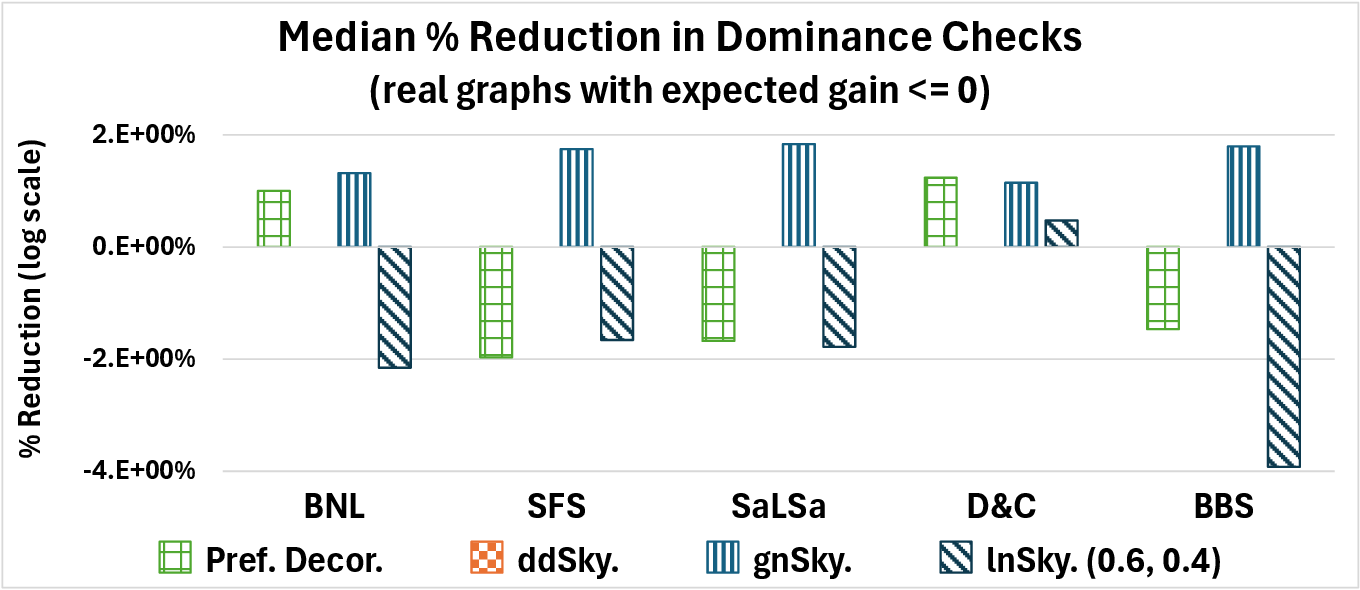}\\
(a) Synthetic causal graphs & (b) Real-world causal graphs\\
\end{tabular}
}
\caption{Median cost reduction \underline{(the higher, the better)}; for these experiments, the expected gain is negative (i.e., we do not expect to see benefits from de-correlation) -- 200K data}\label{fig:relative-negative_dom}
\end{figure}

\begin{figure}[t]
\centerline{
\begin{tabular}{cc}
\includegraphics[width=0.47\columnwidth]{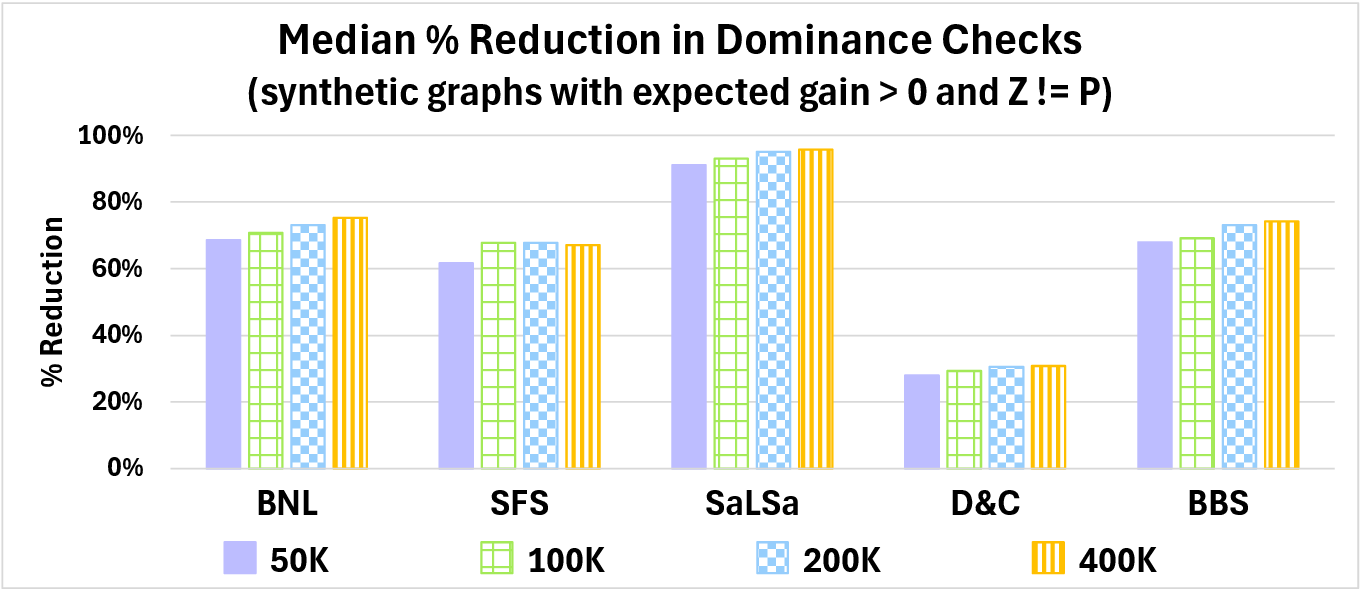}&
\includegraphics[width=0.47\columnwidth]{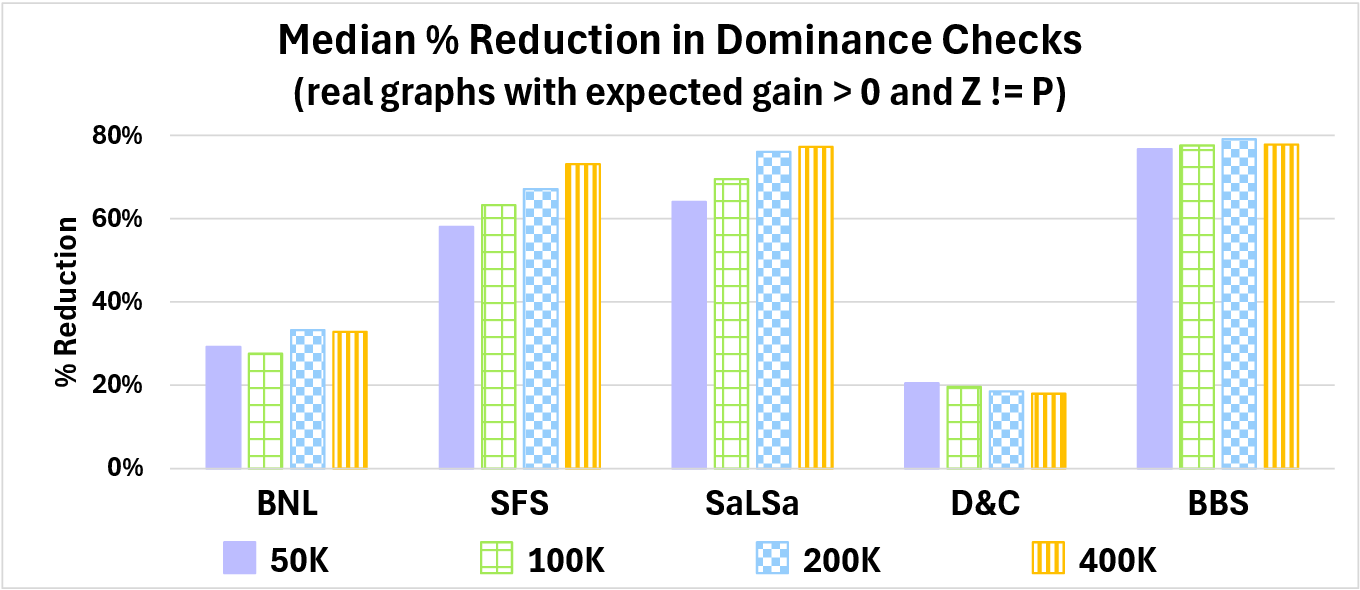}\\
\multicolumn{1}{c}{ (a) Synthetic causal graphs} & \multicolumn{1}{c}{ (b) Real-world causal graphs}\\
\end{tabular}
}
\caption{Median cost reduction \underline{(the higher, the better)} for different numbers of tuples when using \default{{\em \replace{lnSkyline}{lnSky}(0.6,\nsbp0.4)}}; for these experiments, the expected gain is positive and larger than the gain predicted for the preference attribute set}\label{fig:tuples-var_dom}
\end{figure}

\begin{figure}[t]
\centerline{
\begin{tabular}{cc}
\includegraphics[width=0.47\columnwidth]{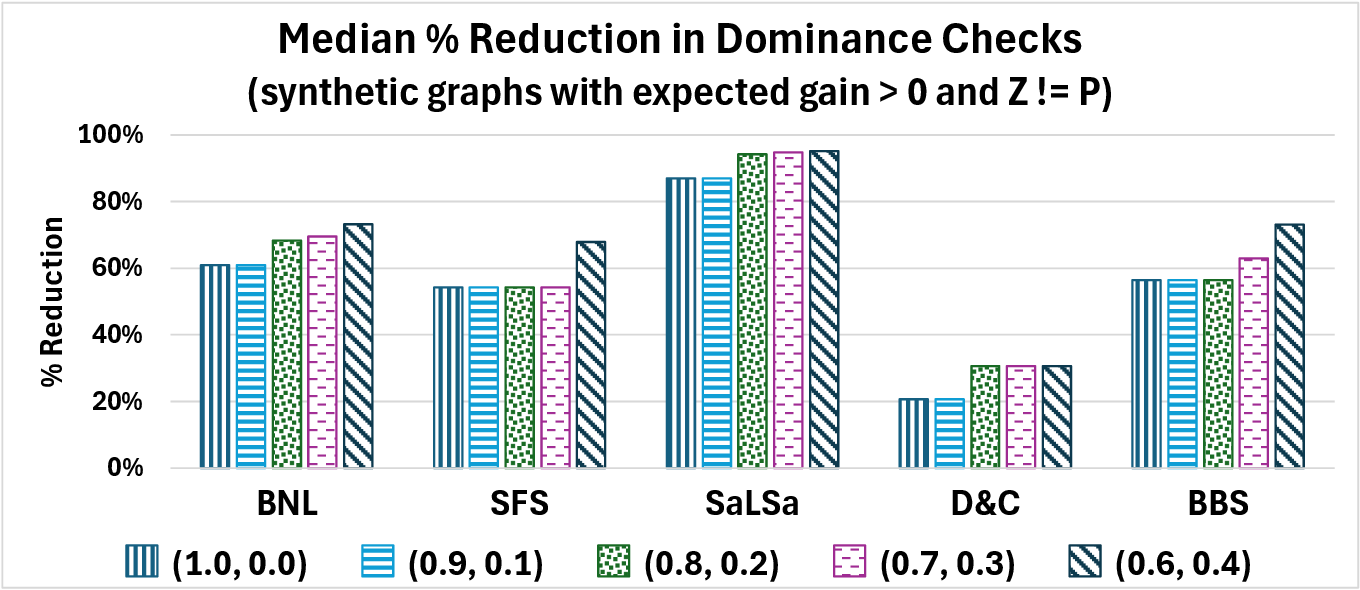} &\includegraphics[width=0.47\columnwidth]{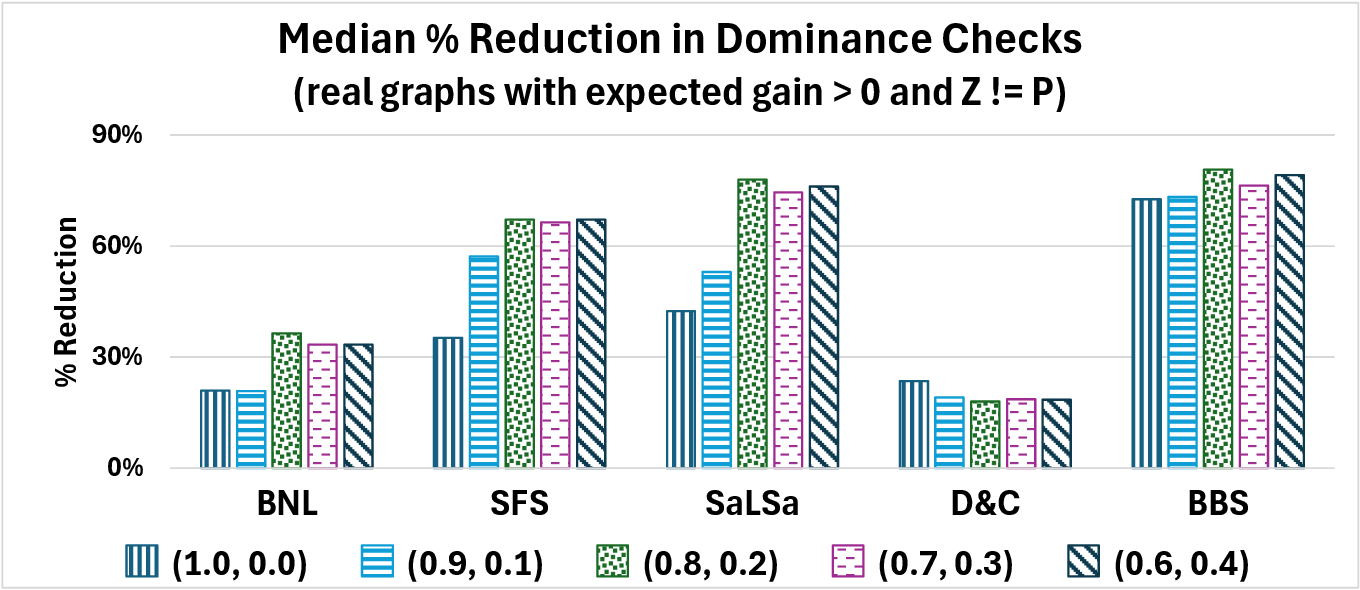}\\
(a) Synthetic causal graphs & (b) Real-world causal graphs
\end{tabular}
}
\caption{ Median cost reduction \underline{(the higher, the better)} for different $\langle\lambda_o,\lambda_b\rangle$ pairs; for these experiments, the expected gain is positive and larger than the gain predicted for the preference attribute set -- 200K data}\label{fig:leakage-var_dom}
\end{figure}

\section{Appendix - C: Computing Conditioned Gain Values}\label{sec:appendix_a}
As outlined in Section~\ref{sec:opt}, for data optimized selective de-correlation for skyline search, we  need a mechanism to  quantify the degrees of leakage and noise as a function of the structure of the causal graph. 
More specifically, let $p$ be a path between two preference variables $X$ and $Y$ and $\mathcal{Z}$ be the set of conditioning attributes. We compute 
the conditioning gain for the path given the conditioning set $\mathcal{Z}$ as
\[c\_gain(p,{\mathcal{Z}}) = corr_{p,\mathcal{Z}}(X,Y) - corr_p(X,Y),\]
where
\begin{itemize}[leftmargin=*]
\item $corr_p(X,Y)$ is the correlation between $X$ and $Y$, implied by the causal path $p$,  before the conditioning step, and
\item $corr_{p,\mathcal{Z}}(X,Y)$; i.e., the correlation between $X$ and $Y$, implied by the causal path $p$,  after the conditioning of attributes in $\mathcal{Z}$.    
\end{itemize}
Given these, the conditioning gain for the path given the conditioning set $\mathcal{Z}$ can be computed as
\[c\_gain(p,{\mathcal{Z}}) = corr_{p,\mathcal{Z}}(X,Y) - corr_p(X,Y).\]
In this section, we discuss how to compute these terms. In particular, we will make the simplifying assumption that the individual variables have normal distributions and sets of variables have multi-variate Gaussian distributions.
\subsection{Computing the Unconditioned Correlation}

Let us first consider computation of the unconditioned correlation, $corr_p(X,Y)$ between attributes $X$ and $Y$ on the path $p$ (we will drop the subscript $p$, when it can be treated as implied).
Let $X = (U_0)$ and $Y = (U_m)$ be two Gaussian variables that are related to each other through a sequence of intermediary Gaussian variables, $U_1, \ldots, U_{m-1}$, that define the path,
\[\Pi = X \mathop{-}_{\{f,b\}} U_1  \mathop{-}_{\{f,b\}}  U_2 \mathop{-}_{\{f,b\}} \ldots \mathop{-}_{\{f,b\}} U_{m-1} 
\mathop{-}_{\{f,b\}}  Y,\]
where each arrow is pointing either forward or backward.
\default{}
Let each forward edge $U_i \rightarrow U_{i+1}$ represent a linear relationship where the  amount $\alpha_{i,i+1} U_i + \epsilon_{i,i+1}$, is contributed to $U_{i+1}$ by $U_i$; here  $U_i$, $U_{i+1}$, and $\epsilon_{i,i+1}$ are Gaussians.
Similarly each backward edge $U_{i-1} \leftarrow U_{i}$ represents a linear relationship where the amount $U_{i} = \alpha_{i,i-1} U_i + \epsilon_{i,i-1}$ is contributed to $U_{i-1}$ by $U_i$; here, $U_i$, $U_{i-1}$, and $\epsilon_{i,i-1}$ are Gaussians.

Given the above, in order to leverage the  formula 
\[corr(X,Y) = \frac{cov(X,Y)}{\sqrt{var(X)}\times \sqrt{var(Y)}}\]
for the computation of $corr(X,Y)$, we need to compute the terms $var(X)$, $var(Y)$, and $cov(X,Y)$.

\subsubsection{Computing Covariance, $\mathbf{cov(X,Y)}$}
In order to compute $cov(X,Y)$, we need to consider three scenarios:

\paragraph{\bf {Path without Forks or Colliders}} If the path does not contain any forks or colliders, we have a chain of mediators between $X$ and $Y$. If $X$ is the cause, then
\[
cov(X,Y) = \left( \prod_{i=0}^{m-1} \alpha_{i,i+1}\right)\times var(X);
\]
else if $Y$ is the cause, then
\[
cov(X,Y) = \left( \prod_{i=0}^{m-1} \alpha_{i+1,i}\right)\times var(Y).
\]

\paragraph{\bf {Path without Colliders}} 
If the path does not contain colliders, then the path can contain at most one fork. Let $U_f$ be the fork of the form "$\leftarrow U_f \rightarrow$" between $X$ and $Y$. Given this, we have
\[
cov(X,Y) = \left( \prod_{i=0}^{f-1} \alpha_{i,i+1}\right)\times var(U_f)
\times \left( \prod_{i=f}^{m-1} \alpha_{i+1,i}\right).
\]

\paragraph{\bf {Path with Colliders}} If any of the intermediary variable is a collider of the form "$\rightarrow U_c \leftarrow$", then $cov(X,Y)$ implied by the path $p$ is 0.

\subsubsection{Computing Variance, $\mathbf{var(U_i)}$}
For $U_i \in \{X = U_0, \ldots, U_{m-1}, Y=U_m\}$, we have the following situations:
\begin{itemize}[leftmargin=*]
\item if $U_i$ has no incoming edges on $p$, then $var(U_i)$ is the measured variance of the variable $U_i$.
\item if $U_i$ has incoming edges, then
\begin{eqnarray*}
    var(U_i) &=& [U_{i-1}\rightarrow U_i] \times \left( \alpha_{i-1,i}^2 \times var(U_{i-1}) + \epsilon_{i-1,i}\right) + \\
     &&[U_{i}\leftarrow U_{i+1}] \times \left(\alpha_{i+1,i}^2 \times var(U_{i+1}) + \epsilon_{i+1,i}.\right). 
\end{eqnarray*}
  
\end{itemize}
Note that, this gives us a set of constraints from which we can obtain the variances for all variables on the path, including $var(X) = var(U_0)$ and $var(Y) = var(U_m)$.

\default{}

\subsection{Computing the Conditioned Correlation}

In order to compute the conditioned correlation, $corr_{p,\mathcal{Z}}(X,Y)$, we need to compute the  conditioned variances and covariances on the path $p$.
%
Let ${\mathcal{Z}} =\{Z_1,\ldots, Z_k\} \subseteq \{U_0, \ldots, U_m\}$ be a set of conditioning variables and let ${\mathcal{S}} = \{S_1,\ldots, S_k \}$ be such that $\forall S_i$, $S_i \subseteq Z_i$ is a non-empty subset of $Z_i$, obtained through  clustering, such that $S_i \sim {\mathcal{N}}(\mu_{S_i}, V_{S_i})$ preserves $\rho_{U_i}$ of the overall variance; i.e., $var(U_i | \mathcal{S}) = \rho_{U_i} \times var(Z_i)$ (see Section~\ref{sec:varscale}).
In order to compute $corr_{p,\mathcal{Z}}(X,Y)$ using the equation
\[corr_{p,\mathcal{Z}}(X,Y) = corr_p(X,Y|{\mathcal{S}}) = \frac{cov_p(X,Y|{\mathcal{S}})}{\sqrt{var_p(X|{\mathcal{S}})}\times \sqrt{var_p(Y|{\mathcal{S}})}},\]
we need to compute the terms $var_p(X|{\mathcal{S}})$, $var_p(Y|{\mathcal{S}})$, and $cov_p(X,Y | {\mathcal{S}})$. 
Note that, as before, we will drop the subscript $p$, when it can be treated as implied.

\subsubsection{Computing Conditional Variances, $\mathbf{var(X|\mathcal{S})}$ and $\mathbf{var(Y|\mathcal{S})}$}

We first discuss how to compute $var(X | \mathcal{S})$ relying on the relationship between $X$ and the conditioning variables on the path\footnote{The term $var(Y | \mathcal{S})$ can be similarly computed.}. We need to consider three cases:

\paragraph{\bf (Case 1).} If $X$ is causally independent of the conditioning variables on the path, then $var(X|\mathcal{S}) = var(X)$.

\paragraph{\bf (Case 2).} Let  $U_1$  be the conditioned variable closest to $X$ on the path
and let there be a linear relationship of the form $X = \alpha U_1 + \epsilon$ between $X$ and $U_1$.
From the definition of conditional variance, we have
\[var(X) = E[var(X|U_1)] + var(E[X|U_1]),\]
which can be re-written as 
\[var(X | \mathcal{S} ) = E[var(X|U_1,\mathcal{S})|\mathcal{S}] + var(E[X|U_1,\mathcal{S}]|\mathcal{S}).\]
Since $(X \bot \mathcal{S} | U_1)$, this can be simplified to
\[var(X | \mathcal{S} ) = \underbrace{E[var(X|U_1)|\mathcal{S}]}_{term1} + \underbrace{var(E[X|U_1]|\mathcal{S})}_{term2}.\]
Again, since we have  $(X \bot \mathcal{S} | U_1)$, we can further simplify $term1$ as
\begin{eqnarray*}
    E[var(X|U_1)|\mathcal{S}] &=& var(X| U_1)\\
    &=& var(X) - \frac{cov(X,U_1)^2}{var(U_1)}.
\end{eqnarray*}
To compute $term2$, we can write the conditional expectation $E[X|U_1]|\mathcal{S}$ as
\[
\mu_{X|\mathcal{S}} + 
\frac{cov(X,U_1|\mathcal{S})}{var(U_1|\mathcal{S})}
\times (U_1 | \mathcal{S} - \mu_{U_1|\mathcal{S}}).
\]
where $\mu$ denotes the mean values. Taking the variance of this expectation gives us the following for $term2$:
\begin{eqnarray*}
    var(E[X|U_1]|\mathcal{S}) & = & \frac{cov(X,U_1 | \mathcal{S})^2}{var(U_1 | \mathcal{S})^2}\times var(U_1 | \mathcal{S})\\
    &=& \frac{cov(X,U_1 | \mathcal{S})^2}{var(U_1 | \mathcal{S})}.
\end{eqnarray*}
Combining $term1$ and $term2$, we can compute $var(X | \mathcal{S} )$ as
\begin{eqnarray*}
    var(X | \mathcal{S}) &=&  var(X) - \frac{cov(X,U_1)^2}{var(U_1)} + \frac{cov(X,U_1 | \mathcal{S})^2}{var(U_1 | \mathcal{S})}.
\end{eqnarray*}
As stated earlier,  $var(U_1 | \mathcal{S}) = \rho_{U_1} \times var(Z_1)$, as discussed in  Section~\ref{sec:varscale}. We discuss the computation of the conditional covariances in Section~\ref{sec:condcov}.

\paragraph{\bf (Case 3)}  Let $\mathfrak{U}_X \subseteq\{U_1,\ldots,U_n\}$ be the set of conditioned variables closest to $X$ with a linear relationship of the form $X = \alpha_1 U_1 + \ldots +\alpha_n U_n +\epsilon$ among $X$, $U_1$, $\ldots$, and $U_n$. Intuitively, the set $\mathfrak{U}_X$ are the set of variables such that $X \bot \mathcal{S}| \mathfrak{U}_X$.
%
From the definition of conditional variance, we have
\[var(X) = E[var(X|\mathfrak{U}_X)] + var(E[X|\mathfrak{U}_X]),\]
which can be re-written as 
\[var(X | \mathcal{S} ) = E[var(X|\mathfrak{U}_X,\mathcal{S})|\mathcal{S}] + var(E[X|\mathfrak{U}_X,\mathcal{S}]|\mathcal{S}).\]
Since $(X \bot \mathcal{S} | \mathfrak{U}_X)$, this can be simplified to
\[var(X | \mathcal{S} ) = \underbrace{E[var(X|\mathfrak{U}_X)|\mathcal{S}\mathfrak{U}_X]}_{term1} + \underbrace{var(E[X|\mathfrak{U}_X]|\mathcal{S})}_{term2}.\]
Again, since we have  $(X \bot \mathcal{S} | \mathfrak{U}_X)$, we can further simplify $term1$ as
\begin{eqnarray*}
    E[var(X|\mathfrak{U}_X)|\mathcal{S}] &=& var(X|\mathfrak{U}_X)\\
    &=&var(X) - \left(\vec{c}_X \;\; \mathbf{C}^{-1}_{\mathfrak{U}_X} \;\;\; \vec{c}^T_X\right),
\end{eqnarray*}
where 
\begin{itemize}[leftmargin=*]
    \item $\vec{c}_X = [cov(X,U_1)\ldots cov(X,U_n)]$ and
    \item  $\mathbf{C}_{\mathfrak{U}_X}$ is a covariance matrix, such that
\begin{eqnarray*}
    \mathbf{C}_{\mathfrak{U}_X} = 
    \begin{bmatrix}
        var(U_1)   & cov(U_1,U_2)  &\ldots & cov(U_1,U_n)\\
        cov(U_2,U_1) & var(U_2) & \ldots&cov(
        U_2,U_n)\\
     \ldots   &\ldots &\ldots & \ldots \\
       cov(U_n,U_1) &cov(U_n,U_2)&\ldots & var(U_n)
    \end{bmatrix}.
\end{eqnarray*}    
\end{itemize}
In order to compute $term2$, we  need to formulate $E[X|\mathfrak{U}_X]|\mathcal{S}$:
\[ \mu_{X|\mathcal{S}} + 
\left(
\vec{c}_{X|\mathcal{S}}\;\; \mathbf{C}^{-1}_{\mathfrak{U}_X | \mathcal{S}}\;\; 
\left(
\begin{bmatrix} U_1 | \mathcal{S} \\ \ldots\\ U_n | \mathcal{S} 
\end{bmatrix}
-
\begin{bmatrix} \mu_{U_1|\mathcal{S}} \\ \ldots\\\mu_{U_n | \mathcal{S}} 
\end{bmatrix}
\right) 
\right),
\]
where $\vec{c}_{X|\mathcal{S}} = [cov(X,U_1|\mathcal{S})\;\; \ldots \;\; cov(X,U_n|\mathcal{S})]$ and  \default{$\mathbf{C}_{\mathfrak{U}_X | \mathcal{S}}$} is a covariance matrix, such that
\begin{eqnarray*}
    \mathbf{C}_{\mathfrak{U}_X | \mathcal{S}} = 
\begin{bmatrix}
        var(U_1| \mathcal{S})   & cov(U_1,U_2| \mathcal{S})  &\ldots & cov(U_1,U_n| \mathcal{S})\\
        cov(U_2,U_1| \mathcal{S}) & var(U_2| \mathcal{S}) & \ldots&cov(
        U_2,U_n| \mathcal{S})\\
     \ldots   &\ldots &\ldots & \ldots \\
       cov(U_n,U_1| \mathcal{S}) &cov(U_n,U_2| \mathcal{S})&\ldots & var(U_n| \mathcal{S})
    \end{bmatrix}.
\end{eqnarray*} 
Note that above $\forall_{i\neq j}\; cov(U_i,U_j| \mathcal{S}) = 0$, due to clustering based conditioning which erases the covariance among the conditioned variables. Thus, we can simplify $E[X|\mathfrak{U}_X]|\mathcal{S}$ as
\[  \mu_{X|\mathcal{S}} + 
\frac{cov(X,U_1| \mathcal{S})}{var(U_1| \mathcal{S})}\times (U_1 | \mathcal{S}) + \ldots + 
\frac{cov(X,U_n| \mathcal{S})}{var(U_n| \mathcal{S})}\times (U_n | \mathcal{S}) + constant.
\]
Given the above, $term2$ (i.e.,  \default{$var(E[X|\mathfrak{U}_X]|\mathcal{S})$}) can be computed as
\begin{eqnarray*}
var(E[X|\mathfrak{U}_X]|\mathcal{S}) &=&
\frac{cov(X,U_1| \mathcal{S})^2}{var(U_1| \mathcal{S})} + \ldots + 
\frac{cov(X,U_n| \mathcal{S})^2}{var(U_n| \mathcal{S})}.
\end{eqnarray*}
Putting $term1$ and $term2$ together we obtain $var(X | \mathcal{S})$ as

\begin{eqnarray*}
    var(X | \mathcal{S}) &=& var(X) - \left(\vec{c}_X \;\; \mathbf{C}^{-1}_{\mathfrak{U}_X} \;\;\; \vec{c}^T_X\right) +\\
    &&\;\;\;\;\frac{cov(X,U_1| \mathcal{S})^2}{var(U_1| \mathcal{S})} + \ldots + 
    \frac{cov(X,U_n| \mathcal{S})^2}{var(U_n| \mathcal{S})}.
\end{eqnarray*}
We discuss the computation of the conditional covariances next.

\subsubsection{Computing Conditional Covariance $\mathbf{cov(X,U_h|\mathcal{S})}$}\label{sec:condcov}
Let $U_h \in {\ U_1,\ldots,U_{m-1}, Y}$; we first discuss how to compute $cov(X, U_h | \mathcal{S})$ relying on the relationship between $X$ and the conditioning variables on the path\footnote{The term $cov(Y,U_h | \mathcal{S})$ can also be similarly computed.}. We need to consider three cases:

\paragraph{\bf (Case 1).} If $X$ is causally independent of $U_h$, ignoring the conditioning variables, then $cov(X,U_h|\mathcal{S}) = 0$.

\paragraph{\bf (Case 2).} Let $U_h = U_1$, such that there is a linear relationship of the form $X = \alpha U_h + \epsilon$ between $X$ and $U_h$. In this case, since $Z\bot \epsilon$, we have
\begin{eqnarray*}
cov(X,U_h | \mathcal{S}) & = & cov(X,U_1 | \mathcal{S})\\
&=& cov(X,U_1) \times \frac{var(U_1 | \mathcal{S})}{var(U_1)}.
\end{eqnarray*}

\paragraph{\bf (Case 3).} Let $\mathfrak{U}_X \subseteq\{U_1,\ldots,U_n\}$ be the set of conditioned variables closest to $X$ with a linear relationship of the form $X = \alpha_1 U_1 + \ldots +\alpha_n U_n +\epsilon$ among $X$, $U_1$, $\ldots$, and $U_n$. Intuitively, the set $\mathfrak{U}_X$ are the set of variables such that $X \bot \mathcal{S}| \mathfrak{U}_X$.
Similarly, let $\mathfrak{U}_h$ be the set of variables such that $U_h \bot \mathcal{S}| \mathfrak{U}_h$.
From the definition of conditional covariance, we have
\begin{eqnarray*}
    cov(X,U_h) &=& E[cov(X,U_h|\mathfrak{U}_X,\mathfrak{U}_h)] +\\
    &&\;\;\;\;cov(E[X|\mathfrak{U}_X,\mathfrak{U}_h],E[U_h|\mathfrak{U}_X,\mathfrak{U}_h]).
\end{eqnarray*}
We can further write this as
\begin{eqnarray*}
    cov(X,U_h|\mathcal{S}) &=& E[cov(X,U_h|\mathfrak{U}_X,\mathfrak{U}_h,\mathcal{S})|\mathcal{S}] + \\&&\;\;\;\;cov(E[X|\mathfrak{U}_X,\mathfrak{U}_h,\mathcal{S}],E[U_h|\mathfrak{U}_X,\mathfrak{U}_h,\mathcal{S}]|\mathcal{S}).
\end{eqnarray*}
Since, $X,U_h \bot \mathcal{S} | \mathfrak{U}_X,\mathfrak{U}_h$, we can rewrite the above equation as
\begin{eqnarray*}
    cov(X,U_h|\mathcal{S}) &=&
    \underbrace{E[cov(X,Y|\mathfrak{U}_X,\mathfrak{U}_h)|\mathcal{S}]}_{term1} + \\
    &&\;\;\;\;
\underbrace{cov(E[X|\mathfrak{U}_X,\mathfrak{U}_h],E[U_h|\mathfrak{U}_X,\mathfrak{U}_h]|\mathcal{S})}_{term2}.
\end{eqnarray*}
Again, since we have $X,U_h \bot \mathcal{S} | \mathfrak{U}_X,\mathfrak{U}_h$, we can simplify $term1$ as 
\begin{eqnarray*}
    E[cov(X,U_h|\mathfrak{U}_X,\mathfrak{U}_h)|\mathcal{S}] &=&cov(X,U_h|\mathfrak{U}_X,\mathfrak{U}_h)\\
    &=&cov(X,U_h) - \left(\vec{c}_X \;\; \mathbf{C}^{-1}_{\mathfrak{U}_X,\mathfrak{U}_h} \;\;\; \vec{c}^T_h\right),
\end{eqnarray*}
where, given the set of conditioning variables $\{U^1, \ldots, U^r\} = \mathfrak{U}_X\cup\mathfrak{U}_h$, we have 
\begin{itemize}
    \item $\vec{c}_X = [cov(X,U^1)\ldots cov(X,U^r)]$,
    \item $\vec{c}_h = [cov(U_h,U^1)\ldots cov(U_h,U^r)]$, and
    \item  $\mathbf{C}_{\mathfrak{U}_X,\mathfrak{U}_h}$ is a covariance matrix, such that
\begin{eqnarray*}
    \mathbf{C}_{\mathfrak{U}_X,\mathfrak{U}_h} = 
    \begin{bmatrix}
        var(U^1)   & cov(U^1,U^2)  &\ldots & cov(U^1,U^r)\\
        cov(U^2,U^1) & var(U^2) & \ldots&cov(
        U^2,U^r)\\
     \ldots   &\ldots &\ldots & \ldots \\
       cov(U^r,U^1) &cov(U^r,U^2)&\ldots & var(U^r)
    \end{bmatrix}.
\end{eqnarray*}    
\end{itemize}
In order to compute $term2$, we first need to formulate $E[X|\mathfrak{U}_X,\mathfrak{U}_h]|\mathcal{S}$ and $E[U_h|\mathfrak{U}_X,\mathfrak{U}_h]|\mathcal{S}$.
We can compute $E[X|\mathfrak{U}_X,\mathfrak{U}_h]|\mathcal{S}]|\mathcal{S}$ as
\[ \mu_{X|\mathcal{S}} + 
\left(
\vec{c}_{X|\mathcal{S}}\;\; \mathbf{C}^{-1}_{\mathfrak{U}_X,\mathfrak{U}_h | \mathcal{S}}\;\; 
\left(
\begin{bmatrix} U^1 | \mathcal{S} \\ 
\ldots \\
U^r | \mathcal{S} 
\end{bmatrix}
-
\begin{bmatrix} \mu_{U^1|\mathcal{S}} \\ \ldots\\\mu_{U^{n} | \mathcal{S}} 
\end{bmatrix}
\right) 
\right),
\]
where $\vec{c}_{X|\mathcal{S}} = [cov(X,U^1|\mathcal{S})\ldots cov(X,U^r|\mathcal{S})]$ and $\mathbf{C}_{\mathfrak{U}_X,\mathfrak{U}_h  | \mathcal{S}}$ is a covariance matrix, such that
\begin{eqnarray*}
    \mathbf{C}_{\mathfrak{U}_X,\mathfrak{U}_h|\mathcal{S}} = 
    \begin{bmatrix}
        var(U^1|\mathcal{S})   & cov(U^1,U^2|\mathcal{S})  &\ldots & cov(U^1,U^r|\mathcal{S})\\
        cov(U^2,U^1|\mathcal{S}) & var(U^2|\mathcal{S}) & \ldots&cov(
        U^2,U^r|\mathcal{S})\\
     \ldots   &\ldots &\ldots & \ldots \\
       cov(U^r,U^1|\mathcal{S}) &cov(U^r,U^2|\mathcal{S})&\ldots & var(U^r|\mathcal{S})
    \end{bmatrix}.
\end{eqnarray*}    
Note that above $cov(U^i,U^j| \mathcal{S}) = 0$, due to clustering based conditioning which erases the covariance among the conditioned variables. Thus, we can simplify $E[X|\mathfrak{U}_X,\mathfrak{U}_h]|\mathcal{S}$ as
\[  \mu_{X|\mathcal{S}} + 
\frac{cov(X,U^1| \mathcal{S})}{var(U^1| \mathcal{S})}\times (U^1 | \mathcal{S}) +\ldots+
\frac{cov(X,U^r| \mathcal{S})}{var(U^r| \mathcal{S})}\times (U^r | \mathcal{S}) + constant.
\]
We can similarly write $E[U_h|\mathfrak{U}_X,\mathfrak{U}_h]|\mathcal{S}$ as
\[  \mu_{U_h|\mathcal{S}} + 
\frac{cov(U_h,U^1| \mathcal{S})}{var(U^1| \mathcal{S})}\times (U^1 | \mathcal{S}) +\ldots+
\frac{cov(U^h,U^r| \mathcal{S})}{var(U^r| \mathcal{S})}\times (U^r | \mathcal{S}) + constant.
\]
Given the above, $term2$ (i.e., $cov(E[X|\mathfrak{U}_X,\mathfrak{U}_h],E[U_h|\mathfrak{U}_X,\mathfrak{U}_h]|\mathcal{S})$) can be computed as
\begin{eqnarray*}
&=&
\frac{cov(X,U^1| \mathcal{S})\times cov(U^h,U^1| \mathcal{S})}{var(U^1| \mathcal{S})^2} \times var(U^1| \mathcal{S})\;\; + \ldots + \\
&& 
\frac{cov(X,U^r| \mathcal{S})\times cov(U^h,U^r| \mathcal{S})}{var(U^r| \mathcal{S})^2} \times var(U^r| \mathcal{S})\;\; + cov\_terms,
\end{eqnarray*}
where $cov\_terms$ account for the conditional covariances among the variables in $\{U^1, \ldots, U^r\} = \mathfrak{U}_X\cup\mathfrak{U}_h$.

Once again, since we have $cov(U^i,U^j| \mathcal{S}) = 0$, for $i\neq j$, the above equation for $term2$ simplifies to 
\begin{eqnarray*}
&=&
\frac{cov(X,U^1| \mathcal{S})\times cov(U_h,U^1| \mathcal{S})}{var(U^1| \mathcal{S})}  +\\
&& 
\ldots + \frac{cov(X,U^r| \mathcal{S})\times cov(U_h,U^r| \mathcal{S})}{var(U^r| \mathcal{S})} 
\end{eqnarray*}
Note that the terms $var(U^1| \mathcal{S}), \ldots,var(U^r| \mathcal{S})$ can be  computed as described in Section~\ref{sec:varscale}. 
Thus,  in order to compute $cov(X,U_h| \mathcal{S})$ by combining $term1$ and $term2$ obtained above, we only need to compute the terms 
$cov(X,U^1| \mathcal{S})$, 
$cov(U_h,U^1| \mathcal{S})$, $\ldots$,
$cov(X,U^r| \mathcal{S})$, and
$cov(U_h,U^r| \mathcal{S})$ -- note that the $cov(U_h,U^i| \mathcal{S})$ terms are equal to $0$ unless $U^h = Y$.

\subsection{Computing the Variance Scaling Factor, $\rho$}\label{sec:varscale}
\subsubsection{Variance Scaling for Truncated Gaussians}\label{sec:truncgauss}
Let $Z \in \mathcal{Z}$  be a Gaussian random variable, \default{$Z \sim \mathcal{N}(\mu_Z, V_Z)$}, with mean, $\mu_Z$ and variance $var(Z) = V_Z$. Let $S$ be a subset of $Z$ with the values in the range of $[l,u]$. We can calculate the truncated variance $var(Z|S)$ as follows:
\begin{eqnarray*}
var(Z|S) &=&  var(Z\; |\; l \leq Z \leq u)\\
&=&var(Z)\times 
\left(
1 
\default{+ \frac{
U\times f(U)  - L\times f(L) 
}
{
\Phi(L) - \Phi(U)} 
}
- \mu_{cond}^2 \right) , 
\end{eqnarray*}
where 
$f(x)$ is the PDF of the standard normal distribution (i.e.,  \default{$\mathcal{N}(0,1)$}):  
\[
f(x) = \frac{1}{\sqrt{2\pi}}\times exp\left( -\frac{x^2}{2}\right),
\]
$\Phi(x)$ is the cumulative distribution function (CDF) of the standard normal distribution evaluated at $x$,
\default{
\[
\Phi(x) = \frac{1}{\sqrt{2\pi}}\times \int_{-\infty}^x exp\left( -\frac{t^2}{2}\right)\,dt,
\]
}
$L$ and $U$ are the boundaries scaled onto the standard normal distribution,
\begin{eqnarray*}
L = \frac{l - \mu_Z}{\sqrt{V_Z}}\;\;\textrm{and}\;\; U = \frac{u-\mu_Z}{\sqrt{V_Z}},
\end{eqnarray*}
and, \default{$\mu^2_{cond}$ is the conditional mean of the variance}, truncated at boundaries $L$ and $U$:
\[
\default{
\mu_{cond} = \frac{f(L) - f(U)}{\Phi(U)-\Phi(L)}}.
\]
In other words, given the truncation boundaries $l$ and $u$ of $S$, the scaling factor, $\rho$, for the variance of set $S$ relative to the variance of $Z$, can be obtained as 
\[
\rho =  \left(
1 
\default{+ \frac{
U\times f(U) - L\times f(L)
}
{
\Phi(L) - \Phi(U)
}}
- \mu_{cond}^2 
\right).
\]

\begin{figure*}[t]
\centerline{
\begin{tabular}{cc}
\includegraphics[width=0.42\textwidth]{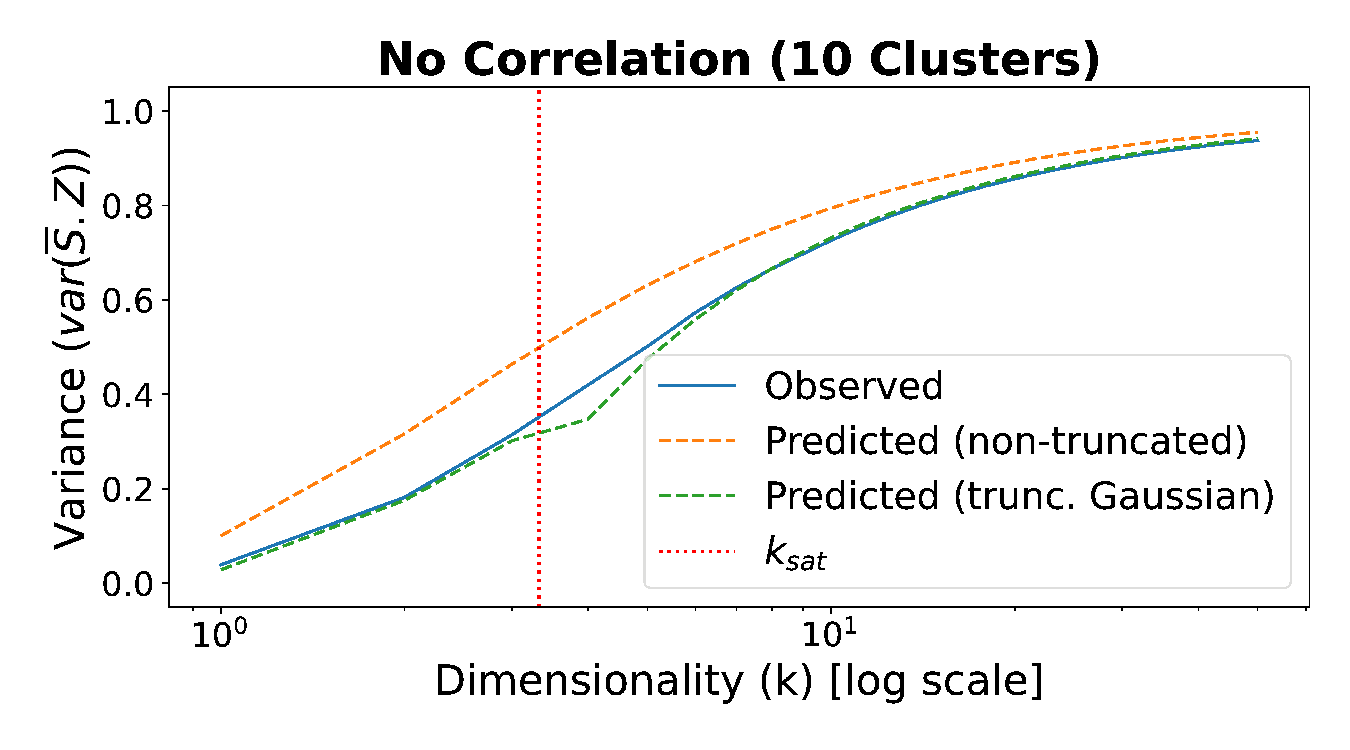} & 
\includegraphics[width=0.42\textwidth]{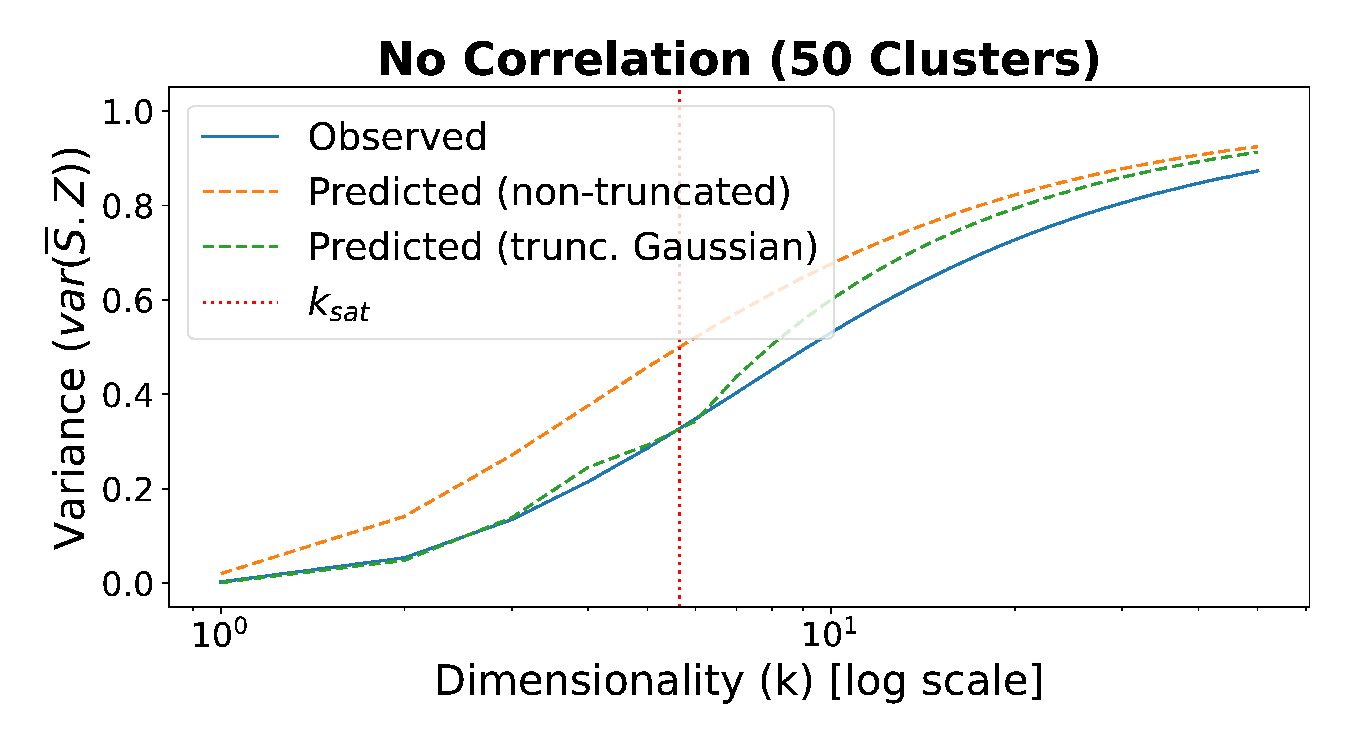}\\
\multicolumn{2}{c}{(a) No correlation among dimensions}\\
\includegraphics[width=0.42\textwidth]{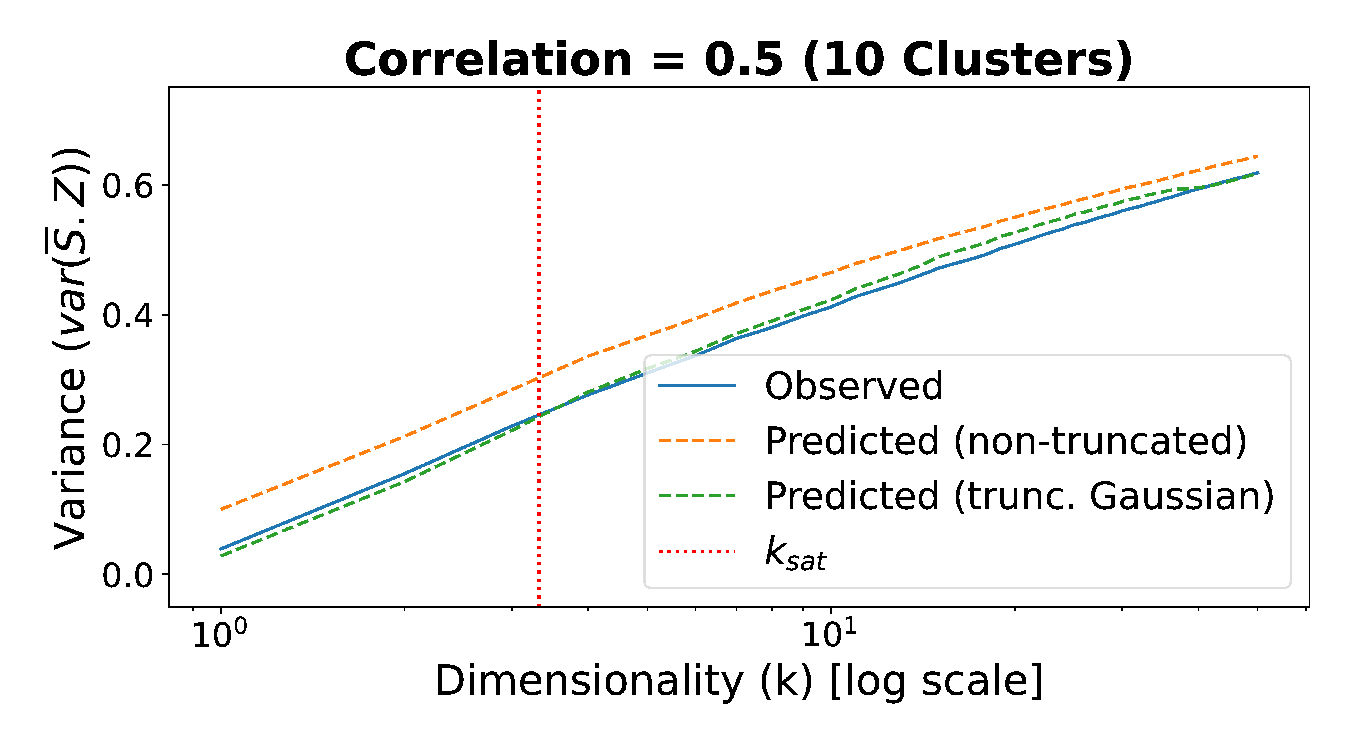} & 
\includegraphics[width=0.42\textwidth]{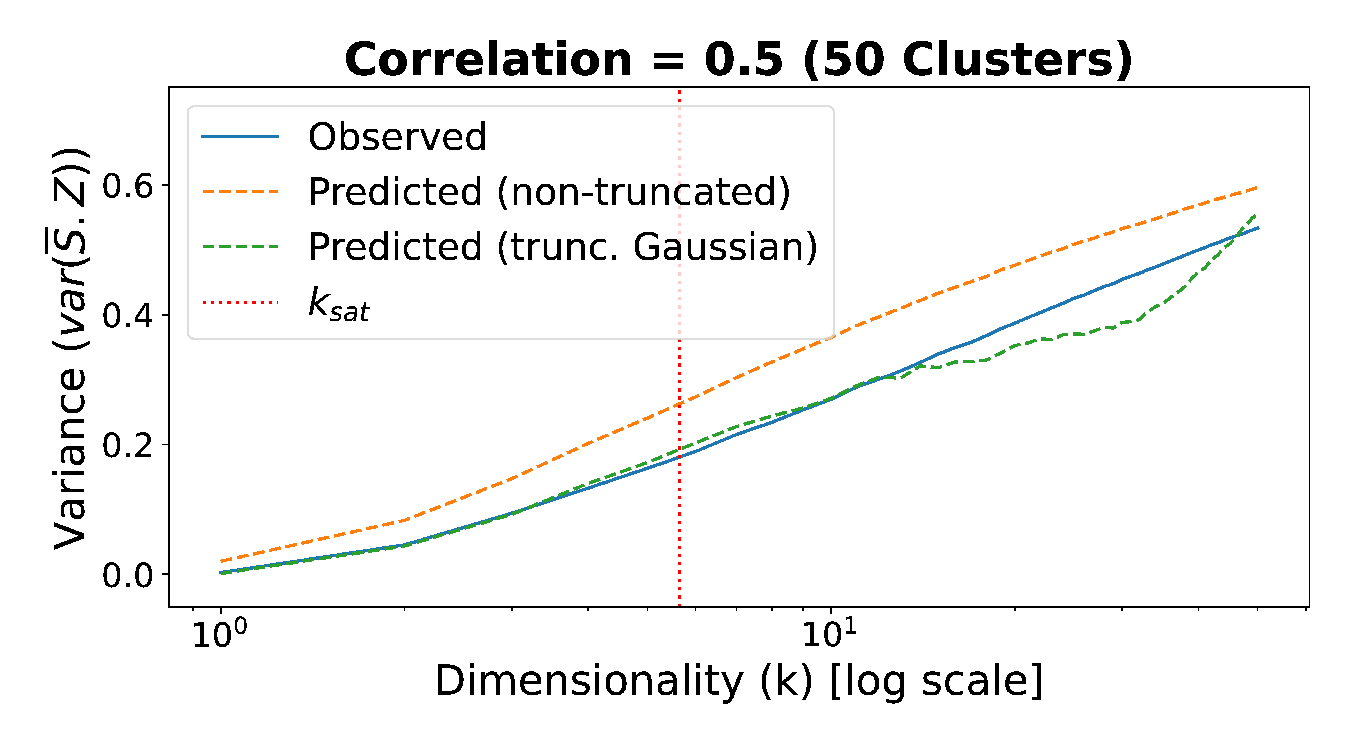}\\
\multicolumn{2}{c}{(b) Average correlation = 0.5, with diverse correlations spread across dimensions drawn from ${\mathcal{N}}(\mu=0.5, \sigma=0.2)$}\\
\includegraphics[width=0.42\textwidth]{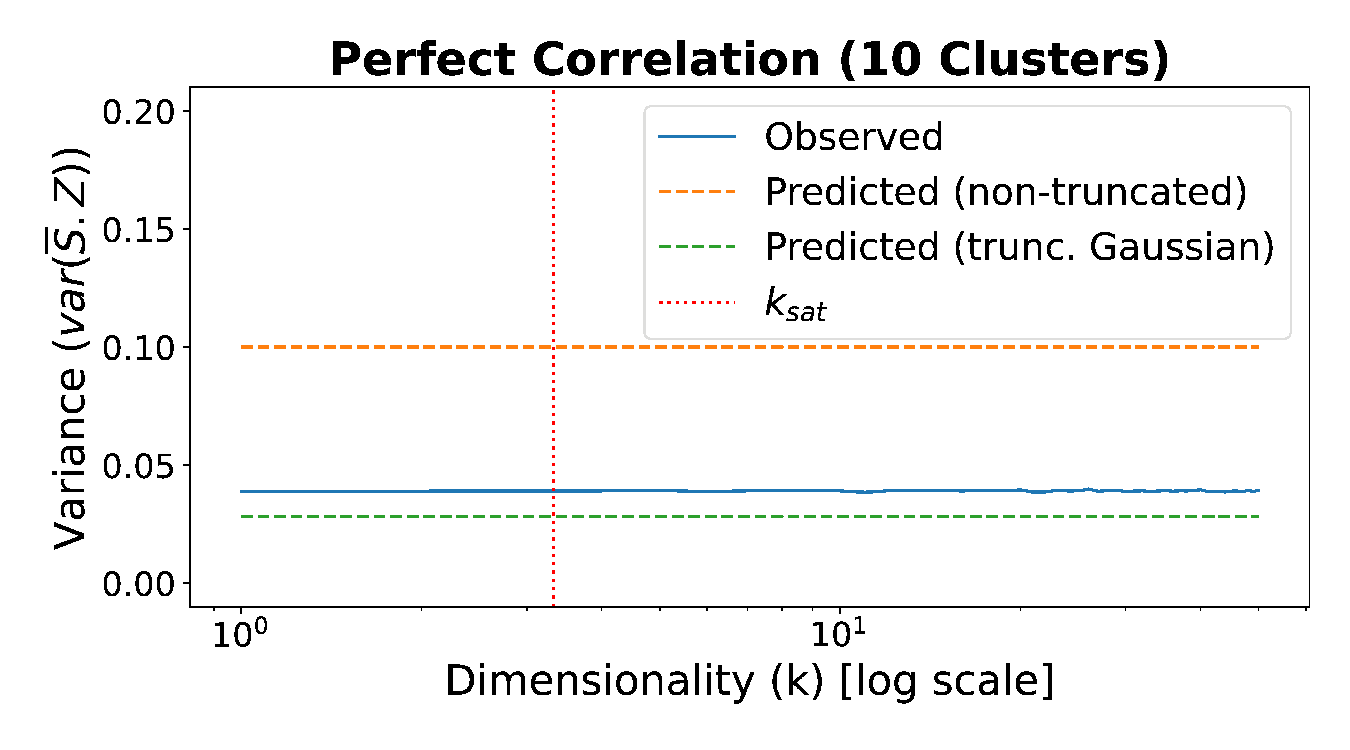} & 
\includegraphics[width=0.42\textwidth]{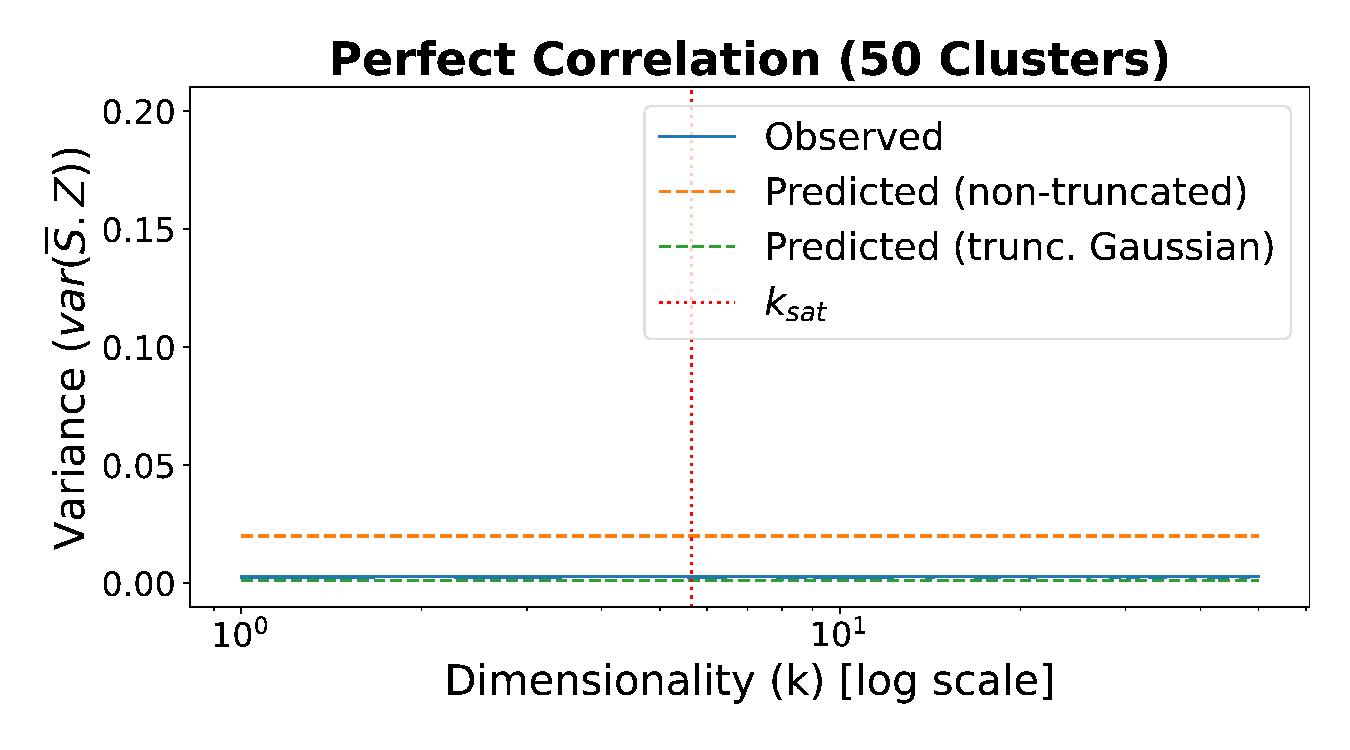}\\
\multicolumn{2}{c}{(c) Perfect correlation among dimensions}\\
\end{tabular}}
\caption{Predicted vs. observed $var(\overline{\mathcal{S}}.Z)$ as a function of the number of dimensions, for 10 and 50 clusters, for data with (a) no correlation, (b) 0.5 correlation, and (c) perfect correlation among data dimensions (the threshold $k_{sat}$ is $log_2{10}$ and $log_2{50}$ for 10 and 50 clusters, respectively) -- as we see in these charts, the non-truncated approximation overestimates the variances, whereas, the proposed truncated Gaussian based model accurately predicts variances, especially in low dimensional settings, which are more likely to occur in practical skyline query settings}\label{fig:var-estimate}
\end{figure*}

\subsubsection{Variance Scaling for Clustered Attribute Sets}
The above formulation indicates that, when we are given the truncation boundaries $l$ and $u$, we can compute a variance scaling factor $\rho$ for $var(Z|S)$. However, in the context of our problem, we are seeking  $var(Z|\mathcal{S})$, where $\mathcal{S}$ is nothing but just one of the {\em multi-attribute} clusters obtained through clustering of the values of the attribute set $\mathcal{Z}$. This means that, we need to compute an {\em average} variance scaling factor, without having a priori knowledge of the multi-attribute cluster $\mathcal{S}$.  

\paragraph{\bf Preliminaries.}
Let $D$ be a data set and let $k = |{\mathcal{Z}}|$ denote the number of conditioning attributes in the given conditioning attribute set,  $\mathcal{Z}$. Let $D_\mathcal{Z} = \prod_{\mathcal{Z}}(D)$ denote the projection of $D$ on the conditioning attribute set and let $\overline{D}_\mathcal{Z}$ be the version of the projected set, such that each attribute is variance normalized; i.e., 
\[\forall_{Z \in \mathcal{Z}}\forall_{d \in D_\mathcal{Z}}\;\; \bar{d}.Z = \frac{d.Z - \mu_Z}{\sqrt{var(Z)}}.
\]
Given the above, let $\overline{\mathfrak{S}}$ denote the set of clusters of $\overline{D}$ obtained through a K-Means-style clustering algorithm (i.e., seeking tight clusters around a centroid) and let $\overline{\mathcal{S}} \in \overline{\mathfrak{S}}$ be one of the resulting clusters. Finally, let $\mathcal{S} \subseteq D_\mathcal{Z}$ be the non-normalized data cluster corresponding to $\overline{\mathcal{S}} \in 
\overline{D}_\mathcal{Z}$.

Since the clusters are obtained on variance normalized data using a clustering algorithm that seeks tight groupings around a centroid, for the analysis we present in this subsection, we will assume that $\overline{\mathfrak{S}}$ consists of (hyper-)spherical clusters -- indicating that for all $\overline{\mathcal{S}}\in\overline{\mathfrak{S}}$, we have $var(\overline{\mathcal{S}}.{Z_i}) = var(\overline{\mathcal{S}}.{Z_j})$ and  $cov(\overline{\mathcal{S}}.{Z_i}, \overline{\mathcal{S}}.{Z_j}) \sim 0$ for $i \neq j$.
Finally, to simplify the analysis, we will assume that for all $\overline{\mathcal{S}}_a, \overline{\mathcal{S}}_b\in\overline{\mathfrak{S}}$, we have $var(\overline{\mathcal{S}}_a)=var(\overline{\mathcal{S}}_b)$.
Given the above, (a) the clusters in $\overline{\mathfrak{S}}$ are similarly shaped to each other and (b) each cluster is similarly shaped across all of its attributes. Therefore, in the rest of the analysis, we can simply refer to $\overline{\mathcal{S}}.Z$, without worrying which cluster in $\overline{\mathcal{S}}\in \overline{\mathfrak{S}}$ or which attribute in $Z \in \mathcal{Z}$ we are considering.



\paragraph{\bf Impact of the Dimensionality and Data Correlation} To investigate the impact of dimensionality of the conditioning attribute set (i.e., the value of $|\mathcal{Z}|$) on the overlap between the clusters (which in turn impacts the value of $var(\overline{\mathcal{S}}.Z)$), let us first consider the case where the data is perfectly correlated.

\underline{Perfect Correlation among Dimensions}  \default{When} the correlation of the data is 1.0, since all the data are on the same line, the data will behave as if $k=1$ and clustering process will return non-overlapping partitions on that line. This would imply that we could approximate the variance as follows:
\[var(\overline{\mathcal{S}}.Z) \sim c^{-1}\times {var(\overline{D}_\mathcal{Z}.Z}).\]
This approximation, however does not take into account that the clusters need to be modeled as truncated Gaussians and we need to compute the variance scaling factor using the truncated Gaussian's discussed in Section~\ref{sec:truncgauss}. In particular, assuming the clustering algorithm returns $c$ equi-width clusters, then, considering that $\sim 99\%$ of the points are within $\pm$ 3 standard deviations) of $\mu$, assuming that the overlap between cluster boundaries are negligible, and assuming that $c$ is an even number,  for all \default{$-\frac{c}{2} \leq i < \frac{c}{2}$, we have the following boundary pair:
\[
l_i=\mu + \frac{3\times \sqrt{V}}{{c}/{2}}\times i\;\;;\;\; u_i = \mu + \frac{3\times \sqrt{V}}{{c}/{2}}\times (i+1)
\]}
Since we are working with pre-normalized data, we have $\mu =0$ and $V = var(\overline{D}_\mathcal{Z}.Z)$. 
Given these, we can then numerically compute $var(\overline{\mathcal{S}}.Z)$ as the average of the variations of the (truncated Gaussian) clusters:
\[var_{PC}(\overline{\mathcal{S}}.Z) \sim c^{-1} \times \sum_{i=-\frac{c}{2}}^{\frac{c}{2} -1} VAR_{trunc\_Gauss}(c,i, var(\overline{D}_\mathcal{Z}.Z)),\]
where $VAR_{trunc\_Gauss}(c,i, V)$ denotes the truncated Gaussian variance for the $i^{th}$ segment of the Gaussian ($-\frac{c}{2} \leq i < \frac{c}{2}$)  with variance $V$ segmented into $c$ equi-bin partitions.
The case for when $c$ is odd can be trivially extended from above.

\paragraph{\underline{No Correlation among Dimensions}}  

Let us next consider the case where the data $\overline{D}_\mathcal{Z}$ are  distributed in the $k = |\mathcal{Z}|$ dimensional space, such that $\mathcal{Z}$ are not correlated.
In this case, the space will be covered by $c = |{\mathfrak S}|$ 
Gaussians (each corresponding to a cluster).
This would imply that we would have $\sim c^\frac{1}{k}$ layers of clusters along each dimension and that the combined variation of the  $c^\frac{1}{k}$ clusters along a given dimension should be
proportional to the variation of the data along that dimension:
\[var(\overline{\mathcal{S}}.Z) \sim c^{-\frac{1}{k}}\times {var(\overline{D}_\mathcal{Z}.Z}).\]
this analysis, however, has two limitations: (a) firstly, it does not take into account that the clusters need to be modeled as truncated Gaussians and (b) does not take into account the situations where $c^{\frac{1}{k}} <2$, i.e., there are not sufficient clusters to divide each dimension to at least two partitions. Next, we discuss how to handle these two limitations.

First, let us assume that $c^{\frac{1}{k}} \geq2$, i.e., there are sufficient clusters to divide each dimension into two. Since we are considering the situation where the dimensions of the data are not correlated, the covariance matrix, $\Sigma$, of the data will be diagonal and, since the data has been pre-normalized,
\default{we would have $\sim c^\frac{1}{k}$ layers of clusters along each dimension. Therefore, we can then numerically compute $var(\overline{\mathcal{S}}.Z)$ as the average of the variations of $c^\frac{1}{k}$ (truncated Gaussian) clusters:}
\default{}
\[var_{NC}(\overline{\mathcal{S}}.Z) \sim
c^{-\frac{1}{k}} \times \sum_{i=-\frac{\sqrt[k]{c}}{2}}^{\frac{\sqrt[k]{c}}{2} -1} VAR_{trunc\_Gauss}(\sqrt[k]{c}, i, var(\overline{D}_\mathcal{Z}.Z)).\]
\default{}
\noindent The case for when $c^{\frac{1}{k}}$ is odd can be trivially extended from above.


Next we consider when the number of clusters is not sufficient to partition  each of the $k$ dimensions; i.e., $c^{\frac{1}{k}} < 2$. In this case, necessarily, some of the dimensions will be left non-partitioned. Let this threshold beyond which some dimensions are saturated and cannot be partitioned be $k_{sat} = log_2(c)$. Let \default{$k_{p} \leq round(k_{sat})$}  be the number of partitioned dimensions and $k_n$ be the number of non-partitioned dimensions ($k_p + k_n = k$). Assuming that the K-means treats all dimensions similarly and does not over-partition certain dimensions, while under partitioning others, this means that we will have \default{$k_{p} = round(k_{sat})$} dimensions partitioned into two and \default{$k_n = k - round(k_{sat})$}  dimensions left non-partitioned. Consequently, we have
\default{}
\begin{eqnarray*}
var_{NC}(\overline{\mathcal{S}}.Z) &\sim&
k_p \times
\frac{1}{2}\times \sum_{i=-\frac{1}{2}}^{\frac{1}{2} - 1} VAR_{trunc\_Gauss}(2, i, var(\overline{D}_\mathcal{Z}.Z))\\
&&+\;\;k_n \times var(\overline{D}_\mathcal{Z}.Z).
\end{eqnarray*}
\default{}

\underline{General Case.} 
In practice, the dimensions of the data space are neither perfectly correlated nor completely independent. A naive approach to estimate $var(\overline{\mathcal{S}}.Z)$ 
would be to take the correlation-weighted average of {\em no correlation} and {\em perfect correlation scenarios};
%
however, this approach would two major limitations: (a) firstly, we would need to combine all pairwise correlations in the data to a single weight value and (b) since it assumes a linear relationship between the variance of the clusters and the average correlation of the data, which is incorrect in practice, it would not take properly into account clusters that need to be modeled as truncated Gaussians.
Next we propose a general formulation that addresses these limitations.

Let $\mathbf{C}$ be the covariance of $\overline{D}_\mathcal{Z}$.
%
We can eigen-decompose $\mathbf{C}$ to obtain two matrices, $U$ and $\mathbf{C}^*$:
\[\mathbf{C} = \mathbf{U}\; \mathbf{C}^*\; \mathbf{U}^T, \]
where $\mathbf{U}^T$ is an ortho-normal transformation that transforms the data $\overline{D}_\mathcal{Z}$ into $\overline{D}^*_\mathcal{Z}$, such that all variances are concentrated on the diagonal; i.e., $\mathbf{C}^*[i,i] = var(\overline{D}^*_\mathcal{Z}.\mathbb{Z}_i)$, where $\mathbb{Z}_i$ is a dimension in the transform space,  and for $i\neq j,\;\;\mathbf{C}^*[i,j] = 0$. 

Since the transformation is ortho-normal it preserves distances, which implies that the K-means in the transform space and in the original space, will return the same clusters. Moreover, since the covariance matrix is diagonal in the transform space, the effect of clustering on variances will be similar to that of the {\em no correlation} scenario considered above. The only difference is that, in this case, different dimensions have different variances -- therefore, we do not expect to have the same number of partitions along each and every dimension. Instead, we expect the number of partitions, $c^*_\mathbb{Z}$, along a given dimension, $\mathbb{Z}_i$, in the transform space, will be proportional to its variance:
%
\[c_{\mathbb{Z}}^{*} \propto
var(\overline{D}_\mathcal{Z}^{*}.\mathbb{Z}).
\]
More generally, for $\mathbb{Z}_i$ and $\mathbb{Z}_j$ in the transform space, we have
\[\frac{c_{\mathbb{Z}_i}^{*}}{c_{\mathbb{Z}_j}^{*}} \sim  \frac{var(\overline{D}_\mathcal{Z}^{*}.\mathbb{Z}_i)}{var(\overline{D}_\mathcal{Z}^{*}.\mathbb{Z}_j)} = \frac{\mathbf{C}^*[i,i] }{\mathbf{C}^*[j,j] } \]
and, given the number, $c$,  of data clusters, we have
\[c = \prod_{i=1}^{k'} c_{\mathbb{Z}_i}^{*},\]
from which we can obtain the number of partitions for each dimension $\mathbb{Z}_i$ ($1 \leq i \leq k'$; $k' \leq k)$ in the transformed space. 

%
%
%
%

As before, let $\overline{\mathfrak{S}}$ denote the set of clusters of $\overline{D}$ obtained through a K-Means-style clustering algorithm (i.e., seeking tight clusters around a centroid) and let $\overline{\mathcal{S}} \in \overline{\mathfrak{S}}$ be one of the resulting clusters. Since the transformation is ortho-normal (i.e., preserving distances), we will get the same clusters in the original and transformed spaces. Therefore, assuming $c_\mathbb{Z}^{*}$ is even, for the given cluster $\overline{\mathcal{S}}$, we can compute the variance along dimension $\overline{\mathcal{S}}$ in the transform space as
\[var(\overline{S}.\mathbb{Z}) \sim \frac{1}{c_\mathbb{Z}^{*}}\times \sum_{i=-\frac{c_\mathbb{Z}^{*}}{2}}^{\frac{c_\mathbb{Z}^{*}}{2} - 1} VAR_{trunc\_Gauss}(c_\mathbb{Z}^{*}, i, var(\overline{D}_\mathcal{Z}^{*}.\mathbb{Z})).\]
\noindent The case for when $c_\mathbb{Z}^{*}$ is odd can be trivially extended from above.

Finally, we need to transform the variances back  to the original space to obtain $var(\overline{S}.Z_i)$ from $var(\overline{S}.\mathbb{Z}_i)$ for $1 \leq i \leq k$.
Let $\mathbf{C}^*_{\overline{S}}$ be the covariance matrix corresponding to the cluster $\overline{S}$ in the transform space; i.e.,
$\mathbf{C}^*_{\overline{S}}[i,i] = var(\overline{S}^*_\mathcal{Z}.\mathbb{Z}_i)$, where $\mathbb{Z}_i$ is a dimension in the transform space, and for $i\neq j,\;\;\mathbf{C}^*_{\overline{S}}[i,j] = 0$.
We obtain the corresponding cluster covariance matrix in the original space as
\[\mathbf{C}_{\overline{S}} = \mathbf{U}\; \mathbf{C}^*_{\overline{S}}\; \mathbf{U}^T, \]
from which we can obtain the value of $var(\overline{S}.Z_i)$ as follows:
\[var(\overline{S}.Z_i) = \mathbf{C}_{\overline{S}}[i,i].\]


Figure \ref{fig:var-estimate} shows the  fit of the predictions by this approach to the observed  standard deviations, for different numbers of clusters and dimensions and for data with different average correlations. We note that the fit is very good, especially for lower dimensionalities, which we generally expect to see in practical skyline queries.

\end{document}